\documentclass[mnsc,nonblindrev]{informs3_hide}

\OneAndAHalfSpacedXI %mnsc
\usepackage{hyperref}
\usepackage{cleveref}
\crefname{subsection}{subsection}{subsections}
\usepackage{tikz}
\usetikzlibrary{positioning}
\usepackage{bbm}
\usepackage{nicefrac}
\usepackage{cancel}

\newcommand{\bI}{\mathbbm{1}}

\newcommand{\bR}{\mathbb{R}}

\newcommand{\OPT}{\mathsf{OPT}}
\newcommand{\Rev}{\mathsf{REV}}
\newcommand{\echar}{\mathsf{end}}
\newcommand{\pre}{\mathsf{pre}}

\newcommand{\cL}{\mathcal{L}}
\newcommand{\cM}{\mathcal{M}}
\newcommand{\cT}{\mathcal{T}}
\newcommand{\vu}{\mathbf{u}}
\newcommand{\vx}{\mathbf{x}}
\newcommand{\vZero}{\mathbf{0}}
\newcommand{\vell}{\boldsymbol{\ell}}
\newcommand{\vcL}{\boldsymbol{\cL}}

\newcommand{\hordist}{2.5}
\newcommand{\vertdist}{1.5}

\usepackage{natbib}
 \bibpunct[, ]{(}{)}{,}{a}{}{,}%
 \def\bibfont{\small}%
\TheoremsNumberedThrough     % Preferred (Theorem 1, Lemma 1, Theorem 2)
\EquationsNumberedThrough    % Default: (1), (2), ...
\MANUSCRIPTNO{MS-RMA-21-01875.R2} % When the article is logged in and DOI assigned to it,
\begin{document}
%%%%%%%%%%%%%%%%

%\setcounter{page}{1}
%\DOI{10.1287/xxxx.0000.0000}%

\RUNAUTHOR{Ma}
\RUNTITLE{When is Assortment Optimization Optimal?}

% \TITLE{Bundling Information Goods of Decreasing Value}
\TITLE{When is Assortment Optimization Optimal?}

\ARTICLEAUTHORS{%
\AUTHOR{Will Ma}
\AFF{Graduate School of Business, Columbia University, New York, NY 10027, \EMAIL{wm2428@gsb.columbia.edu}}
%\AUTHOR{Author2}
%\AFF{Author2 affiliation, \EMAIL{}, \URL{}}
} % end of the block

\ABSTRACT{%

Assortment optimization concerns the problem of selling items with fixed prices to a buyer who will purchase at most one. Typically, retailers select a subset of items, corresponding to an ``assortment'' of brands to carry, and make each selected item available for purchase at its brand-recommended price. Despite the tremendous importance in practice, the best method for selling these fixed-price items is not well understood, as retailers have begun experimenting with making certain items available only through a lottery.

In this paper we analyze the maximum possible revenue that can be earned in this setting, given that the buyer's preference is private but drawn from a known distribution. In particular, we introduce a Bayesian mechanism design problem where the buyer has a random ranking over fixed-price items and an outside option, and the seller optimizes a (randomized) allocation of up to one item. We show that allocations corresponding to assortments are suboptimal in general, but under many commonly-studied Bayesian priors for buyer rankings such as the MNL and Markov Chain choice models, assortments are in fact optimal. Therefore, this large literature on assortment optimization has much greater significance than appreciated before---it is not only computing optimal assortments; it is computing the \textit{economic limit of the seller's revenue} for selling these fixed-price substitute items.

We derive several further results---a more general sufficient condition for assortments being optimal that captures choice models beyond Markov Chain, a proof that Nested Logit choice models cannot be captured by Markov Chain but can to some extent be captured by our condition, and suboptimality gaps for assortments when our condition does not hold. Finally, we show that our mechanism design problem provides the tightest-known LP relaxation for assortment optimization under the ranking distribution model.

}%

%\KEYWORDS{}

%\KEYWORDS{}
\HISTORY{This version from May 20th, 2022.  Forthcoming in \textit{Management Science}.}

\maketitle
%%%%%%%%%%%%%%%%%%%%%%%%%%%%%%%%%%%%%%%%%%%%%%%%%%%%%%%%%%%%%%%%%%%%%%

% Samples of sectioning (and labeling) in MSOM
% NOTE: (1) \section and \subsection do NOT end with a period
%       (2) \subsubsection and lower need end punctuation
%       (3) capitalization is as shown (title style).
%
%\section{Introduction.}\label{intro} %%1.
%\subsection{Duality and the Classical EOQ Problem.}\label{class-EOQ} %% 1.1.
%\subsection{Outline.}\label{outline1} %% 1.2.
%\subsubsection{Cyclic Schedules for the General Deterministic SMDP.}
%  \label{cyclic-schedules} %% 1.2.1
%\section{Problem Description.}\label{problemdescription} %% 2.

% Text of your paper here

\section{Introduction}

A ubiquitous question faced by online and brick-and-mortar retailers alike is which \textit{assortment} of products to carry.
On one hand, carrying a wide variety of products allows the retailer to satisfy more customers' needs, capturing greater market share.
On the other hand, carrying cheap brands can cannibalize the sales of high-end brands, hurting the retailer's margins.
Correctly designing the assortment of products has a tremendous impact on a retailer's bottom line \citep{kok2008assortment}.

This well-studied problem can be formalized as follows.
There is a universe of substitute products, and each product $j$ has a fixed price $r_j$ that has been pre-determined by the brand or the competition.
The buyer has an a-priori \textit{ranking} for buying those products at those prices, which including a ``no-purchase'' option represented by $j=0$ with $r_j=0$.
After seeing a subset, or assortment, of products $S$, the buyer chooses their most-preferred option among $S\cup\{0\}$, paying the associated fixed price.
%We note that this ranking model captures the aforementioned tradeoff between increasing market share vs.\ reducing cannibalization, and
%The ranking can be interpreted as the products sorted in decreasing order of \textit{utility} $v_j-r_j$, where $v_j$ denotes some underlying cardinal valuation for product $j$.
The \textit{assortment optimization} problem is then: given exogenous prices $r_j$ and a distribution of rankings for the buyer, compute an $S$ to offer that maximizes expected revenue.

Our paper asks---can the seller \textit{go beyond} the revenue of the optimal assortment?
In practice, some designer fashion retailers\footnote{An example is End Clothing; see their weekly launches on \texttt{endclothing.com}.} have begun making select products available only through randomized lotteries.
These lotteries occur for new products being launched, and users who sign up agree that if they ``win'' a product, they will be charged the fixed price of that product.
These lotteries do not discount the price (important for protecting the brand value of the high-end products); therefore, they can be directly compared to assortments, and in fact can earn strictly more revenue.
%Perhaps calling it “fixed-price” lottery is confusing.  Actually, in End’s lotteries the _amount paid by the buyer_ is precisely NOT fixed.  Whereas in traditional lotteries, the amount paid by the buyer IS fixed.  The key here is that the amount paid is perfectly coupled with the realization of the item bought.

\begin{example}[General Suboptimality of Assortments] \label{eg:initial}
There are four items $A,B,C,D$ with fixed prices $r_A=2,r_B=r_C=r_D=1$.
The ranking is uniformly drawn from $\{(BA),(CA),(DA),(CB),(DB),(DC)\}$, where we have expressed each ranking as a \textit{list} of items sorted in decreasing order of preference, omitting the items ranked below 0.
On this example, it is easy to check that one cannot do better with an assortment than offering $\{A,B\}$ or $\{A,B,C\}$, both of which have expected revenue \textbf{7/6}.
On the other hand, consider a \textit{top-$k$ lottery}, in which the buyer submits up to $k$ distinct items,
and then randomly receives one of them with probability $1/k$ each
(if the buyer reports $s$ items where $s<k$, then they receive option 0 with probability $1-s/k$).
It can be checked that a top-2 lottery on this example has expected revenue
\begin{align} \label{eqn:initialEgCalc}
\frac{1.5+1.5+1.5+1+1+1}{6}=\textbf{7.5/6},
\end{align}
greater than that of any assortment.
To explain~\eqref{eqn:initialEgCalc}, note that the buyer will always end up submitting 2 items, making a purchase w.p.~(with probability) 1, and paying an average of 1.5 if their list contains $A$ and paying 1 otherwise.
In essence, the top-2 lottery has allowed the retailer to capture full market share while reducing the cannibalized sales of the high-end item $A$.\Halmos
\end{example}

\Cref{eg:initial} describes a specific ``top-$k$'' lottery whose revenue is higher than that of any assortment.
We now use mechanism design to capture the maximum revenue that can be earned through \textit{any} lottery or selling method, when the buyer behaves strategically given their private ranking.

\subsection{Formulation of Mechanism Design Problem} \label{sec:introMDDef}

Note that the tension in our problem lies in the mismatch between the seller's preferences (given by the fixed values $r_j$) and the buyer's preferences (given by their random ranking).
The seller cannot simply dictate a product $j$ with a high value of $r_j$ onto the buyer since the buyer's ranking is private and contains an outside option 0.
In order to elicit this private information, the seller can design a direct mechanism, which asks the buyer to report a ranking, and then (randomly) assigns them an item to purchase (possibly 0) based on this report.

The buyer should be incentivized to truthfully report their ranking and ``participate'' in the mechanism, i.e.\ buy whatever item assigned.
To define what this means, we denote the buyer's ranking using an ordinal list $\ell=(\ell_1,\ldots,\ell_s)$ that has been truncated at the number of items $s$ preferred over 0, with the ordering of items not on the list ignored.
We impose that truthfully participating in the mechanism must maximize the buyer's probability of receiving one of their $k$ most-preferred items, \textit{simultaneously}
for all $k=1,\ldots,s+1$.
Since the buyer's $s+1$'st most-preferred item is 0, which they can get w.p.~1 by not participating,
the mechanism under this definition must w.p.~1 assign the buyer either an item on their list or item 0.
In other words, \textit{ex-post individual rationality} is automatically imposed on the mechanism.

This is a standard definition of randomized incentive-compatibility for ordinal preferences, but can be seen as fairly strong---it equivalently imposes that for \textit{any} underlying cardinal utilities the buyer may have, truthfully reporting the induced ordinal list must guarantee that their expected utility is maximized.
One could instead ask the buyer to directly report their cardinal utilities, or consider other mechanism design formulations, which we discuss in \textbf{\Cref{sec:modelingAss}}.
Nonetheless, we adopt this definition because it allows us to focus purely on list distributions, not requiring the buyer to possess or know their cardinal utilities, and it still leaves room for a rich class of ordinal mechanisms as we discuss below.
Hereafter we refer to mechanisms satisfying this incentive-compatibility constraint and the implied ex-post individual rationality as ``IC''.

Examples of IC mechanisms include top-$k$ lotteries, generalized top-$k$ lotteries where each item can be submitted a different number of times, \textit{budget-additive mechanisms}\footnote{In \Cref{sec:budgetAdditive}, we establish a correspondence between IC allocations and the rich space of monotone submodular functions.  Budget-additive mechanisms are then the allocations induced by budget-additive submodular functions.}, and many others.
An assortment is also an IC mechanism, with an extremely simple implementation---show the buyer a subset $S$ and let them choose.
In analyzing when assortments are optimal, we are analyzing when this simple implementation is as powerful as eliciting preferences.

\subsection{Description of Results} \label{sec:outlineResults}

Our main technique for comparing assortments to general mechanisms is to upper-bound the latter using an optimal stopping problem, through \textbf{\Cref{sec:mechDesignGen,sec:seqImp,sec:monStopPol}}.
Policies for this optimal stopping problem must further satisfy monotonicity constraints on their decisions across different sample paths, which come from the incentive-compatibility constraints of the mechanism.

We show in \textbf{\Cref{sec:markovChainMainResult}} that mechanisms are no better than assortments whenever the buyer's list distribution is generated by a \textit{Markov Chain} choice model, which includes the \textit{Multi-Nomial Logit (MNL)} choice model.
The proof is quite simple given our upper bound, as we show that even the best optimal stopping policy without monotonicity constraints corresponds to an assortment.

We then show that assortments are optimal under a more general sufficient condition, motivated as follows.
One way to view a Markov Chain choice model is---given any two potential prefixes for the buyer's list that end with the \textit{same} item, the conditional probabilities of the buyer choosing any item from any assortment disjoint from these prefixes are \textit{equal}.
As a concrete example, if we take the customers whose lists begin with $(CB)$, and compare them to the customers whose lists begin with $(B)$, then these two groups have the same choice probabilities on all assortments not containing items $B$ or $C$.
Our sufficient condition relaxes this property satisfied by Markov Chain choice models in the following way---if one of the two prefixes \textit{contains the other}, then the former is allowed to have \textit{higher} conditional choice probabilities.
On this example, since prefix $(CB)$ contains $(B)$, the probability of the customer purchasing e.g.\ item $A$ from assortment $\{A\}$ is allowed to be higher conditional on prefix $(CB)$ rather than $(B)$.
Note that \Cref{eg:initial}, in which assortments were not optimal, goes directly against this condition---there the conditional probability of the customer purchasing item $A$ was higher starting from $(B)$ than $(CB)$, being 100\% instead of 0\%.
Our sufficient condition is formalized in \textbf{\Cref{sec:condition}}, and in 
\textbf{\Cref{sec:complexPf}}
we explain and prove why it causes assortments to be optimal, which is heavily based on our monotonicity constraint on the optimal stopping policy. 

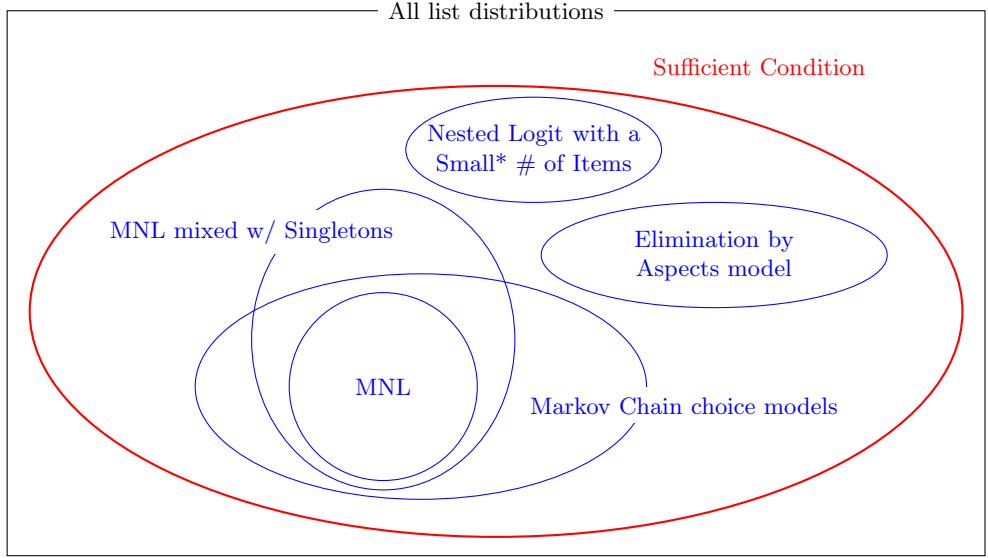
\begin{figure}
\caption{Hierarchy of choice models for which assortments are optimal
%, including references which emphasize the assortment optimization problem on those choice models.
}
\label{fig:hierarchy}
\begin{center}
\begin{tikzpicture}[scale=1]
\draw (-1,0) rectangle (12,7.25);
\node[fill=white] at (5.5,7.25) {\footnotesize All list distributions};

\draw[red, thick] (5.5,3.25) ellipse (6.2 and 3);
\node[red] at (9,6.5) {\footnotesize Sufficient Condition};

%\filldraw[red] (3.75,6.75)node[left]{\footnotesize \shortstack{assortments are \textbf{suboptimal}\\ (\Cref{eg:initial})}} circle (0.07);

\draw[blue] (4.5,2.25) ellipse (3 and 1.5);
\node[blue,fill=white] at (8,2) {\footnotesize Markov Chain choice models};

\draw[blue] (8.4,4)node{\footnotesize \shortstack{Elimination by\\ Aspects model}} ellipse (2.3 and 0.7);

\draw[blue] (6,5.4)node{\footnotesize \shortstack{Nested Logit with a\\ Small* \# of Items}} ellipse (1.7 and 0.7);

\draw[blue] (4,2.25)node{\footnotesize MNL} circle (1.25);

\draw[blue] (4,2.875) ellipse (1.75 and 2);
\node[blue,fill=white] at (2.25,4.3) {\footnotesize MNL mixed w/ Singletons};
\end{tikzpicture}
\end{center}
\footnotesize
*See \Cref{sec:assortment} for the detailed specification.
\normalsize
\end{figure}

Knowing that assortments are optimal under this sufficient condition allows us to capture additional distributions, such as a mixture of MNL with singleton lists (a choice model of interest as demonstrated by \citet{cao2020revenue}), Tversky's Elimination by Aspects model \citep{tversky1972elimination}, and special cases of Nested Logit with a small number of items.
Details can be found in \textbf{\Cref{sec:assortment}}.
We emphasize that none of these distributions could have been captured by a Markov Chain alone, as illustrated in \Cref{fig:hierarchy}.
In fact, to our knowledge our paper is the first\footnote{
It has been pointed out to us that this result was concurrently derived by \citet{li2022non}, who also present further bounds on the inapproximability of Nested Logit using Markov Chain, and vice versa.
} to show that a Nested Logit choice model cannot be captured by any Markov Chain.
Given this fact, our sufficient condition could also be of general interest, since it unifies (at least for a small number of items) these two choice models that have been studied frequently but disparately.

Finally, we investigate the power that mechanisms can have over assortments for general list distributions.
First, we show in \textbf{\Cref{sec:budgetAdditive}} that a simple subclass of mechanisms defined by \textit{budget-additive} submodular functions, which captures all the examples of lotteries given so far, can outperform assortments by a factor of at most $e\approx2.71$.
In other words, assortments are \textit{approximately} optimal if we only compare against simple lotteries.
We also provide a lower bound of $e/2$ for this factor.
We show in \textbf{\Cref{sec:bdedListLength}} that assortments are also approximately optimal if customers are ``picky'', having \textit{short lists} as in the model of \citet{feldman2019assortment}.
In stark contrast, we show in \textbf{\Cref{sec:generalDistNegResult}} that assortments can be \textit{suboptimal by an unbounded factor}, without any restrictions on the lottery or list length.
Finally, we show in \textbf{\Cref{sec:tighterLP}} that our mechanism design problem provides a Linear Programming (LP) relaxation tighter than the state-of-the-art from \citet{bertsimas2019exact}, which is useful for computing optimal assortments.

In the~\namecref{sec:multBuyer} we discuss two extensions---a generalized formulation of our problem with multiple buyers competing for a limited inventory of the items (\textbf{\Cref{sec:multBuyer}}), and a related robust mechanism design formulation where the buyer's utilities are cardinal but only the induced ordinal distribution is known to the seller (\textbf{\Cref{sec:robust}}).

\subsection{Relation to Existing Areas of Literature}

\subsubsection*{Assortment optimization.}
We believe our paper brings significant new understanding to the area of assortment optimization.
First, we show that better selling methods \textit{could} exist out there, even in the form of a simple top-2 lottery.
Yet, this is not possible for commonly-used choice models such as Markov Chain and MNL.
That is, once a firm has settled on using MNL to make decisions (as common in practice; see \citet{feldman2021customer}), it no longer needs to look beyond assortments.

More broadly, our work shows that the vast literature on computing optimal assortments for MNL/Markov Chain has much greater significance than appreciated before---they are actually finding the maximum revenue among \textit{all possible selling methods}, not just the specific selling method of offering an assortment.
Selections from this literature on MNL/Markov Chain have analyzed the general Markov Chain choice model \citep{blanchet2016markov,feldman2017revenue,desir2020constrained,ma2020revenue}, the special case of MNL \citep{talluri2004revenue,rusmevichientong2010dynamic}, the special case of ``Out-trees/In-trees'' \citep{honhon2012optimal}, and the special case of the attention-based consideration set model \citep{gallego2017attention}.
Our broader significance also applies to the assortment optimization algorithm of \citet{cao2020revenue} for MNL mixed with singletons, which satisfies our sufficient condition.
We note that our result about optimality of assortments under the sufficient condition is non-constructive---so although it also captures a Markov Chain mixed with singletons, which is an open problem on the frontier of \Cref{fig:hierarchy}, our work does not imply an assortment optimization algorithm for this choice model.
Nonetheless, our tighter LP relaxation should be able to speed up state-of-the-art assortment optimization algorithms from \citet{bertsimas2019exact}, which we believe would be interesting to explore in future work.

\subsubsection*{Bayesian mechanism design.}
Like our work, papers in this area \citep[see][]{chawla2014bayesian} study the maximum revenue obtainable from selling a set of items, given Bayesian priors on the buyers.
One main difference is that throughout these papers, pricing is a decision variable, and importantly, lotteries can be \textit{discriminatively} priced.
The fact that lotteries can increase revenue at all when item pricing is fixed, as in our \Cref{eg:initial}, was previously unknown.

The other main difference is that buyers in this area are modeled using cardinal valuation distributions, which are generally incomparable to our ordinal ranking models.
The questions we ask about (sub)optimality of assortments are very much in line with this area though, whose central questions include identifying simple optimal mechanisms under certain conditions \citep[e.g.][]{myerson1981optimal,thanassoulis2004haggling,alaei2013simple,
haghpanah2015reverse,
haghpanah2019pure,bergemann2021optimality}, and bounding the suboptimality of simple mechanisms for different valuation classes/distributions \citep[e.g.][]{chawla2007algorithmic,
chawla2010multi,
chawla2015power,briest2015pricing,babaioff2020simple}.
We believe that relating the results for these valuation classes/distributions to our results for ranking models would be interesting future work.
Many of these papers also allow for multiple buyers; our problem admits a natural generalization of this form, which we describe in \Cref{sec:multBuyer}.

\subsubsection*{Mechanism design without money.}
Like our work, papers in this area \citep[see][]{schummer2007mechanism} design mechanisms that elicit ordinal preferences from agents to allocate a (randomized) outcome, and our
%Problems in this area include social choice (voting, facility location), resource allocation (house allocation, rationing), and matching; our
randomized ordinal incentive-compatibility constraint is inherited from this area (originally appearing in \citet{gibbard1977manipulation}).
However, a main difference in our problem is that the benefit to the designer from each outcome $j$ is given by the explicit value $r_j$, whereas in these problems the designer seeks a socially beneficial outcome whose desirability depends on the preferences of the agents.
These problems only have tension when there are multiple agents, who disagree on the best global outcome or want the same resource for themselves.
By contrast, our problem has \textit{tension even for a single agent}, due to the selfish designer with unaligned interests.
To the best of our knowledge, our single-agent problem has no analogue in this area.

Much of the papers in this area also differ from ours in that they seek mechanisms with desirable properties (e.g.\ Pareto-efficiency, stability, non-dictatorialness, non-bossiness) without optimizing an objective.
Our paper complements a growing literature on applying optimization to matching markets \citep{ashlagi2014improving,arnosti2015lottery,ashlagi2016optimal,bodoh2020optimizing,feigenbaum2020dynamic,celebi2020priority,shi2021optimal} that designs policy levers often motivated by the application of school choice.
Although this literature does optimization with a Bayesian prior, it again differs in that there are multiple (in fact usually a continuum of) agents who desire the same resource.
We mention that \citet{shi2021optimal} also connects assortment optimization to mechanism design, using it as a subroutine for optimizing student priorities.
%Another relevant point is that he proposes a definition of cardinal utility for Markov Chain choice models, which could be used to formulate alternate IC constraints in our problem.
Finally, we should note that optimization of mechanisms without money nor a prior can also be studied, under the framework of \citet{procaccia2013approximate}.
%The other main difference in our problem is that we are optimizing a clearly-defined objective over a prior.
%By contrast, mechanisms in this area are prior-free, and often characterized as the \textit{unique feasible solution} satisfying some conditions---e.g.\ dictatorships for voting \citep{gibbard1973manipulation,satterthwaite1975strategy}; Gale's top-trading-cycles mechanism for house allocation \citep{shapley1974cores,roth1982incentive,ma1994strategy}; the uniform allocation rule for rationing \citep{sprumont1991division}; serial dictatorships for one-sided allocation \citep{svensson1999strategy}.

\subsubsection*{Further related work.}
Connections between assortment optimization and mechanism design have also been previously made by \cite{ma2020revenue}.
However, he focuses solely on deterministic mechanisms, in which case the single-buyer problem is equivalent to assortment optimization, and the non-triviality comes from having multiple buyers in a \textit{service-constrained} environment (as introduced in \citet{alaei2013simple}).
By contrast, we focus on a single buyer and randomized mechanisms, which to our surprise, has not been studied until now.  Other papers that introduce incentive constraints to assortment optimization include \citet{saban2021procurement,balseiro2021incentive}.
They study different problems, in which private information is held by the supply side.

Finally, our work is part of a burgeoning literature that models more sophisticated levers which a firm can use to increase revenue beyond assortments.
This includes deliberately decreasing the attractiveness of certain products \citep{berbeglia2021refined},
increasing the attractiveness of products through advertising \citep{wang2021advertising},
and comparing the revenue of the optimal assortment to that of personalized assortments \citep{el2021joint} or a clairvoyant \citep{gallego2021limits}.
We note that for MNL, \citet{gallego2021limits} show assortments to be within a factor of 2 of the clairvoyant who sees the buyer's realization; we show that this factor improves to 1 (i.e.\ assortments are optimal) if we compare to the revenue of lotteries instead.
We should also contrast our work with \citet{wang2020randomized}, which propose \textit{randomized} assortments as a lever to be robust against estimation error.  This is different than offering \textit{lotteries}, as studied in our paper.  When implemented in practice, our mechanisms, e.g.\ top-$k$ lotteries, are more similar to the opaque products described in \citet{elmachtoub2015retailing,elmachtoub2021power}.

\section{Problem Definition and Sequence of Relaxations} \label{sec:probDefn}

There is a seller with $n$ items, denoted by $N=\{1,\ldots,n\}$.
Each item $j\in N$ must be sold at its commonly-known, exogenously-given price $r_j\ge0$.

There is a single buyer who is willing to purchase at most one item from $N$.
The private type of the buyer is given by an ordered subset of $N$, indicating the items they are willing to purchase, in decreasing order of preference.
%For example, if the type is $(1,3)$, then the buyer's first choice is to pay $r_1$ for item~1, second choice is to pay $r_3$ for item~3, and third choice is to purchase nothing at all; the buyer is not willing to consider other items.
We will refer to the type as a \textit{list} $\ell$, which is assumed to be a strictly ordered subset of $N$.

The buyer's type is drawn from a commonly-known \textit{list distribution}, given by a support $\cL$ of ordered subsets and a probability $p(\ell)$ for each list $\ell\in\cL$.
We will refer to the list distribution using $p$, which satisfies $\sum_{\ell\in\cL}p(\ell)=1$.

\begin{definition}[List Notation] \label{def:opsOrderdList}
We define the following notation on lists $\ell$:
\begin{itemize}
\item Let $|\ell|$ denote the length of a list $\ell$.  For example, if $\ell=(1,3)$, then $|\ell|=2$.
\item For $k=1,\ldots,|\ell|$, let $\ell_k$ denote the $k$'th item on $\ell$, and $\ell_{1:k}$ denote the sublist formed by the first $k$ items given in order.  For example, if $\ell=(1,3,2)$, then $\ell_1=1$, and $\ell_{1:2}=(1,3)$.
\item If we apply set-specific operations to a list or sublist, then we are referring to the set of items on that list or sublist.  For example, if $\ell=(1,3,2)$, then $\ell\cap\{2,3\}=\{2,3\}$ and $\ell_{1:2}\cap\{2,3\}=\{3\}$.
\end{itemize}
\end{definition}

\begin{definition}[Assortment $S$, $\Rev\lbrack S\rbrack$, and $\OPT^S$]
An \textit{assortment} $S$ is a subset of items which is offered to the buyer.
For each $j\in S$, the buyer chooses to purchase item $j$ if and only if $j$ appears on their randomly-drawn list and is earliest (i.e.\ most preferred) among items in $S$ on their list.
We let $\Pr[j\succeq S_0]$ denote the probability of this event, suppressing the dependence on $p$ but adding a subscript ``0'' as a reminder that $j$ has to be preferred over the ``no-purchase'' option.
The expected revenue collected by an assortment $S$ can then be expressed as
\begin{align} \label{eqn:asstRev}
\Rev[S]:=\sum_{j\in S}r_j\Pr[j\succeq S_0].
\end{align}
Given item prices $r_1,\ldots,r_n$ and list distribution $p$, the \textit{assortment optimization} problem is to find a subset $S\subseteq N$ which maximizes~\eqref{eqn:asstRev}.  We let $\OPT^S$ denote its optimal objective value.
\end{definition}

\subsection{Mechanism Design Generalization of Assortment Optimization} \label{sec:mechDesignGen}

A (direct revelation) mechanism takes in a reported list, and then possibly using randomness, assigns up to one item for the buyer to purchase.
As discussed in \Cref{sec:introMDDef}, we impose the incentive constraint that for any $k$, truthfully participating in the mechanism must maximize the buyer's probability of receiving one of their $k$ most-preferred items.
This implies that the mechanism cannot assign an item less-preferred than the ``no-purchase'' option.
We hereafter make no distinction between the buyer's reported list vs.\ true list and assume that they will pay the price of any item they are assigned.

\begin{definition}[Mechanism $x$, $\Rev\lbrack x\rbrack$, and $\OPT^x$] \label{def:x}
For any list $\ell$ and item $j\in\ell$, let $x_j(\ell)$ specify the probability of the mechanism assigning $j$ given report $\ell$.  Our mechanism design problem is then formulated by the following LP:
\begin{align}
\max &&\sum_{\ell\in\cL}p(\ell)\sum_{j\in\ell}r_jx_j(\ell) \nonumber
\\ \text{s.t.} &&\sum_{j\in\ell_{1:k}}x_j(\ell) &\ge\sum_{j\in\ell_{1:k}
\cap\ell'
}x_j(\ell') &\forall\ell\in\cL,k\le|\ell|,\ell'\in\cL \label{constr:IC}
\\ &&\sum_{j\in\ell}x_j(\ell) &\le1 &\forall\ell\in\cL \label{constr:sumToOne}
\\ &&x_j(\ell) &\ge0 &\forall\ell\in\cL,j\in\ell \label{constr:nonNeg}
\end{align}
We refer to a \textit{mechanism} using $x$ and say that it is \textit{IC} if it feasibly satisfies~\eqref{constr:IC}--\eqref{constr:nonNeg}.
We let $\Rev[x]$ denote the objective at a given $x$ and use $\OPT^x$ to refer to the optimal objective value of the LP.
\end{definition}

Note that the mechanism only needs to be defined on lists $\ell$ in the buyer's support $\cL$, and the feasible region depends only on $\cL$, with the probabilities $p(\ell)$ appearing in the objective.
Constraints~\eqref{constr:sumToOne} ensure that the buyer is assigned at most one item, with $1-\sum_{j\in\ell}x_j(\ell)$ understood as the probability of the buyer receiving nothing after reporting $\ell$.

\begin{proposition}[Relating Assortments $S$ to Mechanisms $x$] \label{prop:S_to_x}
For any assortment $S\subseteq N$,
%setting
\begin{align} \label{eqn:setXFromS}
x_j(\ell)&=
\begin{cases}
1, &\text{if $j$ appears before any other item in $S$ on $\ell$} \\
0, &\text{otherwise}
\end{cases}
&\forall\ell\in\cL,j\in\ell
\end{align}
defines an IC mechanism with $\Rev[x]=\Rev[S]$.  Therefore, $\OPT^S\le\OPT^x$.
On the other hand, any deterministic IC mechanism can be described by~\eqref{eqn:setXFromS} for some corresponding assortment $S$.
\end{proposition}

\Cref{prop:S_to_x} is proved in \Cref{sec:defPfsRelaxations},
but we only need the forward direction, which is straight-forward, to proceed.
The statement as a whole is our analogue of the fact that in classical single-buyer auctions, a mechanism is deterministic if and only if it corresponds to posted pricing.
Here, a deterministic mechanism corresponds to an assortment.

\subsection{Discussion of Modeling Assumptions} \label{sec:modelingAss}

Before proceeding, we discuss three modeling assumptions made in our problem formulation.

First, we have implicitly assumed the buyer to have an intrinsic, privately-known ranking (or equivalently, utility) for every item in the universe $N$.
In practice, the buyer
may need to search to discover their own ranking, or their
utilities could be influenced by the assortment shown.
Consequently, there are limits to using rankings/utilities to model customer behavior \citep{jagabathula2018limit}, and there exist models out there that capture search \citep{wang2018impact} or inconsistent utilities \citep{chen2020decision,chen2019use}.
Nonetheless, for simplicity we assume the buyer's behavior to be fully governed by some ranking/utilities, all throughout this paper.
This is also a pervasive assumption in mechanism design.

Second, having assumed the ranking/utility model, we impose on the mechanism a fairly strong IC constraint.
We provide some reasons for this modeling choice below.
Alternate models, e.g. one where the buyer reports cardinal utilities and the mechanism only needs to maximize the buyer's expectation, could certainly be interesting future work.
\begin{enumerate}
\item Mechanisms in the less-constrained model may not be ex-post individually rational, which we believe to be necessary in practice, since otherwise the buyer could just not pay for their assigned item after seeing an unfavorable realization. Thus, individual rationality would have to be imposed w.p.~1, which is inconsistent with the utility maximization that is only imposed in expectation.
\item Relatedly, it appears difficult for assortments to be competitive with mechanisms in alternate models where ex-post individual rationality may not hold.\footnote{
In models without ex-post individual rationality, mechanisms can offer items that are \textit{never} ranked above 0 as part of a lottery, and sell them with positive probability.
Meanwhile, an assortment can never sell such items.
Thus, whenever such an item has an unbounded price, the gap between lotteries and assortments would be unbounded.
} 
\item It is cognitively easier for buyers to report an ordinal ranking than decide on cardinal utilities.\footnote{
Although the buyer's preference under uncertainty can be robustly described as maximizing the expectation of some underlying von Neumann-Morgenstern utilities, it is difficult for the buyer to know what these are.
}
This also allows us to make a characterization based purely on the ranking distribution, allowing us to capture choice models that are defined purely combinatorially (e.g.\ Markov Chain).
\item In \Cref{sec:robust} we discuss how our IC constraint leads to a ``robust'' mechanism.
\end{enumerate}

%Additionally, in \Cref{sec:robust} we show how our formulation provides a feasible,
%but not necessarily optimal,
%solution to a \textit{robust} mechanism design problem where.
%in which the buyer's true utility is cardinal but only the induced ordinal distribution is known.

Finally, we note that the optimality of posting a single assortment in our model suggests that the seller cannot benefit by showing different items to different buyers.
However, it is not to suggest that an online retailer cannot benefit by showing \textit{personalized assortments} to different users based on available information such as IP address.
Crucially, our model assumes the buyers to be indistinguishable to the seller, except through preferences that they strategically disclose.
If buyers can be segmented based on IP address, then our result should be applied at the segment level, determining for each indistinguishable segment whether the online retailer should show a single assortment or ask the buyers from that segment to report preferences.
%Papers that ask the different question of ``what is the power of personalization'' are referenced in \Cref{sec:furtherRW}.

\subsection{Sequential Implementation of IC Mechanisms} \label{sec:seqImp}

\Cref{prop:S_to_x} showed that $\OPT^S\le\OPT^x$, and in this paper we are interested in identifying conditions under which $\OPT^S=\OPT^x$.
To aid in upper-bounding $\OPT^x$, our goal in this \namecref{sec:seqImp} is to describe every IC mechanism as a sequential exchange between the buyer and seller.

To begin, we rewrite $\Rev[x]$ as $$\sum_{\ell\in\cL}p(\ell)\sum_{k=1}^{|\ell|}r_{\ell_k}x_{\ell_k}(\ell),$$ where we are now summing over the items on list $\ell$ in decreasing order of preference.
We aim to express each variable $x_{\ell_k}(\ell)$ as a difference in an increasing set function.

\begin{definition}[Function $f$, $\Rev\lbrack f\rbrack$, and $\OPT^f$]
For an increasing set function $f:2^N\to[0,1]$, i.e.\ one which satisfies $f(S)\le f(S')$ for all $S\subseteq S'\in 2^N$, define its ``revenue'' as
$$
\Rev[f]:=\sum_{\ell\in\cL}p(\ell)\sum_{k=1}^{|\ell|}r_{\ell_k}(f(\ell_{1:k})-f(\ell_{1:k-1})).
$$
Let $\OPT^f$ denote the supremum value of $\Rev[f]$ over all increasing set functions $f:2^N\to[0,1]$.
\end{definition}

\begin{proposition}[Relating Mechanisms $x$ to Functions $f$] \label{prop:x_to_f}
For any IC mechanism $x$,
%setting
\begin{align}
f(S) &=\max_{\ell\in\cL}\sum_{j\in S\cap\ell}x_j(\ell) &\forall S\subseteq N \label{eqn:fFromx}
\end{align}
defines an increasing set function $f:2^N\to[0,1]$ with
\begin{align} \label{eqn:diffF}
x_{\ell_k}(\ell) &=f(\ell_{1:k})-f(\ell_{1:k-1}) &\forall\ell\in\cL,k=1,\ldots,|\ell|
\end{align}
and $\Rev[f]=\Rev[x]$.  Therefore, $\OPT^x\le\OPT^f$.
\end{proposition}

\Cref{prop:x_to_f} is proved in \Cref{sec:defPfsRelaxations}, but is easy to understand in the following way.
Given any mechanism $x$, defining $f(S)$ as in~\eqref{eqn:fFromx} represents the maximum probability the buyer could have of being assigned an item in $S$.
Clearly, this probability is increasing in $S$ and takes values in $[0,1]$.
Finally, the IC constraints~\eqref{constr:IC} can then be used to establish identity~\eqref{eqn:diffF}.

An important consequence of \Cref{prop:x_to_f} is that every IC mechanism $x$ can be implemented by the following sequential procedure.
First, the buyer draws their most-preferred item $\ell_1$ and reports it to the mechanism.
The mechanism assigns this item w.p.\ $f(\{\ell_1\})$, with $f$ defined according to~\eqref{eqn:fFromx}, in which case this probability equals $x_{\ell_1}(\ell)$ (since $f(\emptyset)=0$).
If $\ell_1$ is not assigned, then the buyer draws and reports their second-most-preferred item $\ell_2$.
The mechanism can assign this item according to an independent coin flip w.p.\ $\frac{f(\{\ell_1,\ell_2\})-f(\{\ell_1\})}{1-f(\{\ell_1\})}$, to ensure that the ex-ante probability of $\ell_2$ being assigned is $f(\{\ell_1,\ell_2\})-f(\{\ell_1\})$, equaling $x_{\ell_2}(\ell)$ by~\eqref{eqn:diffF}.
This process repeats until either an item is assigned, or the buyer's list terminates when they attempt to draw their next-most-preferred item, in which case there is no recourse to sell the buyer a previously-reported item.

As emphasized by this sequential description, the probability of a list $\ell$ receiving its $k$'th-most-preferred item under an IC mechanism can depend only on the first $k$ items on the list, and nothing afterward.
We will refer to these initial items as a \textit{prefix}.

\begin{definition}[Prefix Notation]
A \textit{prefix} $\rho$ is an ordered subset of $N$, referring to all lists with the same (ordered) beginning as $\rho$.
The probability of a prefix $\rho$ occurring is
\begin{align*}
\Pr[\rho]:=\sum_{\ell\in\cL:\ell_{1:|\rho|}=\rho}p(\ell)
\end{align*}
where we overload the $\Pr[\cdot]$ operator to indicate the probability that the randomly-drawn list begins with $\rho$.
We say that a prefix $\rho$ is \textit{realizable} if $\Pr[\rho]>0$.
Any operations we defined for ordered lists $\ell$ in \Cref{def:opsOrderdList} will also apply to prefixes $\rho$.
We let $\rho_{\echar}:=\rho_{|\rho|}$ denote the \textit{endpoint} item of $\rho$, and
$\rho_{\pre}:=\rho_{1:|\rho|-1}$ denote the sequence of items on $\rho$ before its endpoint.
\end{definition}

By exchanging sums in the original definition of $\Rev[f]$ and letting $\rho$ denote $\ell_{1:k}$, we can now simplify the revenue of function $f$ as
\begin{align}
\Rev[f]
&=\sum_{\rho}r_{\rho_{\echar}}(f(\rho)-f(\rho_{\pre}))\left(\sum_{\ell\in\cL:\ell_{1:|\rho|}=\rho}p(\ell)\right) \nonumber
\\ &=\sum_{\rho}r_{\rho_{\echar}}(f(\rho)-f(\rho_{\pre}))\Pr[\rho]. \label{def:revFAgain}
\end{align}

\subsection{Relaxing to Monotone Stopping Policies} \label{sec:monStopPol}

Through \Cref{prop:x_to_f}, one can upper-bound $\OPT^x$ by the maximum value of expression~\eqref{def:revFAgain} over increasing set functions $f:2^N\to[0,1]$.
However, this relaxed optimization problem is still difficult to analyze, because the value of $f$ on any set $S$ has consequences for many different prefixes $\rho$ in the sum in~\eqref{def:revFAgain}.
Therefore, in this \namecref{sec:monStopPol} we re-interpret~\eqref{def:revFAgain} as the reward collected by an optimal stopping policy, and introduce a relaxed ``cross-path'' constraint under which the optimal stopping decisions can be analyzed.
We now describe the optimal stopping problem, first defining a tree diagram which probabilistically generates the buyer's list.

\begin{definition}[Tree Diagram, Transition Probabilities $q$] \label{def:qAndTree}
Consider a list distribution which is given by its realizable prefixes $\rho$ and their positive probabilities $\Pr[\rho]$.
We define its \textit{tree diagram} as follows.
There is a \textit{node} for every realizable prefix $\rho$, \textit{labeled} by the endpoint $\rho_{\echar}$.
There is an additional node labeled ``root'', and an imaginary ``terminal'' node 0.
For every realizable $\rho$, there is an arc from the node for prefix $\rho_{\pre}$ to the node for $\rho$, with transition probability
$$
q(\rho):=\Pr[\rho]/\Pr[\rho_{\pre}]
$$
where $\rho_{\pre}$ is understood to be ``root'', with $\Pr[\rho_{\pre}]=1$, if $|\rho|=1$.
Transition probability $q(\rho)$ can be interpreted as the likelihood of the next item being $\rho_{\echar}$ conditional on a random list starting with $\rho_{\pre}$.
Note that the sum of outgoing transition probabilities from the node for any $\rho$ is $\sum_{\rho':\Pr[\rho']>0,\rho'_{\pre}=\rho}\Pr[\rho']/\Pr[\rho]$, which is upper-bounded by 1.
\end{definition}

\begin{definition}[Optimal Stopping Problem] \label{def:optStopProb}
Given the tree diagram for a list distribution, define the following \textit{optimal stopping problem}.
Starting from ``root'', the player transitions along the tree diagram according to the outgoing probabilities from the current node.
Upon arriving at a node labeled $j$, the player can either stop to end the game with reward $r_j$, or proceed with the next transition.
The game could also end with 0 reward upon transitioning to the ``terminal'' node, which occurs with probability $1-\sum_{\rho':\Pr[\rho']>0,\rho'_{\pre}=\rho}\Pr[\rho']/\Pr[\rho]$ (which is non-negative) when starting from the node for prefix $\rho$.
\end{definition}

An example of a tree diagram, with labeled transition probabilities, can be found at the beginning of \Cref{sec:pfMain}.
The optimal stopping problem from \Cref{def:optStopProb} based on tree diagrams will be used to upper-bound $\OPT^x$.
However, we do not want to allow for any online stopping policy, i.e.\ the optimal one obtained from dynamic programming, because that would be too loose---an example of this can also be found at the beginning of \Cref{sec:pfMain}.
Therefore, we still maintain a simpler ``monotonicity'' constraint on the stopping decisions across sample paths, dependent on the history.
Importantly though, we do allow the policy to indicate its stopping decisions based on the label $j$ of the current node, which affects the stopping reward $r_j$ (contrast this with the function $f$, which had to be defined on sets $S$, without being able to favor stopping on one item over another).

\begin{definition}[Monotone Stopping Policy $\phi$, $\Rev\lbrack\phi\rbrack$, and $\OPT^{\phi}$]
A \textit{monotone stopping policy} $\phi$ is defined by $n$ boolean-valued functions $\phi_j:2^{N\setminus\{j\}}\to\{0,1\}$, one for each item $j$, which are \textit{increasing}, in that $\phi_j(H)\le\phi_j(H')$ for all $H\subseteq H'\in 2^{N\setminus\{j\}}$.
For each $j\in N$ and (unordered) history $H\subseteq N\setminus\{j\}$, the value $\phi_j(H)\in\{0,1\}$ indicates whether to deterministically stop on a node labeled $j$ after encountering exactly the set of labels $H$ (in any order).
Define the ``revenue'' of $\phi$
\begin{align} \label{eqn:revPhi}
\Rev[\phi]:=&\sum_{\rho}r_{\rho_{\echar}}\Pr[\rho]\left(\phi_{\rho_{\echar}}(\rho_{\pre})\prod_{k=1}^{|\rho|-1}(1-\phi_{\rho_k}(\rho_{1:k-1}))\right).
\end{align}
Let $\OPT^{\phi}$ denote the maximum value of $\Rev[\phi]$ over all monotone stopping policies $\phi$.
\end{definition}

Assortments $S$ corresponds to monotone stopping policies for which \textit{every $\phi_j$ is a constant function}---the all-1 function if $j\in S$, and the all-0 function otherwise.
For a general deterministic stopping policy $\phi$, objective $\Rev[\phi]$ represents the expected reward collected by $\phi$ in the aforementioned optimal stopping problem.
To see this, consider any realizable prefix $\rho$.
Since $\phi$ takes values in $\{0,1\}$, the large parentheses in~\eqref{eqn:revPhi} evaluates to 1 if and only if $\phi_{\rho_{\echar}}(\rho_{\pre})=1$ (i.e.\ $\phi$ would stop on $\rho_{\echar}$ after visiting a sequence of nodes labeled $\rho_1,\ldots,\rho_{k-1}$) and $\phi_{\rho_k}(\rho_{1:k-1})=0$ for all $k<|\rho|$ (i.e.\ $\phi$ would not have stopped earlier in this sequence).
Therefore,~\eqref{eqn:revPhi} counts the contribution from all prefixes $\rho$ which the policy would reach and stop on (collecting reward $r_{\rho_{\echar}}$), thereby equaling the total expected reward of $\phi$.

\begin{proposition}[Relating Functions $f$ to Monotone Stopping Policies $\phi$] \label{prop:f_to_phi}
The optimization problem for $\OPT^f$ always has an optimal solution which is an increasing boolean set function.  Moreover, for any increasing boolean $f:2^N\to\{0,1\}$, setting
\begin{align} \label{eqn:phiFromf}
\phi_j(H) &=f(H\cup\{j\}) &\forall j\in N,H\subseteq N\setminus\{j\}
\end{align}
defines a monotone stopping policy $\phi$ with $\Rev[\phi]=\Rev[f]$.
Therefore, $\OPT^f\le\OPT^{\phi}$.
\end{proposition}

\Cref{prop:f_to_phi} is proved in \Cref{sec:defPfsRelaxations}.
The key observation is that the constraints enforcing $f$ to be increasing describe a totally unimodular system, which allows us to restrict our attention \textit{boolean}-valued functions $f$.
To show $\Rev[\phi]=\Rev[f]$, one has to check that the linear term $f(\rho)-f(\rho_{\pre})$ in~\eqref{def:revFAgain} equals the non-linear expression $f(\rho)\prod_{k=1}^{|\rho|-1}(1-f(\rho_{1:k}))$ when $f$ is boolean.

\section{Assortments are Optimal for Markov Chain choice models} \label{sec:markovChainMainResult}

In \Cref{sec:probDefn} we defined a sequence of relaxed optimization problems satisfying
$
\OPT^S\le\OPT^x\le\OPT^f\le\OPT^{\phi}.
$
In this section we use the relaxation $\OPT^{\phi}$ to show that assortments are optimal for Markov Chain choice models, presenting a refined proof that only works on Markov Chains.
In the subsequent \Cref{sec:pfMain} we derive a different and more general proof based on a sufficient condition.

First we provide a definition\footnote{Markov Chain choice models were originally introduced in \citet{blanchet2016markov}.  Although usually defined in terms of assortment purchase probabilities (i.e.\ given by $\{\Pr[j\succeq S_0]:S\subseteq N,j\in S\}$), a Markov Chain choice model has a natural interpretation as a distribution over lists, as defined in \citet{ma2020revenue}.} of Markov Chain choice models.

\begin{definition}[Markov Chain choice model] \label{def:mc}
There is a Markov Chain with states $N\cup\{0\}$ and a probability $\sigma_{jj'}$ for transitioning
from any state $j\in N$
to any state $j'\in N\cup\{0\}$,
where 0 is a terminal state with no outgoing transitions.
Start at each state $j$ with probability $\lambda_j$ (where $\sum_j\lambda_j=1$) and then transition along the Markov Chain according to probabilities $\sigma_{jj'}$ (where $\sum_{j'\in N\cup\{0\}}\rho_{jj'}=1$ for all $j\in N$).
Every time a state $j\in N$ is visited for the \textit{first} time, add $j$ to the end of the list.
The list terminates upon reaching state 0, which is assumed to occur w.p.~1.
\end{definition}

To show that assortments are optimal for Markov Chain choice models, we do not even need the constraint that $\phi$ is monotone.
More precisely, we show that among \textit{all} stopping policies defined by $n$ functions $\phi_j:2^{N\setminus\{j\}}\to\{0,1\}$, there always exists an optimal one corresponding to an assortment, i.e.\ one for which every $\phi_j$ is either the all-1 or all-0 function.
Intuitively, this is because the state for a list generated by a Markov Chain should be fully captured by the current item $j$, and hence one should either always want to stop on $j$ (in which case $\phi_j(H)=1$ for all $H\subseteq N\setminus\{j\}$) or never want to stop on $j$ (in which case $\phi_j(H)=0$ for all $H$).
However, note that the preceding argument does not immediately work, because the state does depend on the history $H$, in that items in $H$ cannot appear in the future of the list.
To rectify this argument, we must consider a further relaxation of stopping policies that allows for ``repeat visits'' and only makes sense for Markov Chain choice models.
This is formalized in the theorem below, whose proof is deferred to \Cref{sec:defPfsMC}.

We note that even though the monotonicity constraint on $\phi$ is derived from incentive compatibility, a policy $\phi$ without the monotonicity constraint is not as powerful as a mechanism without incentive compatibility constraints.  Indeed, the latter achieves the first-best revenue where the buyer's list is revealed to the seller up-front, while the former must still decide whether to stop on each item without knowing the future of the buyer's list.

\begin{theorem} \label{thm:mcmr}
For a list distribution arising from a Markov Chain choice model, given any item prices $r_1,\ldots,r_n$, assortments are optimal within the larger class of IC mechanisms.
\end{theorem}

\section{Assortments are Optimal under Sufficient Condition} \label{sec:pfMain}

In \Cref{sec:markovChainMainResult} we showed assortments to be optimal for Markov Chain choice models, by upper-bounding $\OPT^{\phi}$ using a ``repeat visits'' optimal stopping problem that only made sense on Markov Chains.
In this section we derive a different proof for assortments being optimal, for general list distributions satisfying a sufficient condition that includes Markov Chain choice models.

To motivate our sufficient condition for general list distributions, we start by identifying a property satisfied by all list distributions generated from a Markov Chain.
Take any Markov Chain choice model, and consider the tree diagram (see \Cref{def:qAndTree}) describing its list distribution.
Take any two prefixes $\rho,\rho'$ with $\rho_{\echar}=\rho'_{\echar}$ that have nodes in this tree, and consider the distribution of suffixes starting from those nodes.
We claim that these suffixes are ``similar'' in the following sense.
Take any assortment $S$ which is \textit{disjoint from $\rho\cup\rho'$}.
Then for any item $j\in S$, the probability of $j$ being ``chosen'' starting from either node is identical, equal to the probability of starting from state $\rho_{\echar}=\rho'_{\echar}$ in the Markov Chain and visiting state $j$ before any other state in $S$ or 0.

\begin{figure}
\caption{
Markov Chain and tree diagram representations of the same list distribution, used to illustrate our sufficient condition.
}
\label{fig:mnl}
\begin{center}
\begin{tikzpicture}
\node[draw,circle] (Amc) at (0-6.5,0) {$A$};
\node[draw,circle,ultra thick] (Bmc) at (2-6.5,2) {$B$};
\node[draw,circle] (Cmc) at (4-6.5,0) {$C$};
\node (Astart) at (0-6.5,3) {\footnotesize $\lambda_A=\nicefrac{1}{4}$};
\node (Bstart) at (2-6.5,3) {\footnotesize $\lambda_B=\nicefrac{1}{4}$};
\node (Cstart) at (4-6.5,3) {\footnotesize $\lambda_C=\nicefrac{1}{4}$};
\node (0start) at (6-6.5,3) {\footnotesize $\lambda_0=\nicefrac{1}{4}$};
\node[draw,circle] (0) at (2-6.5,-2) {$0$};

\draw[->] (Amc) edge node[fill=white]{\footnotesize $\nicefrac{1}{3}$} (0);
\draw[->] (Bmc) edge (0);
\draw[->] (Cmc) edge node[fill=white]{\footnotesize $\nicefrac{1}{3}$} (0);
\node[fill=white] at (2,-1) {\footnotesize $\nicefrac{1}{3}$};
\draw[->] (Astart) edge (Amc);
\draw[->] (Bstart) edge (Bmc);
\draw[->] (Cstart) edge (Cmc);
\draw[->] (0start) edge[bend left=40] (0);

\draw[->] (Amc) edge[bend right=10]node[fill=white]{\footnotesize $\nicefrac{1}{3}$} (Cmc);
\draw[->] (Cmc) edge[bend right=10]node[fill=white]{\footnotesize $\nicefrac{1}{3}$} (Amc);
\draw[->] (Amc) edge[bend right=20]node[fill=white]{\footnotesize $\nicefrac{1}{3}$} (Bmc);
\draw[->] (Bmc) edge[bend right=20]node[fill=white]{\footnotesize $\nicefrac{1}{3}$} (Amc);
\draw[->] (Bmc) edge[bend right=20]node[fill=white]{\footnotesize $\nicefrac{1}{3}$} (Cmc);
\draw[->] (Cmc) edge[bend right=20]node[fill=white]{\footnotesize $\nicefrac{1}{3}$} (Bmc);

\node[draw,rectangle] (Start) at (5,4) {root};

\node[draw,circle] (A) at (1,2) {$A$};
\node[draw,circle,ultra thick] (B) at (5,2) {$B$};
\node at (6.2,2) {$\pmb{\rho'=(B)}$};
\node[draw,circle] (C) at (9,2) {$C$};

\node[draw,circle,ultra thick] (AB) at (0,0) {$B$};
\node at (0.8,0) {$\pmb{\rho=}$};
\node at (0.8,-0.5) {$\pmb{(AB)}$};
\node[draw,circle] (AC) at (2,0) {$C$};
\node[draw,circle] (BA) at (4,0) {$A$};
\node[draw,circle] (BC) at (6,0) {$C$};
\node[draw,circle] (CA) at (8,0) {$A$};
\node[draw,circle] (CB) at (10,0) {$B$};

\node[draw,circle] (ABC) at (0,-2) {$C$};
\node[draw,circle] (ACB) at (2,-2) {$B$};
\node[draw,circle] (BAC) at (4,-2) {$C$};
\node[draw,circle] (BCA) at (6,-2) {$A$};
\node[draw,circle] (CAB) at (8,-2) {$B$};
\node[draw,circle] (CBA) at (10,-2) {$A$};

\draw[->](Start)--node[fill=white]{\footnotesize $\nicefrac{1}{4}$}(A);
\draw[->](Start)--node[fill=white]{\footnotesize $\nicefrac{1}{4}$}(B);
\draw[->](Start)--node[fill=white]{\footnotesize $\nicefrac{1}{4}$}(C);

\draw[->](A)--node[fill=white]{\footnotesize $\nicefrac{1}{3}$}(AB);
\draw[->](A)--node[fill=white]{\footnotesize $\nicefrac{1}{3}$}(AC);
\draw[->](B)--node[fill=white]{\footnotesize $\nicefrac{1}{3}$}(BA);
\draw[->](B)--node[fill=white]{\footnotesize $\nicefrac{1}{3}$}(BC);
\draw[->](C)--node[fill=white]{\footnotesize $\nicefrac{1}{3}$}(CA);
\draw[->](C)--node[fill=white]{\footnotesize $\nicefrac{1}{3}$}(CB);

\draw[->](AB)--node[fill=white]{\footnotesize $\nicefrac{1}{2}$}(ABC);
\draw[->](AC)--node[fill=white]{\footnotesize $\nicefrac{1}{2}$}(ACB);
\draw[->](BA)--node[fill=white]{\footnotesize $\nicefrac{1}{2}$}(BAC);
\draw[->](BC)--node[fill=white]{\footnotesize $\nicefrac{1}{2}$}(BCA);
\draw[->](CA)--node[fill=white]{\footnotesize $\nicefrac{1}{2}$}(CAB);
\draw[->](CB)--node[fill=white]{\footnotesize $\nicefrac{1}{2}$}(CBA);
\end{tikzpicture}
\end{center}
\end{figure}
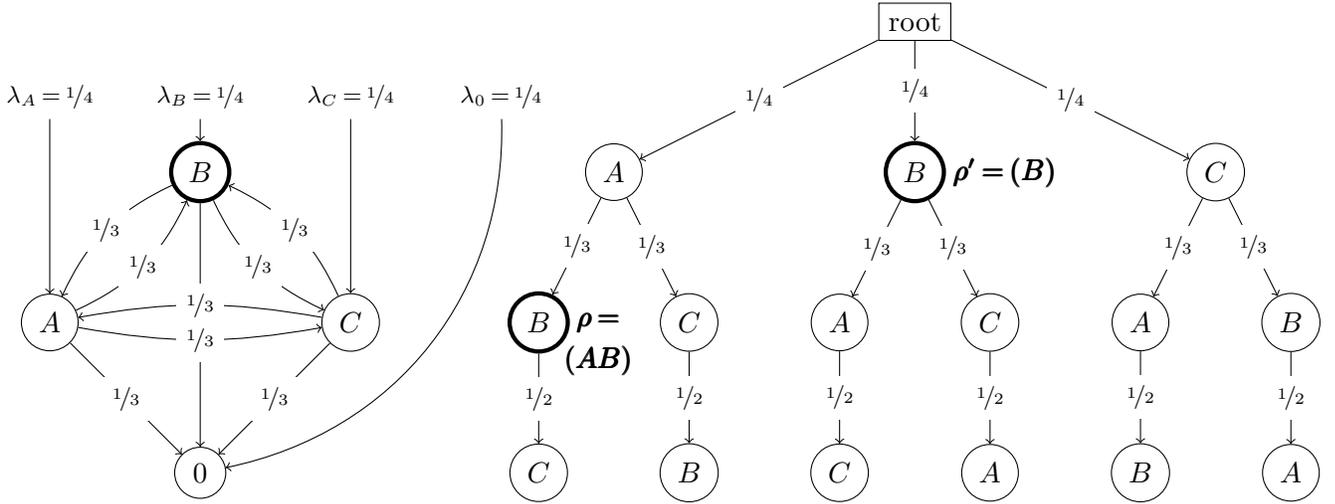

As a concrete example, consider the Markov Chain choice model and its tree diagram in \Cref{fig:mnl}, which describes an MNL choice model (see \Cref{sec:assortment}) with three items $A,B,C$ whose weights equal that of the no-purchase option.
Consider the prefixes $\rho=(AB)$ and $\rho'=(B)$ whose nodes are highlighted in the tree diagram.
Taking assortment $S=\{C\}$, it is easy to see that the probability of encountering $C$ in the suffix from $\rho$ is $1/2$, while the probability of encountering $C$ in the suffix from $\rho'$ is $\frac{1}{3}\cdot\frac{1}{2}+\frac{1}{3}=1/2$, both equal to the probability of visiting $C$ before 0 when starting from state $B$ in the Markov Chain.
Note that the quantifier ``disjoint from $\rho\cup\rho'$'' on the assortment $S$ was important.
If $S$ was allowed to be the assortment $\{A,C\}$, then the probability of $C$ being chosen decreases from $\frac{1}{3}\cdot\frac{1}{2}+\frac{1}{3}$ to $\frac{1}{3}$ when starting from node $\rho'$.

Our sufficient condition can then be thought of as a more lenient version of this property satisfied by Markov Chain choice models, in that if one of the prefixes is contained within the other as a set, say $\rho'\subseteq\rho$, then the choice probabilities starting from $\rho$ are permitted to be higher, for any assortment $S$ disjoint from $\rho\cup\rho'$ and any item $j\in S$.
This allows us to capture additional choice models, such as Markov Chain mixed with singleton lists.
To illustrate this, consider the list distribution which is $(ABC)$ or $(B)$ each with 50\% likelihood, which is a 50-50 mixture of the deterministic list $(ABC)$ (generatable from a Markov Chain) with the singleton list $(B)$.  This does not satisfy the equal choice probabilities property, since the probability of choosing item $C$ from assortment $\{C\}$ is 1 when starting from prefix $(AB)$, and 0 when starting from prefix $(B)$.  However, because the prefix $(AB)$ setwise contains prefix $(B)$, such a difference is permitted.

We now proceed to formalize our sufficient condition, in \Cref{sec:condition}, and prove that assortments are optimal under it, in \Cref{sec:complexPf}.
This proof is also based on showing that stopping policies $\phi$ are no better than assortments.
However, unlike the case in \Cref{sec:markovChainMainResult}, this is a different proof that makes use of the monotonicity constraints on $\phi$.
In fact, it is easy to see that non-monotone stopping policies are better than assortments on the 50-50 mixture between lists $(ABC)$ and $(B)$ described above---suppose $r_C=2,r_B=1,r_A=0$.  Then the best assortments $\{B\}$ and $\{C\}$ both earn expected revenue 1, while a non-monotone stopping policy can stop on prefix $(B)$ without stopping on prefix $(AB)$, earning expected revenue $\frac{2+1}{2}=1.5$.

\subsection{Sufficient Condition} \label{sec:condition}

We first formalize what it means for all choice probabilities to be higher when starting from one prefix instead of another.
For a realizable prefix, i.e.\ a prefix with non-zero probability of occurring, we refer to its conditional distribution of suffixes as its \textit{future}, and define a notion of domination.
% between futures.

\begin{definition}[Domination of Futures]
\label{def:domFuture}
Given realizable prefixes $\rho,\rho'$, the future from $\rho$ is said to \textit{dominate} the future from $\rho'$ if for all $S\subseteq N\setminus(\rho\cup\rho')$ and items $j\in S$,
%we have that
\begin{align}
\label{eqn:domFuture}
\Pr[j\succeq S_0|\rho] &\ge\Pr[j\succeq S_0|\rho'].
\end{align}
\end{definition}

\begin{example}[Domination] \label{eg:domination}
Consider the following futures with two items $A$ and $B$:
\begin{itemize}
\item Future~1, which is $(A)$ w.p.~1/2 and $()$ w.p.~1/2 (where $()$ denotes the empty future);
\item Future~2, which is $(A)$ w.p.~1/2 and $(B)$ w.p.~1/2;
\item Future~3, which is $(BA)$
w.p.~1/2 and $()$ w.p.~1/2.
\end{itemize}
Assume that the prefixes under consideration contain neither items $A$ nor $B$.
It is trivial to see that Future~2 dominates Future~1.
%, but not vice versa.
On the other hand, Future~3 does not dominate Future~1, because the probability of selling item $j=A$ under assortment $S=\{A,B\}$ is not higher for Future~3.
Also note that Future~2 dominates Future~3, because it is always easier to sell items when $A$ and $B$ appear on different branches of the possible suffix realizations.\Halmos
\end{example}

%\begin{remark}
In our preceding example, Future~1 did not dominate Future~2, and Future~3 did not dominate Future~1, only because the prefixes contained neither $A$ nor $B$.
By contrast,
if $\rho=(B)$ and $\rho'=()$, then $\rho$ with Future~1 would actually dominate $\rho'$ with Future~2, because comparisons for item $B$ (which lies in $\rho\cup\rho'$) are ignored.
Similarly, $\rho=(B)$ with Future~3 would dominate $\rho'=()$ with Future~1, because $\{A,B\}$ would no longer be a valid choice for $S$.
%\end{remark}

Having defined domination between futures, we now introduce our sufficient condition.

\begin{definition}[History-monotone Futures] \label{def:monotoneTiered}
A list distribution is said to have \textit{history-monotone futures} if for all realizable prefixes $\rho,\rho'$ with $\rho_{\echar}=\rho'_{\echar}$,
whenever $\rho_{\pre}\nsubseteq\rho'_{\pre}$ (i.e.\ $\rho_{\pre}$ viewed as a set is not contained within $\rho'_{\pre}$), the future from $\rho$ dominates the future from $\rho'$.
\end{definition}

For every pair of prefixes with the same endpoint, our sufficient condition imposes a relationship between their futures, based on the containment between the prefixes.  Note that since $\rho_{\echar}=\rho'_{\echar}$, the condition $\rho_{\pre}\nsubseteq\rho'_{\pre}$ is equivalent to $\rho\nsubseteq\rho'$.  There are three possibilities for the containment:
% between $\rho$ and $\rho'$:
\begin{enumerate}
\item \textbf{Neither of $\rho,\rho'$ are contained within the other.}
An example of this is $\rho=(AC),\rho'=(BC)$.
In this case, 
\Cref{def:monotoneTiered} says that the futures from $\rho$ and $\rho'$ must dominate each other, i.e.\ the choice probabilities for all $S$ and $j$ must be identical when conditioned on either prefix $\rho$ or $\rho'$.  Note that this does not\footnote{
As an example, let $D,E,F$ be three items distinct from $A,B,C$.  Suppose one suffix distribution is $(DEF)$ or $(ED)$ with equal probability, while another is $(DE)$ or $(EDF)$ with equal probability.  These suffix distributions are different, yet induce the same choice probabilities for all assortments $S\subseteq\{D,E,F\}$.
} impose the suffixes starting from $\rho$ and $\rho'$ to be identically distributed.
\item \textbf{One of $\rho,\rho'$ is contained within the other.}
An example of this is $\rho=(AC),\rho'=(C)$.
In this case, since $\rho\nsubseteq\rho'$, \Cref{def:monotoneTiered} says that the future from $\rho$ must dominate the future from $\rho'$, i.e.\ the choice probabilities for all $S$ and $j$ must be higher when conditioned on prefix $\rho$.
\item \textbf{Both of $\rho,\rho'$ are contained within the other.}
An example of this is $\rho=(ABC),\rho'=(BAC)$.
In this case, \Cref{def:monotoneTiered} imposes \textit{no relationship} between the futures from $\rho$ and $\rho'$.
\end{enumerate}

On the whole, cases~1--3 above impose that \textit{larger} histories (in the sense of not being contained within other histories) must have \textit{dominating} futures.  This is why our sufficient condition is called \textit{history-monotone futures}.

\begin{example}[Distribution without History-monotone Futures] \label{eg:suffCondFalse}
The simplest example of a list distribution that does \textit{not} have history-monotone futures requires 3 items $A,B,C$, with the random list being $(BA)$ or $(BC)$ with equal probability.  Consider prefixes $\rho=(CB)$ and $\rho'=(B)$, which have the same endpoint.  According to history-monotone futures, the larger prefix $(CB)$ must have the dominating future.  However, letting $j=A$ and $S=\{A\}$, the choice probability $\Pr[j\succeq S_0|(CB)]$ equals 0, while $\Pr[j\succeq S_0|(B)]$ equals 1.  Since the choice probability from the larger prefix $(CB)$ is strictly smaller, this list distribution does not have history-monotone futures.\Halmos
\end{example}

The list distribution from the introductory \Cref{eg:initial}, which demonstrated assortments to not be optimal, can also be checked to not have history-monotone futures.
In fact, we now show that assortments must be optimal for any list distribution with history-monotone futures.

\subsection{Statement and Proof of Main Result} \label{sec:complexPf}

\begin{theorem} \label{thm:mr}
For a list distribution with history-monotone futures, given any item prices $r_1,\ldots,r_n$, assortments are optimal within the larger class of IC mechanisms.
\end{theorem}

\textbf{Intuition behind history-monotone futures.}
History-monotone futures are sufficient for assortments to be as powerful as mechanisms, by the following intuition.  Recall that the revenue of mechanisms can be upper-bounded by that of monotone stopping policies $\phi$, and $\phi$'s that stop on either all or none of the prefixes with the same endpoint correspond to assortments.
Therefore, if a general $\phi$ were to do strictly better, then for some prefixes $\rho,\rho'$ with the same endpoint $\rho_{\echar}=\rho'_{\echar}$, it must continue from $\rho'$ while stopping on $\rho$, and benefit from doing so because it earns strictly greater profit in the future from $\rho'$ than $\rho$.
However, strictly greater profit requires the future from $\rho'$ to \textit{not be dominated by} $\rho$, which under the condition of history-monotone futures, can occur only if $\rho'_{\pre}\supseteq\rho_{\pre}$.
Therefore, the monotonicity constraints on $\phi$ would impose $\phi_{\rho_{\echar}}(\rho_{\pre})\le\phi_{\rho'_{\echar}}(\rho'_{\pre})$, i.e.\ $\phi$ can only stop on $\rho$ (which has the worse future) if it also stops on $\rho'$ (which has the better future).
This means $\phi$ needs to compromise between stopping on either both or none of the prefixes $\rho,\rho'$, and by repeating this argument for all pairs of prefixes with the same endpoint, $\phi$ should indeed for each item want to stop on either all of none of its prefixes.

\textbf{Proof sketch.}
Our goal is to formalize the intuitive argument above.  The argument actually has an issue though---$\phi$ can earn strictly greater profit by continuing from $\rho'$ than from $\rho$ even if the future from $\rho'$ is dominated by the future from $\rho$.
To see why, consider \Cref{eg:domination} from \Cref{sec:condition}, in which
Future~3 (equaling $(BA)$ w.p.~1/2)
was dominated by
Future~2 (equaling $(A)$ w.p.~1/2, $(B)$ w.p.~1/2).
If $\phi$ only stops on item $A$ when $B$ is in the history (which is valid under the definition of monotone stopping policies), then $\phi$ would collect strictly more revenue from item $A$ by continuing from the dominated Future~3 instead of the dominating Future~2.

Therefore, we instead prove \Cref{thm:mr} using an extremal argument, where among all monotone stopping policies $\phi$ that are strictly better than any assortment, we take the $\phi$ that \textit{maximizes} the number of 1-entries (i.e.\ maximizes the number of (item, history)-pairs $(j,H)$ for which $\phi_j(H)=1$).
We modify $\phi$ to have one more 1-entry without decreasing its revenue, causing a contradiction and implying that such a $\phi$ must not have existed in the first place.
To find a modification, we let $S$ denote the set of items $j$ for which $\phi_j$ is the all-1 function.
We then introduce the concept of an \textit{$S$-adjusted price} for a prefix $\rho$, which is the revenue $r_{\rho_{\echar}}$ minus the future revenue from assortment $S$ conditional on starting from $\rho$.
In light of the issue mentioned above, this concept is useful because if $\rho_{\echar}=\rho'_{\echar}$, then $\rho'$ can only have a strictly lower $S$-adjusted price than $\rho$ (i.e.\ it is worse to stop on) if the future from $\rho'$ is not dominated by the future from $\rho$.

We show that the revenue difference $\Rev[\phi]-\Rev[S]$, which must be positive, can be expressed as a non-negative linear combination of $S$-adjusted prices for prefixes disjoint from $S$.
We then use the relationship between $S$-adjusted prices and domination to show that under history-monotone futures, the set of $S$-disjoint prefixes can be partitioned into ``tiers'', over which the $S$-adjusted prices are monotone.
This monotonicity ultimately helps us find an item $j\notin S$ and history $H$ for which changing $\phi_j(H)$ from 0 to 1 does not decrease $\Rev[\phi]$, completing the proof.

\textbf{Intermediate concepts and results needed.}
We now introduce the results based on $S$-adjusted\footnote{We note that \Cref{def:sAdj} extends the $S$-adjusted prices from \citet{desir2020constrained} to be defined for prefixes instead of items, and \Cref{lem:revAdj} extends their externality-adjustment technique to arbitrary choice models.} prices and partitioning into tiers that are needed for our proof of \Cref{thm:mr}, which is presented afterward.
Proofs of the intermediate results are deferred to \Cref{sec:defPfsMR}.

\begin{definition}[$S$-adjusted Price for a Prefix] \label{def:sAdj}
Consider any assortment $S$ and any realizable prefix $\rho$ disjoint from $S$.  Then the $S$-adjusted price of $\rho$ is defined to be
\begin{align}
\label{eqn:adjustedRev}
r^{S}(\rho) &:=r_{\rho_{\echar}}-\sum_{j'\in S}r_{j'}\Pr[j'\succeq S_0|\rho].
\end{align}
\end{definition}

\begin{lemma}[$S$-adjusted Revenue for a Monotone Stopping Policy] \label{lem:revAdj}
For any monotone stopping policy $\phi$ and assortment $S$,
\begin{align}
\Rev[\phi]-\Rev[S] &=\sum_{\rho:\rho\cap S=\emptyset}r^{S}(\rho)\Pr[\rho]\phi_{\rho_{\echar}}(\rho_{\pre})\prod_{k=1}^{|\rho|-1}(1-\phi_{\rho_k}(\rho_{1:k-1})).
\end{align}
\end{lemma}

In \Cref{lem:revAdj}, the expression $\phi_{\rho_{\echar}}(\rho_{\pre})\prod_{k=1}^{|\rho|-1}(1-\phi_{\rho_k}(\rho_{1:k-1}))$ equals 1 if and only if $\phi$ stops on prefix $\rho$ after not stopping before that.
\Cref{lem:revAdj} then intuitively says that the difference $\Rev[\phi]-\Rev[S]$ equals the expected $S$-adjusted price of the prefix that is stopped on over a random run of $\phi$, where $S$-adjusted price is only defined for prefixes $\rho$ disjoint from $S$.

\begin{lemma}[Partitioning Prefixes into Tiers] \label{lem:tiers}
Suppose the list distribution has history-monotone futures, and for a fixed $S\subsetneq N$ and $j\notin S$, consider the set of realizable prefixes $\rho$ with $\rho_{\echar}=j$ and $\rho\cap S=\emptyset$.
These prefixes can be partitioned into $T^S_j$ tiers, denoted by $\cT^S_j(1),\ldots,\cT^S_j(T^S_j)$, which satisfy (after suppressing the superscript $S$ and subscript $j$):
\begin{enumerate}
\item\label{cond:one} For any two distinct tiers $t,t'$ satisfying $t>t'$ and any prefixes $\rho\in\cT(t),\rho'\in\cT(t')$, we have that $\rho\supsetneq\rho'$ (when viewed as sets) and $\Pr[j'\succeq S_0|\rho]\ge\Pr[j'\succeq S_0|\rho']$ for all $j'\in S$;
\item\label{cond:two} Within any single tier $t\in[T]$, either:
\begin{enumerate}
\item\label{tier:incomparable} For all distinct prefixes $\rho,\rho'\in\cT(t)$, we have that $\rho\nsubseteq\rho'$ and $\rho'\nsubseteq\rho$ (when viewed as sets) and $\Pr[j'\succeq S_0|\rho]=\Pr[j'\succeq S_0|\rho']$ for all $j'\in S$; or
\item\label{tier:allSame} For all prefixes $\rho,\rho'\in\cT(t)$, we have that $\rho$ and $\rho'$ are identical when viewed as sets.
\end{enumerate}
\end{enumerate}
\end{lemma}

\begin{corollary}[Monotonicity of $S$-adjusted Prices] \label{cor:monAdjustedPrice}
For any $j\notin S$,
consider the partitioning of realizable $S$-disjoint prefixes with endpoint $j$ into tiers.
For tiers $t,t'$ and prefixes $\rho\in\cT^S_j(t),\rho'\in\cT^S_j(t')$, the $S$-adjusted prices satisfy $r^S(\rho)\le r^S(\rho')$ whenever either: $t>t'$; or $t=t'$, and $\rho\neq\rho'$ when viewed as sets.
\end{corollary}

In \Cref{sec:egTiers}, we present an example illustrating the intricacies in the definition of history-monotone futures and the resulting tier decomposition.
Note that \Cref{cor:monAdjustedPrice} is immediately implied by cases~\ref{cond:one} and~\eqref{tier:incomparable} of \Cref{lem:tiers},
after applying the definition of $S$-adjusted prices.

\textbf{Proof of \Cref{thm:mr}.}
\Cref{thm:mr} is proved in \Cref{sec:mrPf}, following the proof sketch and intermediate concepts introduced above.

\subsection{Choice Models Satisfying Sufficient Condition}\label{sec:assortment}

In this \namecref{sec:assortment} we describe choice models from the literature that satisfy our sufficient condition, for which assortments are optimal.
All proofs from this \namecref{sec:assortment} are deferred to \Cref{sec:defPfsAssortment}.

We start with the most common MNL choice model, stated in the form of a random list.

\begin{definition}[Multi-Nomial Logit (MNL)] \label{def:mnl}
Consider an urn which starts with balls $j=0,\ldots,n$, each with a weight $w_j>0$.
Let $W$ denote the total weight of balls remaining in the urn.
Sequentially, sample a ball \textit{without} replacement and add its label to the end of the list, where each remaining ball $j$ is drawn with probability $w_j/W$.
The list terminates upon drawing the 0-ball.
\end{definition}

The list distribution from \Cref{def:mnl} is consistent with the closed-form MNL choice probabilities $\Pr[j\succeq S_0]=\frac{w_j}{\sum_{j'\in S\cup\{0\}}w_{j'}}$ denoting the likelihood of item $j$ being chosen when an assortment $S$ is offered \citep{plackett1975analysis,luce1959individual}.
As mentioned in \Cref{sec:markovChainMainResult}, MNL is captured by the larger family of Markov Chain choice models \citep{blanchet2016markov}.  Our proposition below in conjunction with our general result based on the sufficient condition of history-monotone futures provides an alternative proof (to \Cref{thm:mcmr}) that assortments are optimal for MNL and Markov Chain.

\begin{proposition}\label{prop:mc}
A Markov Chain choice model has history-monotone futures.
\end{proposition}

As discussed in \Cref{sec:markovChainMainResult}, Markov Chain already captures a wide range of choice models from the literature.  However, the following is not captured, which still satisfies our sufficient condition.

\begin{definition}[Tversky's Elimination by Aspects model] \label{def:elimByAspects}
Consider the MNL model, where in addition the set $N\cup\{0\}$ has been partitioned into disjoint \textit{nests} $N_0,\ldots,N_m$, with $N_0=\{0\}$.
The list is generated as in the MNL model (where balls $j$ are drawn with probabilities proportional to their weights $w_j$), except upon any ball $j$ in nest $i\in\{1,\ldots,m\}$ being drawn,
the subsequent draws are constrained to balls from the same nest $i$ (one way to implement this is to keep redrawing until the ball drawn is from nest $i$),
all of which must be drawn before proceeding to other nests.
The list terminates upon drawing the 0-ball.
\end{definition}

%The Elimination by Aspects model generalizes MNL when each item is in its own nest.
The Elimination by Aspects model \citep{tversky1972elimination} is not captured by a Markov Chain choice model, since there would have to be a positive transition probability between every pair of items, which would allow nests to be traversed in a non-contiguous fashion.
Despite this fact, we still show that conditional on any two prefixes $\rho,\rho'$ with $\rho_{\echar}=\rho'_{\echar}$, the choice probabilities for assortments disjoint from $\rho\cup\rho'$ are identical, as was the case for Markov Chain.

\begin{proposition}\label{prop:elimByAspects}
The Elimination by Aspects model has history-monotone futures.
\end{proposition}

Next, we consider the mixture of distributions with singleton lists.
As discussed at the start of \Cref{sec:pfMain}, representability by a Markov Chain is not preserved under such a mixture.
Nonetheless, we show that our sufficient condition \textit{is} preserved, using the fact that if $\rho$ contains another prefix $\rho'$ with $\rho_{\echar}=\rho'_{\echar}$, then $\rho$ is permitted to have higher conditional choice probabilities.
Indeed, mixing with singleton lists can only reduce these choice probabilities for singleton prefixes, which are contained in all other prefixes with the same endpoint, thereby preserving our sufficient condition.

\begin{definition}[Mixture with Singletons]
The mixture of any list distribution $p$ and singleton probabilities $\alpha_1,\ldots,\alpha_n$, with $\sum_j\alpha_j\le1$, is defined as follows.
The random list equals the singleton $(j)$ with probability $\alpha_j$, for all $j$.
Otherwise, with probability $1-\sum_j\alpha_j$, the random list is drawn according to $p$.
\end{definition}

\begin{proposition}\label{prop:mixture}
If a list distribution has history-monotone futures, then the mixture of that list distribution with any singletons has history-monotone futures.
\end{proposition}

Finally, we show that special cases of Nested Logit are captured by our sufficient condition.

\begin{definition}[General Nested Logit Choice Model]
Suppose the set $N\cup\{0\}$ has been partitioned into disjoint nests $N_0,\ldots,N_m$, with $N_0=\{0\}$.
Each $j\in N\cup\{0\}$ has a weight $w_j>0$ and each nest $i\in\{1,\ldots,m\}$ has a \textit{dissimilarity parameter} $\gamma_i\in(0,1]$.  When shown assortment $S$, the probability of the buyer choosing an item $j\in S$, if $j$ lies in nest $i$, is given by
\begin{align} \label{eqn:nestedLogitChoiceProbs}
\frac{(\sum_{j'\in S\cap N_i}w_{j'})^{\gamma_i}}{w_0+\sum_{i'=1}^n(\sum_{j'\in S\cap N_{i'}}w_{j'})^{\gamma_{i'}}}\cdot\frac{w_j}{\sum_{j'\in S\cap N_i}w_{j'}}.
\end{align}
The expression~\eqref{eqn:nestedLogitChoiceProbs} is often referred to as the Nested Logit choice probabilities.
\end{definition}

We note that Nested Logit corresponds to MNL if $\gamma_i=1$ for all $i$, and corresponds to Tversky's Elimination by Aspects model as $\gamma_i\to0$ for all $i$ \citep[Ch.4]{train2009discrete}.
We first show that for $\gamma_i\neq1$ the Nested Logit choice probabilities are not captured by a Markov Chain choice model.

\begin{proposition} \label{prop:NLnotMC}
The Nested Logit choice probabilities cannot be captured by a Markov Chain choice model, even when there are only 3 items that have equal weights and are in the same nest.
\end{proposition}

As far as positive results, a general challenge in verifying our sufficient condition is that there is no known description of Nested Logit using a combinatorially-generated random list.
In fact, multiple different list distributions could be consistent with the same set of Nested Logit choice probabilities.
In light of these challenges, we now show two positive results for Nested Logit on a small number of items that construct using brute force a consistent list distribution with history-monotone futures.
These results complement \Cref{prop:NLnotMC} by showing that in these special cases, although Nested Logit cannot be captured by Markov Chain, it still satisfies our sufficient condition.

\begin{proposition} \label{prop:NL3itemRep}
Any Nested Logit choice model with at most 3 items, which are in the same nest, can be represented by a list distribution with history-monotone futures.
\end{proposition}

\begin{proposition} \label{prop:NL4itemRep}
Any Nested Logit choice model with at most 4 items, which have equal weights and are in the same nest, can be represented by a list distribution with history-monotone futures.
\end{proposition}

%All in all, our sufficient condition holds for Markov Chain choice models and the similar Elimination by Aspects model, where conditional on two different prefixes $\rho,\rho'$ with $\rho_{\echar}=\rho'_{\echar}$, the choice probabilities for all assortments disjoint from $\rho,\rho'$ are identical.
%Making use of the fact that conditional choice probabilities are permitted to be higher for ``larger'' prefixes, our sufficient condition is also preserved after mixing any choice model with singletons.
%We note that mixing with a collection of singleton lists\footnote{This is also referred to as an \textit{Independent Demand} model in the literature \citep{gallego2004managing}.} is of broad interest in choice modeling \citep{cao2020revenue}.
These result suggest that Nested Logit can potentially be defined as a randomly-generated list for which the longer the history, the more likely the list is to continue into the future.
To our knowledge, such intuition for Nested Logit was not known before, and could be of future interest as a combinatorial property that unifies Markov Chain and Nested Logit.

%We finish by making a list of choice models that, in their full generality, can be checked to not satisfy our sufficient condition: random-utility models where each item has an arbitrary independent distribution, the Mallows model \citep{desir2016assortment}, and the One-way Substitution and Locational Choice Models \citep{honhon2012optimal}.
%Of course, any choice model that can capture any of the bad \Cref{eg:initial,eg:suffCondFalse} also cannot satisfy our sufficient condition.

\section{Revenue Gaps and Assortment Optimization for General List Distributions} \label{sec:generalLists}

Sections~\ref{sec:markovChainMainResult}--\ref{sec:pfMain} considered specific list distributions for which there was no gap between the revenues of mechanisms and assortments.
In this \namecref{sec:generalLists} we consider general list distributions, and establish:
\begin{itemize}
\item Constant-factor upper and lower bounds on the revenue gap when the mechanism is restricted to be ``budget-additive'', a simple subclass that includes top-$k$ lotteries (\textbf{\Cref{sec:budgetAdditive}});
\item The revenue gap to be bounded when lists have bounded length (\textbf{\Cref{sec:bdedListLength}});
\item The revenue gap to be generally unbounded without any restrictions on the mechanism or the list length (\textbf{\Cref{sec:generalDistNegResult}}).
\end{itemize}

Related to the question of revenue gaps, in \textbf{\Cref{sec:tighterLP}} we show that our new LP motivated by mechanism design provides a strictly tighter relaxation for assortment optimization than existing LP's, which are currently used to compute the optimal assortment via Integer Programming.

\subsection{Bounds on the Revenue Gap for Budget-Additive Mechanisms} \label{sec:budgetAdditive}

To define the subclass of budget-additive mechanisms, we first establish a property on the monotone-increasing set function $f$ that was associated with every mechanism $x$, with $f(S)$ denoting the maximum probability that $x$ assigns an item in $S$ (see \Cref{prop:x_to_f}).
We now show that any monotone \textit{submodular}\footnote{
A submodular function $f:2^N\to\bR$ must satisfy $f(S\cup S')+f(S\cap S')\le f(S)+f(S')$ for all subsets $S,S'\subseteq N$.  
} function $f$ induces an IC mechanism, with the converse also being true under a mild condition.

\begin{proposition}[Submodularity] \label{prop:subm}
For any monotone submodular $f:2^N\to[0,1]$, setting
\begin{align*}
x_{\ell_k}(\ell) &=f(\ell_{1:k})-f(\ell_{1:k-1}) &\forall\ell\in\cL,k=1,\ldots,|\ell|
\end{align*}
defines an IC mechanism.
Conversely, for any IC mechanism $x$, setting
\begin{align*}
f(S) &=\max_{\ell\in\cL}\sum_{j\in S\cap\ell}x_j(\ell) &\forall S\subseteq N
\end{align*}
defines a monotone submodular function from $2^N$ to [0,1], under the condition that all possible lists (even those with zero probability) are included in $\cL$ so that the IC constraints apply to all types.
\end{proposition}

\Cref{prop:subm} is proved in \Cref{sec:defPfsGeneral}, but using its statement we can now define subclasses of mechanisms based on subclasses of monotone submodular functions, since any such function induces an IC mechanism.
A well-studied subclass of monotone submodular functions are the budget-additive functions, in which $f$ takes the form $f(S)=\min\{\sum_{j\in S}w_j,B\}$ for some non-negative weights $w_j$ and a budget $B$.
We assume that $B\in[0,1]$, since we need $f(S)$ to take values in [0,1].

\begin{definition}[Budget-additive Mechanism]
A \textit{budget-additive mechanism} is defined by weights $w_1,\ldots,w_n\ge0$ and a budget $B\in[0,1]$, with $f:2^N\to[0,B]$ denoting the function $f(S)=\min\{\sum_{j\in S}w_j,B\}$.
Then, the probability of the mechanism assigning a type $\ell$ its $k$'th-most-preferred item $\ell_k$ is $x_{\ell_k}(\ell)=f(\ell_{1:k})-f(\ell_{1:k-1})$, 
for all $\ell\in\cL$ and $k=1,\ldots,|\ell|$.
\end{definition}

Since the function $f(S)=\min\{\sum_{j\in S}w_j,B\}$ is easily seen to be submodular, \Cref{prop:subm} implies that budget-additive mechanisms are indeed IC.
The special case of top-$k$ lotteries is captured by setting $w_j=1/k$ for all $j$ and $B=1$.
The special case of an assortment $S$ (equivalently, a deterministic mechanism) is captured by setting $w_j=\bI(j\in S)$ for all $j$ and $B=1$.
% maybe mention that usually we want $B=1$

We now show that assortments can earn at least $1/e$ times the revenue of a budget-additive mechanism, without any assumption on the list distribution.
Our proof entails constructing a random assortment guided by the weights $w_j$, and showing that this distribution of assortments has expected revenue at least $1/e$ times that of the budget-additive mechanism.
This implies that at least one deterministic assortment must have the same revenue guarantee.
We should note that the idea of constructing an assortment through including each item independently at random has been previously used by \citet{aouad2018approximability,feldman2019assortment} for assortment optimization.

\begin{theorem} \label{thm:budgetAdditive}
Given any budget-additive mechanism defined by $w_1,\ldots,w_n\ge0$ and $B\in[0,1]$,
the random assortment in which each item $j$ is included independently with probability
$
1-e^{-w_j}
$
has expected revenue at least $1/e$ times that of the budget-additive mechanism.
\end{theorem}

\proof{Proof of \Cref{thm:budgetAdditive}.}
Take any budget-additive mechanism and let $f$ denote the function $f(S)=\min\{\sum_{j\in S}w_j,B\}$.
By \Cref{prop:x_to_f}, the revenue of the mechanism can be expressed as
\begin{align*}
\Rev[f]=\sum_{\ell\in\cL}p(\ell)\sum_{k=1}^{|\ell|}r_{\ell_k}(f(\ell_{1:k})-f(\ell_{1:k-1})).
\end{align*}
Thus, it suffices to show that for every type $\ell\in\cL$ and position $k\in\{1,\ldots,|\ell|\}$ with $f(\ell_{1:k})-f(\ell_{1:k-1})>0$, the random assortment sells item $\ell_k$ to type $\ell$ with probability at least
\begin{align} \label{eqn:5028}
\frac{1}{e}(f(\ell_{1:k})-f(\ell_{1:k-1})).
\end{align}

To argue this, note that if the buyer's type realizes to $\ell$, then they will choose item $\ell_k$ if and only if $\ell_k$ is included in the assortment and none of their more-preferred items $\ell_1,\ldots,\ell_{k-1}$ are included.  By independence, the probability of this event occurring is
\begin{align} \label{eqn:0981}
(1-e^{-w_{\ell_k}})\prod_{k'=1}^{k-1}e^{-w_{\ell_{k'}}}
&=(1-e^{-w_{\ell_k}})\exp(-\sum_{k'=1}^{k-1}w_{\ell_{k'}}).
\end{align}

Now, by definition of $f$, the coefficient $f(\ell_{1:k})-f(\ell_{1:k-1})$
%=\min\{\sum_{k'=1}^kw_{\ell_{k'}},B\}-\min\{\sum_{k'=1}^{k-1}w_{\ell_{k'}},B\}$, which
can only be non-zero if $\sum_{k'=1}^{k-1}w_{\ell_{k'}}<B$ and $w_{\ell_k}>0$, in which case it equals $\min\{w_{\ell_k},B-\sum_{k'=1}^{k-1}w_{\ell_{k'}}\}$.
Let $m$ denote this quantity.
Then, we can lower-bound the ratio of~\eqref{eqn:0981} to~\eqref{eqn:5028}:
\begin{align*}
\frac{(1-e^{-w_{\ell_k}})\exp(-\sum_{k'=1}^{k-1}w_{\ell_{k'}})}{\frac{1}{e}(f(\ell_{1:k})-f(\ell_{1:k-1}))}
&\ge\frac{(1-e^{-m})\exp(B-\sum_{k'=1}^{k-1}w_{\ell_{k'}}-B)}{\frac{1}{e}m}
\\ &\ge\frac{(1-e^{-m})\exp(m-B)}{\frac{1}{e}m}
\\ &\ge\frac{(1-e^{-m})e^{m-1}}{\frac{1}{e}m}
\end{align*}
where the first inequality holds because $w_{\ell_k}\ge m$,
the second inequality holds because $B-\sum_{k'=1}^{k-1}w_{\ell_{k'}}\ge m$,
and the third inequality holds because $B\le1$.
The final expression can be analytically checked to be at least 1 for all values of $m\in[0,1]$, completing the proof.
\Halmos\endproof

We note that the inclusion probabilities in \Cref{thm:budgetAdditive} were optimized over all functions of the form $1-e^{-\beta w_j}$ with $\beta>0$.  The best guarantee of $1/e$ occurred at $\beta=1$.

Now, to complement \Cref{thm:budgetAdditive}, we show that assortments can earn at most $2/e$ times the revenue of budget-additive mechanisms (more specifically, top-2 lotteries), by constructing a family of instances that scale up and optimize our motivational \Cref{eg:initial} from the Introduction.

\begin{theorem} \label{thm:topKGap}
Consider the following family of instances, where $n$ denotes the number of items and $M$ denotes a fixed large constant.
Items have prices $r_j=M^j$ for all $j=1,\ldots,n$.
The buyer's list always has length at most 2, and the most expensive item on it is $j$ w.p.~$M^{-j}$, for all $j$ (the buyer's list is empty w.p.~$1-\sum_{j=1}^nM^{-j}$).
Conditional on the most expensive item on the buyer's list being $j$ for some $j>1$, item $j$ will always be the buyer's second choice, with the buyer's first choice being an item $j'$ drawn uniformly from $\{1,\ldots,j-1\}$.

On such an instance, a top-2 lottery obtains revenue at least $n/2$.
On the other hand, as $n\to\infty$, no assortment can earn revenue greater than $n/e+O(\log n)$, showing that assortments can obtain no more than $2/e$ times the revenue of the best budget-additive mechanism in the worst case.
\end{theorem}

The proof of \Cref{thm:topKGap} is deferred to \Cref{sec:defPfsGeneral}.
The key step in upper-bounding the revenue earned by any assortment is to show that the best assortment of any size $k$ on this instance consists of the $k$ highest-priced items.

\subsection{Revenue Gap under Bounded List Length} \label{sec:bdedListLength}

In \Cref{sec:budgetAdditive} we showed that the ratio between the revenues of the best assortment and the best \textit{budget-additive} mechanism lies in $[1/e,2/e]$, without any further assumptions on the instance.
In this \namecref{sec:bdedListLength} we show that the revenue ratio between the best assortment and the best \textit{unrestricted} mechanism is lower-bounded by $2/(eL)$, where $L$ denotes the maximum length of any list that can realize.
Although this bound approaches 0 (and is hence meaningless) as $L\to\infty$, we show in \Cref{sec:generalDistNegResult} that this is \textit{necessarily} the case, from a complexity-theoretic perspective.

\begin{theorem} \label{thm:bdedListLength}
Define $L=\max_{\ell:p(\ell)>0}|\ell|$.
Given any mechanism $x$, let $f_j=\max_{\ell\ni j}x_j(\ell)$ denote the maximum probability with which $x$ allocates item $j$, for all $j\in N$.
Then, the random assortment in which each item $j$ is included independently with probability $\varphi(f_j)$, where $\varphi$ is the function
$$
\varphi(a)=\begin{cases}
2a/ L, &a\le1/2; \\
1/ L, &a>1/2;
\end{cases}
$$
has expected revenue at least $2/(eL)$ times that of the mechanism.
\end{theorem}

The proof of \Cref{thm:bdedListLength} is deferred to \Cref{sec:defPfsGeneral}.
Based on our analysis, we designed the function $\varphi:[0,1]\to[0,1]$ to maximize the guarantee
\begin{align} \label{eqn:varphiObj}
\inf_{a\in[0,1]}\frac{\varphi(1-a)}{1-a}(1-\varphi(a))^{ L-1}
\end{align}
subject to the constraints that $\varphi(a)$ is non-decreasing and $\varphi(a)/a$ is non-increasing (two facts needed for our proof).
It is easy to see that no higher value in~\eqref{eqn:varphiObj} is possible, since when $a=1/2$, the maximum value of $\frac{\varphi(1/2)}{1/2}(1-\varphi(1/2))^{ L-1}$ over $\varphi(1/2)\in[0,1]$ is $\frac{2}{ L}(1-\frac{1}{ L})^{ L-1}$.
In fact, the same guarantee has been achieved by \citet{feldman2019assortment} based on a different function and LP, for the assortment optimization problem on lists of bounded length.

\subsection{Negative Results based on Hardness of Assortment Optimization} \label{sec:generalDistNegResult}

In this \namecref{sec:generalDistNegResult}, we consider the computation of optimal (deterministic, budget-additive, general) mechanisms for an arbitrary list distribution that is \textit{input explicitly}.
By this, we mean that for each realizable $\ell\in\cL$, the input contains a floating-point number for $p(\ell)$, with these probabilities summing to 1.\footnote{
Note the contrast with the MNL and Markov Chain choice models considered in earlier sections, for which there was an implicit structural representation of the probabilities for exponentially many lists.
}
Leveraging a strong hardness result for assortment optimization under this input model, we establish: (i) an unbounded revenue gap result for general mechanisms, and (ii) a computational hardness result for budget-additive mechanisms.
%Later, we show that our notion of/ IC mechanisms provides the tightest LP relaxation known to date for assortment optimization under this input model.
To begin, we show that the optimal general mechanism under this input model can be computed in polynomial time.

\begin{proposition}[Computation for General Mechanisms] \label{prop:mechPolyTime}
On an explicitly-input list distribution, an optimal mechanism $x$ and its revenue $\OPT^x$ can be computed in polynomial time.
\end{proposition}

\proof{Proof of \Cref{prop:mechPolyTime}.}
The optimal mechanism and its revenue is described by the LP from \Cref{def:x}.
This LP has $O(n|\cL|)$ variables and $O(n|\cL|^2)$ constraints, which is polynomial in the size of the input, and hence can be solved in polynomial time.
\Halmos\endproof

\begin{theorem}[from \citet{aouad2018approximability}] \label{thm:aouad}
On an explicitly-input list distribution, the optimal assortment revenue $\OPT^S$ is NP-hard to approximate within any constant factor (in fact, any factor sublinear in $n$).
\end{theorem}

\Cref{thm:aouad} combines with \Cref{prop:mechPolyTime} to show below that there cannot be a constant (in fact, sublinear) revenue gap between mechanisms and assortments.  It is also later combined with \Cref{thm:budgetAdditive} to derive negative results for budget-additive mechanisms.

\begin{corollary}[Revenue Gap for General Mechanisms] \label{cor:unbdedGap}
Unless P=NP, on general
list
distributions, the revenue of mechanisms can be an unbounded factor greater than that of assortments.
\end{corollary}

\proof{Proof of \Cref{cor:unbdedGap}.}
Suppose there was a constant $B\ge1$ such that $\OPT^x\le B\cdot\OPT^S$ for all instances.
By \Cref{prop:mechPolyTime}, the value of $\OPT^x$ can be computed in polynomial time.
Since $\frac{1}{B}\cdot\OPT^x\le\OPT^S\le\OPT^x$, this implies that the value of $\OPT^S$ can be approximated within a factor of $B$ in polynomial time, which can only be the case if $P=NP$, by \Cref{thm:aouad}.
\Halmos\endproof

Note the contrast between \Cref{cor:unbdedGap} and the earlier \Cref{thm:budgetAdditive}, which showed the revenue of \textit{budget-additive} mechanisms to be at most a factor of $e\approx2.71$ better than assortments.
These two facts together imply that general mechanisms (based on general submodular functions) can be an unbounded factor better than mechanisms based on budget-additive submodular functions.

%Since assortments correspond to integer solutions in the LP defining $\OPT^x$ (see \Cref{prop:S_to_x}), \Cref{thm:unbdedGap} is saying that this LP has an unbounded \textit{integrality gap}, i.e.\ an unbounded factor between the objective values of fractional vs.\ integer solutions.
%We do not know of an explicit sequence of instances for which this gap tends to $\infty$.
%Note that although the reduction in \citet{aouad2018approximability} goes through the Maximum Independent Set problem, which has a simple sequence of instances whose integrality gap tends to $\infty$, this does not translate to an $\infty$ gap in our problem.
%This is because a solution to the LP relaxation of the Maximum Independent Set problem does not translate into a fractional solution, i.e.\ a mechanism, in the LP defining $\OPT^x$.

\begin{corollary}[Computation for Budget-additive Mechanisms] \label{cor:budgetAdditiveCompHard}
On an explicitly-input list distribution, computing the optimal budget-additive mechanism is NP-hard.
\end{corollary}

\proof{Proof of \Cref{cor:budgetAdditiveCompHard}.}
Let $\OPT^w$ denote the revenue of an optimal budget-additive mechanism.
By \Cref{thm:budgetAdditive}, $\frac{1}{e}\cdot\OPT^w\le\OPT^S$.
Meanwhile, $\OPT^S\le\OPT^w$, since budget-additive mechanisms capture assortments.
Therefore, computing $\OPT^w$ would allow us to approximate $\OPT^S$ within a constant factor, which is an NP-hard problem, by \Cref{thm:aouad}.
\Halmos\endproof

Similarly, there is a contrast between \Cref{cor:budgetAdditiveCompHard} and \Cref{prop:mechPolyTime}, highlighting that computation for the restricted subclass of budget-additive mechanisms is actually harder.
In fact, for any subclass whose revenue gap with assortments is bounded, computation must be NP-hard.

\subsection{A Tighter LP Relaxation for Assortment Optimization} \label{sec:tighterLP}

Finally, we compare our mechanism design LP to existing LP's that are also relaxations of the assortment optimization problem on an explicitly-input list distribution.
To elaborate, assortments are equivalent to deterministic mechanisms (\Cref{prop:S_to_x}), which are equivalent to integer solutions in our LP for $\OPT^x$ (\Cref{def:x}).
Therefore, one could solve its corresponding Integer Program (IP), in which $x$ is restricted to be $\{0,1\}$-valued, to solve the assortment optimization problem.
The empirical runtime for solving such an IP, using heuristics such as Branch-and-Bound, has been observed to depend on how much larger the feasible region of the LP is than that of the IP.
In particular, for two different LP's with the same integer solutions, one is \textit{tighter} if its feasible region is contained within that of the other, in which case its IP tends to solve more quickly.

We now show that our mechanism design LP is tighter than previous LP's which have been formulated specifically to solve this explicit-input assortment optimization problem (also known as the \textit{first-choice product line optimization} problem).
It suffices to show that our LP is contained within the LP of \citet{bertsimas2019exact}, which has already been shown to be tighter than the other LP's of \citet{mcbride1988integer,belloni2008optimizing,feldman2019assortment}.

\begin{definition}[LP of \citet{bertsimas2019exact}] \label{def:BMLP}
In this LP, $x_j(\ell)$ denotes the probability of selling item $j$ to list $\ell$, and $z_j$ is an auxiliary variable that can be interpreted as the probability of item $j$ being included in the assortment.
Constraints~\eqref{eqn:bertMisEasy} enforce that the probability of selling an item cannot exceed the probability of it being in the assortment.
Constraints~\eqref{eqn:bertMisHard} are based on the fact that a list $\ell$ would only purchase an item \textit{less-preferred} to $\ell_k$ if $\ell_k$ is \textit{not} in the assortment.
Finally, constraints~\eqref{eqn:bertMisSumToUnity} ensure that each list is sold at most one item.
\begin{align}
\max &&\sum_{\ell\in\cL}p(\ell)\sum_{j\in\ell}r_jx_j(\ell) \nonumber
\\ && x_{\ell_k}(\ell) &\le z_{\ell_k} &\forall \ell\in\cL,k\le|\ell| \label{eqn:bertMisEasy}
\\ && \sum_{k'>k}x_{\ell_{k'}}(\ell) &\le 1-z_{\ell_k} &\forall \ell\in\cL,k\le|\ell| \label{eqn:bertMisHard}
\\ &&\sum_{j\in\ell}x_j(\ell) &\le1 &\forall\ell\in\cL \label{eqn:bertMisSumToUnity}
\\ &&z_j,x_j(\ell) &\ge0 &\forall\ell\in\cL,j\in\ell \nonumber
\end{align}
\end{definition}

\citet{bertsimas2019exact} show this to be a valid LP relaxation of the assortment optimization problem, in that its integer solutions are equivalent to assortments.
We now show our LP relaxation to be tighter, in that our feasible region is contained within their feasible region (after the $z$-variables have been projected out).

\begin{proposition}[Tighter Relaxation] \label{prop:tighter}
Given any feasible solution $x$ to the LP for $\OPT^x$, one can define $z$-variables so that $x$ is also a feasible solution to the LP from \Cref{def:BMLP}.
\end{proposition}

\proof{Proof of \Cref{prop:tighter}.}
Take a feasible solution $x$ to the LP for $\OPT^x$.  Set $z_j=\max_{\ell\ni j}x_j(\ell)$ for all $j$, after which~\eqref{eqn:bertMisEasy} follows.
Meanwhile,~\eqref{eqn:bertMisSumToUnity} is implied immediately from~\eqref{constr:sumToOne}.
Finally, to establish \eqref{eqn:bertMisHard}, fix any $\ell$ and $k$.
Suppose $z_{\ell_k}=x_{\ell_k}(\ell')$ for some $\ell'\ni\ell_k$.  Then we have
\begin{align*}
z_{\ell_k}=x_{\ell_k}(\ell')\le\sum_{j\in\ell_{1:k}\cap\ell'}x_j(\ell')\le\sum_{j\in\ell_{1:k}}x_j(\ell)\le 1-\sum_{k'>k}x_{\ell_{k'}}(\ell)
\end{align*}
where the first inequality holds because $x_j(\ell')\ge0$ and the second inequality holds by the IC constraints~\eqref{constr:IC}.
This establishes~\eqref{eqn:bertMisHard} and completes the proof.
\Halmos\endproof

It is also evident from the proof of \Cref{prop:tighter} that this containment is strict.
That is,~\eqref{eqn:bertMisEasy}--\eqref{eqn:bertMisHard} from the LP of \citet{bertsimas2019exact} are satisfied as long as $\sum_{j\in\ell_{1:k}}x_j(\ell)\ge x_{\ell_k}(\ell')$ for $\ell'\ni\ell_k$, which is strictly weaker than our IC constraint of $\sum_{j\in\ell_{1:k}}x_j(\ell)\ge\sum_{j\in\ell_{1:k}\cap\ell'}x_j(\ell')$ for all $\ell'$.
However, we should mention that although our LP relaxation is tighter and has fewer variables, it does have quadratically many constraints in $|\cL|$ instead of linearly many, and hence the relaxation itself can be slower to solve if $|\cL|$ is large.

%\subsubsection*{Summary of \Cref{sec:generalLists}.}
%We show that the revenue gap between mechanisms and assortments, although unbounded in the worst case (\textbf{\Cref{cor:unbdedGap}}), is small for short lists (\textbf{\Cref{thm:bdedListLength}}).
%Moreover, our LP describing the optimal mechanism can be solved in polynomial time (\textbf{\Cref{prop:mechPolyTime}}), and is tighter than existing LP relaxations for the assortment optimization problem (\textbf{\Cref{prop:tighter}}).
%Therefore, our mechanism design LP should be able to speed up Integer Programming approaches to assortment optimization, an investigation we leave to future work.
%Additionally, the concept of budget-additive mechanisms provides a relaxation for assortment optimization with a constant-factor revenue gap (\textbf{\Cref{thm:budgetAdditive}}).
%Although this subclass of mechanisms cannot be optimized in polynomial time (\textbf{\Cref{cor:budgetAdditiveCompHard}}), if it can be solved in special cases, then this can further speed up the computation of an (approximately) optimal assortment in those cases.

\section{Summary and Future Questions} \label{sec:conc}

In this paper we introduce a Bayesian mechanism design problem that captures the \textit{best method for selling items} in the assortment optimization environment.
The design is over random allocations of a set of alternatives that each bring a fixed benefit (``price'') to the principal, but the allocation must be incentive-compatible and individually-rational for an agent who has a private ordinal preference and an outside option.
We believe the model to be quite parsimonious, and it arises naturally from the intersection of three large bodies of work: assortment optimization, Bayesian mechanism design, and mechanism design without money.
While we focus on the single-agent problem in this paper, a natural extension is to have multiple agents compete for a limited supply of the items.
This would bring connections to celebrated mechanisms such as Gale's top trading cycles; we provide an example of this connection in \Cref{sec:multBuyer}.

Our main technique is to use an optimal stopping problem, that is much easier to interpret, to upper-bound the revenue of mechanisms.
This leads to a clean proof that assortments are optimal for Markov Chain choice models.
We also introduce the notion of ``cross-path'' monotonicity constraints on the decisions of an optimal stopping policy, which leads to our more general sufficient condition under which assortments are optimal.
This result is proved using an existential, extremal argument, and also extends the externality-adjustment framework of \citet{desir2020constrained} from Markov Chains to general ranking distributions. 
We show that special cases of Nested Logit, among other choice models not captured by a Markov Chain, satisfy our sufficient condition, and believe it could be of general interest for unifying Markov Chain with Nested Logit.

We also derive suboptimality gaps and computational results for general ranking distributions.
From here, we believe the most interesting results for future exploration are the tightened LP relaxation which can speed up assortment optimization on explicitly-input ranking distributions, and the 1-to-1 correspondence between IC mechanisms and monotone submodular functions.
It is also an interesting theoretical question how to explicitly construct a family of counterexamples demonstrating the unbounded gap between assortments and mechanisms, that does not rely on the complexity assumption $P\neq NP$.

Finally, a major implication of our work is that the large literature on MNL and Markov Chain choice models is not just optimizing assortments---it is actually solving the mechanism design problem!
A priori, it is not even clear that such a problem for MNL/Markov Chain can be solved in polynomial time, since the number of types (and hence the size of our mechanism design LP) is exponential in the size of the input.
We hope the results and questions built in this paper will draw more attention to the assortment optimization problem from communities interested in mechanism design, and vice versa.

% Acknowledgments here
\ACKNOWLEDGMENT{%
A one-page abstract of this article is in the Proceedings of the 23rd ACM Conference on Economics and Computation (EC).
We thank Joey Huchette and Juan Pablo Vielma for help with using the JuMP package \citep{DunningHuchetteLubin2017} to computationally check the integrality of polytopes arising from randomized ordinal IC constraints.
We also thank Nick Arnosti, Santiago Balseiro, Irene Lo, and Peng Shi for detailed comments on the presentation of this work, and Yilun Chen, Vineet Goyal, and Velibor Mi\v{s}i\'{c} for insightful discussions.
Finally, we thank the anonymous reviewers for \textit{Management Science} and Department Editor Itai Ashlagi, who invested significant effort in helping us motivate, position, and connect this work.
}% Leave this (end of acknowledgment)

% Bibliography
\bibliographystyle{informs2014}
\bibliography{bibliography}

% FOR SUBMISSION
%\ECSwitch
%\ECDisclaimer
%\ECHead{E-Companion}

% FOR SSRN
\clearpage

% Appendix here
% Options are (1) APPENDIX (with or without general title) or
%             (2) APPENDICES (if it has more than one unrelated sections)
% Outcomment the appropriate case if necessary
%
% \begin{APPENDIX}{<Title of the Appendix>}
% \end{APPENDIX}
%
%   or
%
\begin{APPENDICES}
\crefalias{section}{appendix}

\section{Generalized Formulation with Multiple Buyers} \label{sec:multBuyer}

The seller has one copy of each item in a set $N=\{1,\ldots,n\}$.
Each item $j\in N$ has a fixed price $r_j\ge0$.
There is a set of buyers $M=\{1,\ldots,m\}$, where each buyer has a random list that is assumed to be drawn \textit{independently}.
We let $\cL_i$ denote the support for buyer $i$'s list, and $p_i(\ell_i)$ denote the probability that buyer $i$ has list $\ell_i$, with $\sum_{\ell_i\in\cL_i}p_i(\ell_i)=1$ for all $i$.

A direct mechanism maps each list \textit{profile} $\vell:=(\ell_1,\ldots,\ell_m)$ lying in the set of feasible reports $\vcL:=\cL_1\times\cdots\times\cL_m$ to an allocation matrix $X(\vell)$, where $X_{ij}(\vell)$ denotes the probability of buyer $i$ being assigned item $j$.
Due to only having one copy of each item, and the fact that each buyer cannot be assigned more than one item, $X$ must satisfy the constraints
\begin{align}
\sum_{i:\ell_i\ni j}X_{ij}(\vell) &\le 1 &\forall \vell\in\vcL, j\in N \label{mult:item}
\\ \sum_{j\in\ell_i}X_{ij}(\vell) &\le 1 &\forall \vell\in\vcL, i\in M \label{mult:buyer}
\\ X_{ij}(\vell) &\ge 0 &\forall \vell\in\vcL, j\in\ell_i \label{mult:nonNeg}
\end{align}
Satisfying these constraints guarantees that under any profile $\vell$ that could be reported, the resulting matrix $X(\vell)$ will be implementable, i.e.\ there will be a randomized assignment of items to buyers such that no item gets assigned more than once w.p.~1, no buyer receives more than one item w.p.~1, and the probability that each buyer $i$ receives each item $j$ is exactly $X_{ij}(\vell)$.  This is due to the integrality of the bipartite matching polytope.

Finally, with multiple buyers there are different options for the incentive-compatibility constraint, depending on what the buyers assume about each other.
The most restrictive constraint is dominant-strategy incentive compatibility (DSIC), which imposes for each buyer $i\in M$, and any possibility of what the other agents report $\vell_{-i}:=(\ell_{i'})_{i'\in M\setminus\{i\}}$ lying in $\vcL_{-i}:=\prod_{i'\in M\setminus\{i\}}\cL_{i'}$, that truthful reporting will satisfy the single-buyer IC constraint used throughout this paper (i.e. truthful reporting will maximize buyer $i$'s likelihood of receiving one of their $k$ most-preferred items, simultaneously for all $k$).
This can be formally expressed as
\begin{align} \label{mult:DSIC}
\sum_{j\in(\ell_i)_{1:k}}X_{ij}(\ell_i,\vell_{-i})-\sum_{j\in(\ell_i)_{1:k}\cap\ell'_i}X_{ij}(\ell'_i,\vell_{-i}) &\ge0 &\forall \vell_{-i}\in\vcL_{-i},\ell_i\in\cL_i,k\le|\ell_i|,\ell'_i\in\cL_i
\end{align}
On the other hand, Bayesian incentive-compatibility (BIC) only imposes the single-buyer IC constraint for each $i\in M$ assuming that $i$ knows the distributions for the other buyers and everyone buyer will report truthfully.
It is more relaxed because it only imposes that~\eqref{mult:DSIC} holds after taking an expectation over $\vell_{-i}$:
\begin{align} \label{mult:BIC}
\sum_{\vell_{-i}\in\vcL_{-i}}\left(\sum_{j\in(\ell_i)_{1:k}}X_{ij}(\ell_i,\vell_{-i})-\sum_{j\in(\ell_i)_{1:k}\cap\ell'_i}X_{ij}(\ell'_i,\vell_{-i})\right)\prod_{i'\in M\setminus\{i\}}p_{i'}(\ell_{i'}) &\ge0 &\forall \ell_i\in\cL_i,k\le|\ell_i|,\ell'_i\in\cL_i
\end{align}
We note that under either DSIC or BIC, individual rationality is still implied assuming one uses our strong single-buyer IC constraint, which is why we only have a variable $X_{ij}(\vell)$ if $j\in\ell_i$.

The DSIC mechanism design problem can then be formulated as maximizing the expected revenue
\begin{align}
\sum_{\vell\in\vcL}\left(\sum_{i\in M}\sum_{j\in\ell_i}r_jX_{ij}(\vell)\right)\prod_{i\in M}p_i(\ell_i)
\end{align}
subjective to constraints~\eqref{mult:item}, \eqref{mult:buyer}, \eqref{mult:nonNeg}, and \eqref{mult:DSIC}.
The BIC mechanism design problem is identical, except that constraint~\eqref{mult:DSIC} is replaced with~\eqref{mult:BIC}.

We now provide a simple example with $n=m=2$ for which the deterministic \textit{top-trading-cycles} mechanism, using deliberate item endowments, is optimal even among the larger class of BIC mechanisms.  The top-trading-cycles mechanism is a common one from the literature on mechanism design over allocations (more specifically, the house allocation problem), so the purpose of this example of to provide a connection to the multi-buyer generalization of our problem.  We also show for this example that \textit{serial dictatorships} are suboptimal.

\begin{example}
The seller has one copy of each of two items $A,B$.
There are two buyers, with buyer 1's list equally likely to be $(A)$, $(B)$, or $(AB)$, and buyer 2's list equally likely to be $(A)$, $(B)$, or $(BA)$.
The buyers' lists are drawn independently.
The fixed prices are $r_A=r_B=1$, so the seller's objective is to maximize the expected number of items sold.

We first note that an upper bound on the expected revenue of any mechanism is 16/9.
This is because even if the seller knows the true realizations of buyer lists, they can still only sell one item if both lists realize to $(A)$ or both lists realize to $(B)$, earning expected revenue $2(7/9)+1(2/9)=16/9$.

Now, any \textit{serial dictatorship} mechanism can earn revenue at most 15/9.  Indeed, by symmetry, we can WOLOG assume buyer 1 to go first, choosing their favorite item from $\{A,B\}$, with buyer 2 having the option of purchasing the remaining item.  Then buyer 2 has a 2/3 chance of not purchasing an item when their list realizes to $(A)$ (because buyer 1 takes item $A$ with lists $(A)$ or $(AB)$), and a 1/3 chance of not purchasing when their list realizes to $(B)$ (because buyer 1 takes item $B$ with list $(B)$).
Therefore, the revenue is upper-bounded by $2-\frac{2/3+1/3}{3}=15/9$.

On the other hand, consider the following mechanism.  We initially give item $B$ to buyer 1 and item $A$ to buyer 2.  Then, the buyers trade items only if \textit{both} buyers agree.  At the end, each buyer can choose to purchase the item they have.
We claim that this mechanism is optimal, earning revenue 16/9 which matches the upper bound.
To see why, note that whenever a buyer's list realizes to length 2, they start with their less-preferred item, and hence a trade occurs whenever the other buyer is not willing to purchase their starting item.
It can be checked that this results in both items being sold unless the lists both realize to $(A)$ or both realize to $(B)$.
Therefore, the expected revenue of this mechanism, which is a special case of the \textit{top-trading-cycles} mechanism where buyer 1 is endowed with item $B$ and buyer 2 is endowed with item $A$, is 16/9.
We note that if we reverse the endowments, with buyer 1 starting with $A$ and buyer 2 starting with $B$, then the revenue is significantly worse, being 14/9.\Halmos
\end{example}

\section{Interpretation as a Robust Mechanism} \label{sec:robust}

In this \namecref{sec:robust} we formulate a robust mechanism design problem in which the objective considers the \textit{worst case} cardinal utility distribution consistent with a given ordinal distribution.
We show that the formulation studied throughout this paper provides a feasible solution to the robust problem.  We also construct an example where it does not provide an optimal solution.

The robust mechanism design problem is formulated as follows.
We are given items $j\in N=\{1,\ldots,n\}$ with fixed prices $r_j\ge0$, and a distribution $p$ over lists $\ell\in\cL$, as defined throughout this paper.
The robust problem differs in that the buyer has a true cardinal utility vector $\vu\in\bR^{\{0\}\cup N}$, which is randomly drawn from a distribution such that the induced ordinal distribution is consistent with $p$.
The designer commits to a cardinal mechanism, which is defined by a \textit{menu} $\cM$ of \textit{entries} $\vx\in[0,1]^{\{0\}\cup N}$, where each $\vx$ describes a randomized assignment in which the buyer purchases item $j$ (for fixed price $r_j$) w.p.~$\vx_j$ for all $j=0,1,\ldots,n$, with $\sum_{j=0}^n\vx_j=1$.
The 0 entry defined by $\vx_j=\bI(j=0)$ for all $j=0,\ldots,n$ must lie in $\cM$.
Finally, the buyer chooses the option $\vx\in\cM$ that maximizes their own expected utility $\sum_{j=0}^n\vu_j\vx_j$.
The seller's objective is to maximize the \textit{worst-case} expected revenue, facing the lack of distributional information about $\vu$ other than it is consistent with $p$.

The worst-case revenue for a menu $\cM$ can be expressed as follows.
For each $\ell\in\cL$, let $E^{\cM}(\ell)$ denote the \textit{exposable} entries for list $\ell$, defined as the subset of $\vx$'s in $\cM$ for which there exist a utility vector $\vu\in\bR^{\{0\}\cup N}$ satisfying both consistency with $\ell$, i.e.\
\begin{align} \label{eqn:consistentWithEll}
\vu_{\ell_1}>\cdots>\vu_{\ell_{\echar}}>\vu_0;\ \vu_0>\vu_j\ \forall j\notin\ell,
\end{align}
and expected utility maximization for $\vx$, i.e.\
\begin{align} \label{eqn:expectedUtilityMaximization}
\vx\in\argmax_{\vx'\in\cM}\sum_{j=0}^n\vu_j\vx'_j.
\end{align}
Meanwhile, for each $\vx\in\cM$, let $r(\vx):=\sum_{j=1}^n r_j\vx_j$ denote the seller's expected revenue when entry $\vx$ is chosen.
Using these two definitions, whenever the buyer's utility realizes to be consistent with a $\ell\in\cL$, the buyer will choose the $\vx\in E^{\cM}(\ell)$ that \textit{minimizes} $r(\vx)$.
Therefore, the worst-case revenue of menu $\cM$ is
\begin{align} \label{eqn:robustObj}
\sum_{\ell\in\cL}p(\ell)\inf_{\vx\in E^{\cM}(\ell)}r(\vx).
\end{align}

Two remarks are in order.
In~\eqref{eqn:consistentWithEll}, we impose strictness, because otherwise all lists could be made to choose the 0 entry by setting $\vu_j=0$ for all $j=0,\ldots,n$.
Given this, in~\eqref{eqn:expectedUtilityMaximization} it does not matter whether we impose strictness, because the utility vector $\vu$ can always be perturbed so that the $\argmax$ in~\eqref{eqn:expectedUtilityMaximization} is unique.

\begin{proposition} \label{prop:feasSolnToRobust}
Given any instance defined by $\{r_j:j\in N\}$ and $\{p(\ell):\ell\in\cL\}$, the optimal objective value $\OPT^x$ of the mechanism design problem from \Cref{def:x} provides a lower bound to the robust mechanism design problem of maximizing~\eqref{eqn:robustObj} over menus $\cM$ containing the 0 entry.
\end{proposition}

\proof{Proof of \Cref{prop:feasSolnToRobust}.}
Fix an optimal solution $x$ to our mechanism design problem from \Cref{def:x}.
Based on $x$, we specify a feasible menu $\cM$ for the robust problem whose objective value~\eqref{eqn:robustObj} is at least $\OPT^x$.
For any $\ell\in\cL$, let $x(\ell)$ denote the vector $(1-\sum_{j=1}^n x_j(\ell),x_1(\ell),\ldots,x_n(\ell))$, which we note is a feasible entry for $\cM$ due to LP constraints~\eqref{constr:sumToOne} and~\eqref{constr:nonNeg}.
Moreover, we will let $\vZero=(1,0,\ldots,0)$ denote the 0 entry for $\cM$.

Consider the menu $\cM=\{x(\ell):\ell\in\cL\}\cup\{\vZero\}$.  We claim that for any $\ell\in\cL$, the set $E^{\cM}(\ell)$ is the singleton $\{x(\ell)\}$.
This would complete the proof because $r(x(\ell))=\sum_jr_jx_j(\ell)$, making objective function~\eqref{eqn:robustObj} identical to the objective function from \Cref{def:x}.

We first fix an $\ell\in\cL$ and show that $x(\ell)\in E^{\cM}(\ell)$.  This is easy to see---in fact, take \textit{any} $\vu$ satisfying~\eqref{eqn:consistentWithEll}.  For $\vx'\in\cM$, there are two possibilities: $\vx'=\vZero$, or $\vx'=x(\ell')$ for some $\ell'\in\cL$.  If $\vx'=\vZero$, then it is clear that
\begin{align} \label{eqn:8090}
\sum_{j=0}^n\vu_j \big(x(\ell)\big)_j=\vu_0(1-\sum_{j\in\ell}x_j(\ell))+\sum_{j\in\ell}\vu_jx_j(\ell)\ge\vu_0=\sum_{j=0}^n\vu_j\vx'_j
\end{align}
from the conditions in~\eqref{eqn:consistentWithEll} and hence $x(\ell)$ lies in the $\argmax$ in~\eqref{eqn:expectedUtilityMaximization}.
Otherwise, if $\vx'=x(\ell')$ for some $\ell'\in\cL$, then we can write
\begin{align}
\sum_{j=0}^n\vu_j \big(x(\ell)\big)_j
&=\vu_0+\sum_{k=1}^{|\ell|}(\vu_{\ell_k}-\vu_0)x_{\ell_k}(\ell) \nonumber
\\ &=\vu_0+\sum_{k=1}^{|\ell|}(\vu_{\ell_k}-\vu_{\ell_{k+1}})\sum_{k'=1}^kx_{\ell_{k'}}(\ell) \nonumber
\\ &\ge\vu_0+\sum_{k=1}^{|\ell|}(\vu_{\ell_k}-\vu_{\ell_{k+1}})\sum_{j\in\ell_{1:k}\cap\ell'}x_j(\ell') \nonumber
\\ &=\vu_0(1-\sum_{j\in\ell\cap\ell'}x_j(\ell'))+\sum_{j\in\ell\cap\ell'}\vu_jx_j(\ell') \nonumber
\\ &\ge\sum_{j=0}^n\vu_j\big(x(\ell')\big)_j \label{eqn:7271}
\end{align}
where in the second equality $\ell_{|\ell|+1}$ is understood to be item 0,
in the first inequality we have used IC constraints~\eqref{constr:IC} and the fact that the expressions in parentheses are strictly positive by~\eqref{eqn:consistentWithEll}, and in the final inequality we have used the fact that $\vu_0>\vu_j$ for $j\in\ell'\setminus\ell$ by~\eqref{eqn:consistentWithEll}.
This completes the proof that $x(\ell)\in E^{\cM}(\ell)$.

We now show that $\vx'\notin E^{\cM}(\ell)$ for any $\vx'\in\cM$ distinct from $x(\ell)$.
To see this, suppose for contradiction the existence of a $\vu$ satisfying~\eqref{eqn:consistentWithEll} such that $\vx'$ lies in the $\argmax$ in~\eqref{eqn:expectedUtilityMaximization}.
There are two cases: $\vx'=\vZero$ or $\vx'=x(\ell')$ for some $\ell'\in\cL$, for which the deductions above resulting in either~\eqref{eqn:8090} or~\eqref{eqn:7271} can still be respectively made.  In order for the inequalities to not be strict, we must have either $x_j(\ell)=0$ for all $j\in\ell$ or $x_j(\ell)=x_j(\ell')$ for all $j=0,\ldots,n$ respectively.
However, this implies that $x(\ell)=\vZero$ or $x(\ell)=x(\ell')$ respectively, contradicting the fact that $\vx'$ is distinct from $x(\ell)$.
Therefore, $E^{\cM}(\ell)$ cannot contain any vectors other than $x(\ell)$ and this completes the proof of \Cref{prop:feasSolnToRobust}.
\Halmos\endproof

We now show that \Cref{def:x} is not identical to the robust mechanism design problem, by constructing an instance on which $\OPT^x$ is strictly lower than the optimal objective value of the robust problem.

\begin{example}
There are five items $j=1,\ldots,5$ with $r_1=2,r_2=r_3=r_4=1,r_5=1.5$.
The buyer's ordinal distribution is uniform over the 12 lists
$$
(21),(31),(41),(32),(42),(43),(5),(5),(5),(251),(351),(451).
$$
It can be computed that $\OPT^x=1.3125$ for this instance, with an optimal solution being the \textit{budget-additive mechanism} (see \Cref{sec:budgetAdditive}) with $w_1=w_2=w_3=w_4=0.5,w_5=1$ and $B=1$.
Indeed, this mechanism extracts an expected revenue of 1.5 from the first three lists,
an expected revenue of 1 from lists $(32),(42),(43)$,
an expected revenue of 1.5 from lists $(5),(5),(5)$,
and an expected revenue of 1.25 from the final three lists,
for an overall average of $\frac{1.5+1+1.5+1.25}{4}=1.3125$.

Meanwhile, consider the robust mechanism design problem for the same instance.  A feasible solution is given by the following menu $\cM$, with non-zero entries
$$
(\nicefrac{1}{2},\nicefrac{1}{2},0,0,0),(\nicefrac{1}{2},0,\nicefrac{1}{2},0,0),(\nicefrac{1}{2},0,0,\nicefrac{1}{2},0),(0,\nicefrac{1}{2},\nicefrac{1}{2},0,0),(0,\nicefrac{1}{2},0,\nicefrac{1}{2},0),(0,0,\nicefrac{1}{2},\nicefrac{1}{2},0),(0,0,0,0,1).
$$
The first six lists would choose the first six entries respectively, with the set of exposable entries always being a singleton.
List $\ell=(5)$ would similarly have $E^{\cM}(\ell)=\{(0,0,0,0,1)\}$.
Finally, lists $\ell=(251),(351),(451)$ have $|E^{\cM}(\ell)|=2$, consisting of one entry that deterministically grants their 2nd choice and another entry that grants their 1st and 3rd choices with 50\% probability each.
However, regardless of which entry is chosen, the expected revenue would be 1.5.
Therefore, the objective value~\eqref{eqn:robustObj} for this menu $\cM$ is $\frac{1.5+1+1.5+1.5}{4}=1.375$, strictly greater than $\OPT^x$.\Halmos
\end{example}

\section{Proofs from \Cref{sec:probDefn}} \label{sec:defPfsRelaxations}

\proof{Proof of \Cref{prop:S_to_x}.}
We first prove the forward direction, which is easy to see.
To show that $x$ is IC, note that constraints~\eqref{constr:sumToOne} and~\eqref{constr:nonNeg} are satisfied by definition.  Constraints~\eqref{constr:IC} are satisfied by considering two cases.  First, if $\ell\in\cL$ is such that $x_{\ell_{k'}}(\ell)=1$ for some $k'\le|\ell|$, then the LHS of~\eqref{constr:IC} is 1 (and hence~\eqref{constr:IC} is satisfied) as long as $k\ge k'$.  On the other hand, if $k<k'$, then it must be the case that $\ell_1,\ldots,\ell_k$ are not in the assortment $S$, by definition of $x_j(\ell)$ in~\eqref{eqn:setXFromS}.  Therefore, the RHS of~\eqref{constr:IC} is 0 (and hence~\eqref{constr:IC} is again satisfied).  In the second case, $\ell$ is such that $x_j(\ell)=0$ for all $j\in\ell$.  This means that none of the items $j$ on $\ell$ are in the assortment $S$.  Then no matter what $\ell'$ the buyer reports, it is not possible for them to receive an item on their list, and thus the RHS of~\eqref{constr:IC} is always 0 (and hence~\eqref{constr:IC} is still satisfied). 
Finally, to see that $\Rev[x]=\Rev[S]$, note that $\Pr[j\succeq S_0]$ equals the sum of probabilities $p(\ell)$ over lists $\ell$ such that $x_j(\ell)=1$, by definition, completing the proof of the forward direction.

To prove the converse direction, take any IC mechanism with $x_j(\ell)\in\{0,1\}$ for all $\ell\in\cL$ and $j\in\ell$.
Define $S$ to be the set of $j\in N$ for which there exists an $\ell'$ such that $x_j(\ell')=1$, i.e.\ the set of items that are possible to obtain through some ``lie'' $\ell'$.
To show that~\eqref{eqn:setXFromS} is satisfied, we again consider two cases.
First, if $\ell\in\cL$ is such that $x_{\ell_{k'}}(\ell)=1$ for some $k'\le|\ell|$, then IC constraints~\eqref{constr:IC} imply that $x_{\ell_k}(\ell')=0$ for any $k<k'$ and $\ell'\in\cL$.
In other words, none of the items lying before position $k'$ in list $\ell$ are in the assortment $S$ we defined, and hence $x_{\ell_{k'}}(\ell)=1$ satisfies the description in~\eqref{eqn:setXFromS}.
For the other part of description~\eqref{eqn:setXFromS}, the fact that $x_j(\ell)=0$ for all $j\neq\ell_{k'}$ holds by constraints~\eqref{constr:sumToOne} and~\eqref{constr:nonNeg}.
In the second case, $\ell$ is such that $x_j(\ell)=0$ for all $j\in\ell$.
We can again use constraints~\eqref{constr:IC} to see it must be the case that $j\notin S$ for any $j\in\ell$, completing the proof of the converse direction.
\Halmos\endproof

\proof{Proof of \Cref{prop:x_to_f}.}
It is immediate that $f$ is non-negative and increasing, from the non-negativity constraints~\eqref{constr:nonNeg} on mechanism $x$.
It is also immediate that $f$ is upper-bounded by 1, from constraints~\eqref{constr:sumToOne}.
To establish~\eqref{eqn:diffF}, note that by the LP constraints~\eqref{constr:IC}, the probability of a type $\ell$ receiving one of their $k'$ most-preferred items equals $f$ evaluated on this set of items, i.e.\
\begin{align} \label{eqn:icConstrOnObj}
\sum_{j\in\ell_{1:k'}}x_j(\ell) &=f(\ell_{1:k'}) &\forall\ell\in\cL,k'=1,\ldots,|\ell|.
\end{align}
For any $\ell\in\cL$ and $k=1,\ldots,|\ell|$,
subtracting~\eqref{eqn:icConstrOnObj} with $k'=k-1$ from~\eqref{eqn:icConstrOnObj} with $k'=k$ (or subtracting $f(\emptyset)=0$ if $k=1$), we see that the probability $x_{\ell_k}(\ell)$ of type $\ell$ receiving their $k$-th most preferred item equals $f(\ell_{1:k})-f(\ell_{1:k-1})$, completing the proof.
\Halmos\endproof

\proof{Proof of \Cref{prop:f_to_phi}.}
The optimization problem for $\OPT^f$, using~\eqref{def:revFAgain} as the expression for $\Rev[f]$, can be formulated by the following LP with decision variables $f(S)$:
\begin{align}
\max\sum_{\rho}r_{\rho_{\echar}}(f(\rho)-f(\rho_{\pre}))\Pr[\rho] \nonumber
\\ f(S) &\le f(S\cup\{j\}) &\forall S\subsetneq N,j\notin S \label{constr:increasing}
\\ f(S) &\in[0,1] &\forall S\subseteq N \label{constr:f01}
\end{align}
The matrix formed by the LHS of constraints~\eqref{constr:increasing}
is totally unimodular, because all of its entries lie in $\{-1,0,+1\}$, and every row contains exactly one $+1$ entry and one $-1$ entry \citep{schrijver1998theory}.
Therefore, the feasible region defined by~\eqref{constr:increasing}--\eqref{constr:f01} is integral, and moreover since the objective function is linear, there must exist an integer optimal solution with $f(S)\in\{0,1\}$ for all $S$.  This defines an increasing boolean set function $f$, completing the proof of the first statement.

To show that $\Rev[\phi]=\Rev[f]$, note that when $f$ is increasing and boolean-valued, $f(\rho)-f(\rho_{\pre})=f(\rho)\prod_{k=1}^{|\rho|-1}(1-f(\rho_{1:k}))$.
This in turn equals $\phi_{\rho_{\echar}}(\rho_{\pre})\prod_{k=1}^{|\rho|-1}(1-\phi_{\rho_k}(\rho_{1:k-1}))$, by the definition of $\phi$ in~\eqref{eqn:phiFromf}.
Substituting into the expression for $\Rev[f]$, we see that
\begin{align*}
\Rev[f]=\sum_{\rho}r_{\rho_{\echar}}\phi_{\rho_{\echar}}(\rho_{\pre})\prod_{k=1}^{|\rho|-1}(1-\phi_{\rho_k}(\rho_{1:k-1}))\Pr[\rho],
\end{align*}
completing the proof of the second statement.
\Halmos\endproof

\section{Proofs from \Cref{sec:markovChainMainResult}} \label{sec:defPfsMC}

\proof{Proof of \Cref{thm:mcmr}.}
Our goal is to show that $\OPT^S=\OPT^x$.  Since we have established in general that $\OPT^S\le\OPT^x\le\OPT^{\phi}$, it suffices to show $\OPT^{\phi}\le\OPT^S$.  Under a Markov Chain choice model, consider the following relaxation on the optimal stopping value $\OPT^{\phi}$.
First, instead of only adding an item $j$ to the list when it is visited for the \textit{first} time, we allow for stopping on an item $j$ (ending with reward $r_j)$ \textit{any} time $j$ is visited.
Furthermore, we place no constraints on the stopping policy (whereas before $\phi$ could only make deterministic stopping decisions based on the unordered history $H$, and had to satisfy monotonicity constraints).

This is now an infinite-horizon optimal stopping problem which terminates in finite time w.p.~1.
Since stopping is allowed on repeat visits to items $j$ and there are no longer constraints on the stopping policy, the value-to-go under the optimal stopping policy is fully determined by the current item $j$, which we denote using $V[j]$.  Moreover, these values-to-go must satisfy
\begin{align} \label{eqn:optStopSys}
V[j] &=\max\{r_j,\sum_{j'\in N}\rho_{jj'}V[j']\} &\forall j\in N.
\end{align}
After solving system~\eqref{eqn:optStopSys} which has $|N|$ equations and $|N|$ unknowns, a policy that stops on an item $j$ if and only if $r_j\ge\sum_{j'\in N}\rho_{jj'}V[j']$ is optimal \citep[see][]{bertsekas1995dynamic}.
The expected revenue of this policy is equal to the expected revenue earned by the assortment $S=\{j\in N:r_j\ge\sum_{j'\in N}\rho_{jj'}V[j']\}$.
Therefore, optimal stopping policies are no better than assortments under Markov Chain choice models, completing the proof of \Cref{thm:mcmr}.
\Halmos\endproof

\section{Proofs from \Cref{sec:complexPf}} \label{sec:defPfsMR}

\proof{Proof of \Cref{lem:revAdj}.}
We start with the definition of $\Rev[\phi]$ for an arbitrary monotone stopping policy $\phi$, and expand $\Pr[\rho]$ as $\prod_{k=1}^{|\rho|}q(\rho_{1:k})$ (which holds by the definition of $q$ in \Cref{def:qAndTree}):
\begin{align*}
\Rev[\phi]=&\sum_{\rho}r_{\rho_{\echar}}\phi_{\rho_{\echar}}(\rho_{\pre})q(\rho)\prod_{k=1}^{|\rho|-1}(1-\phi_{\rho_k}(\rho_{1:k-1}))q(\rho_{1:k}).
\end{align*}
We split the sum on the RHS into the contribution from prefixes $\rho$ that do not intersect $S$, and prefixes $\rho'$ that end on $S$ but do not intersect $S$ before the end:
\begin{align}
\Rev[\phi]=&\sum_{\rho:\rho\cap S=\emptyset}r_{\rho_{\echar}}\phi_{\rho_{\echar}}(\rho_{\pre})q(\rho)\prod_{k=1}^{|\rho|-1}(1-\phi_{\rho_k}(\rho_{1:k-1}))q(\rho_{1:k}) \nonumber
\\ &+\sum_{\substack{\rho':\rho_{\echar}'\in S,\\ \rho_{\pre}'\cap S=\emptyset}}\phi_{\rho'_{\echar}}(\rho'_{\pre})\left(r_{\rho_{\echar}'}q(\rho')\prod_{k=1}^{|\rho'|-1}(1-\phi_{\rho_k'}(\rho_{1:k-1}'))q(\rho_{1:k}')\right). \label{eqn:leftOff}
\end{align}
Note that we did not need to account for the contribution from prefixes that intersect $S$ before the end, because policy $\phi$ would have stopped on the first intersection with $S$.

We now rewrite the expression inside the large parentheses in~\eqref{eqn:leftOff} as follows:
\begin{align}
&r_{\rho_{\echar}'}q(\rho')\prod_{k=1}^{|\rho'|-1}(1-\phi_{\rho_k'}(\rho_{1:k-1}'))q(\rho_{1:k}') \nonumber
\\ &=r_{\rho_{\echar}'}\Big(\prod_{k=1}^{|\rho'|}q(\rho_{1:k}')\Big)\Big(1-\sum_{k=1}^{|\rho'|-1}\phi_{\rho_k'}(\rho_{1:k-1}')\prod_{k'=1}^{k-1}(1-\phi_{\rho_{k'}'}(\rho_{1:k'-1}'))\Big) \nonumber
\\ &=r_{\rho_{\echar}'}\prod_{k=1}^{|\rho'|}q(\rho_{1:k}')
-\sum_{k=1}^{|\rho'|-1}\Big(r_{\rho_{\echar}'}\prod_{k'=k+1}^{|\rho'|}q(\rho_{1:k'}')\Big)\phi_{\rho_k'}(\rho_{1:k-1}')q(\rho_{1:k}')
\Big(\prod_{k'=1}^{k-1}(1-\phi_{\rho_{k'}'}(\rho_{1:k'-1}'))q(\rho_{1:k'}')\Big).
\label{eqn:largeParen}
\end{align}
The first equality holds because $\prod_{k=1}^{|\rho'|-1}(1-\phi_{\rho_k'}(\rho_{1:k-1}'))+\sum_{k=1}^{|\rho'|-1}\phi_{\rho_k'}(\rho_{1:k-1}')\prod_{k'=1}^{k-1}(1-\phi_{\rho_{k'}'}(\rho_{1:k'-1}'))$ sums to unity,
describing the mutually exclusive and collectively exhaustive events of policy $\phi$ not stopping on any of the first $|\rho'|-1$ elements of prefix $\rho'$, or stopping on one of them (say the $k$'th element) after not stopping before that (not stopping on elements $k'=1,\ldots,k-1$).

Substituting~\eqref{eqn:largeParen} into the expression inside large parentheses in~\eqref{eqn:leftOff}, we obtain:
\begin{align*}
\Rev[\phi]
=&\sum_{\rho:\rho\cap S=\emptyset}r_{\rho_{\echar}}\phi_{\rho_{\echar}}(\rho_{\pre})q(\rho)\prod_{k=1}^{|\rho|-1}(1-\phi_{\rho_k}(\rho_{1:k-1}))q(\rho_{1:k})
+\sum_{\substack{\rho':\rho_{\echar}'\in S,\\ \rho_{\pre}'\cap S=\emptyset}}r_{\rho_{\echar}'}\prod_{k=1}^{|\rho'|}q(\rho_{1:k}')
\nonumber
\\ &-\sum_{\substack{\rho':\rho_{\echar}'\in S,\\ \rho_{\pre}'\cap S=\emptyset}}
\sum_{k=1}^{|\rho'|-1}\Big(r_{\rho_{\echar}'}\prod_{k'=k+1}^{|\rho'|}q(\rho_{1:k'}')\Big)\phi_{\rho_k'}(\rho_{1:k-1}')q(\rho_{1:k}')
\Big(\prod_{k'=1}^{k-1}(1-\phi_{\rho_{k'}'}(\rho_{1:k'-1}'))q(\rho_{1:k'}')\Big)
\\=&\sum_{\rho:\rho\cap S=\emptyset}r_{\rho_{\echar}}\phi_{\rho_{\echar}}(\rho_{\pre})q(\rho)\prod_{k=1}^{|\rho|-1}(1-\phi_{\rho_k}(\rho_{1:k-1}))q(\rho_{1:k})
+\Rev[S]
\nonumber
\\ &-\sum_{\rho:\rho\cap S=\emptyset}\phi_{\rho_{\echar}}(\rho_{\pre})q(\rho)
\Big(\prod_{k'=1}^{|\rho|-1}(1-\phi_{\rho_{k'}}(\rho_{1:k'-1}))q(\rho_{1:k'})\Big)
\sum_{\substack{\rho':\rho_{\echar}'\in S,\\ \rho_{\pre}'\cap S=\emptyset,\\ \rho'_{1:|\rho|}=\rho}}
\Big(r_{\rho_{\echar}'}\prod_{k'=|\rho|+1}^{|\rho'|}q(\rho_{1:k'}')\Big).
\end{align*}
In the third line above, we have used $\Rev[S]$ to replace a sum from the first line which described the revenue of the policy that stops on assortment $S$ and stops nowhere else.
In the fourth line above, we have let $\rho$ denote $\rho'_{1:k}$ (a prefix disjoint from $S$) and exchanged the sums from the second line, pulling forward the terms which only depend on the first $k$ elements of $\rho'$ and rewriting them in terms of $\rho$.
The inner sum in the fourth line is then over the $\rho'$ which start identically as $\rho$, end on $S$, but do not intersect $S$ before the end.

Having performed this exchange, we can now conveniently merge the sums from the third and fourth lines above to see that
\begin{align*}
&\Rev[\phi]-\Rev[S]
\\ &=\sum_{\rho:\rho\cap S=\emptyset}\Bigg(r_{\rho_{\echar}}-\sum_{\substack{\rho':\rho_{\echar}'\in S,\\ \rho_{\pre}'\cap S=\emptyset,\\ \rho'_{1:|\rho|}=\rho}}
r_{\rho_{\echar}'}\prod_{k=|\rho|+1}^{|\rho'|}q(\rho_{1:k}')\Bigg)\phi_{\rho_{\echar}}(\rho_{\pre})q(\rho)\prod_{k=1}^{|\rho|-1}(1-\phi_{\rho_k}(\rho_{1:k-1}))q(\rho_{1:k})
\\ &=\sum_{\rho:\rho\cap S=\emptyset}\Bigg(r_{\rho_{\echar}}-\sum_{j\in S}r_{j}\Pr[j\succeq S_0|\rho]\Bigg)\Pr[\rho]
\phi_{\rho_{\echar}}(\rho_{\pre})\prod_{k=1}^{|\rho|-1}(1-\phi_{\rho_k}(\rho_{1:k-1})).
\end{align*}
where the second equality uses the fact that $\rho$ is disjoint from $S$ to simplify the expression inside large parentheses, and also the fact that $q(\rho)\prod_{k=1}^{|\rho|-1}q(\rho_{1:k})=\Pr[\rho]$ which holds by definition of $q$.
Substituting in the definition of $S$-adjusted prices, this completes the proof of \Cref{lem:revAdj}.
\Halmos\endproof

\proof{Proof of \Cref{lem:tiers}.}
Fix an assortment $S\subsetneq N$ along with an item $j\notin S$, and consider the realizable prefixes $\rho$ with $\rho_{\echar}=j$ and $\rho\cap S=\emptyset$.
Construct an undirected graph, with a vertex for each such prefix, and an edge between two prefixes $\rho,\rho'$ if and only if $\rho_{\pre}\nsupseteq\rho'_{\pre}$ and $\rho'_{\pre}\nsupseteq\rho_{\pre}$ (neither is contained within the other when viewed as sets).
\Cref{def:monotoneTiered} implies that the futures from adjacent prefixes $\rho,\rho'$ in this graph must dominate each other, and since $S$ intersects neither $\rho$ nor $\rho'$, it must be that inequality~\eqref{eqn:domFuture} holds with equality:
\begin{align} \label{eqn:graphConnComp}
\Pr[j'\succeq S_0|\rho] &=\Pr[j'\succeq S_0|\rho'] &\forall j'\in S.
\end{align}

Consider the connected components of this graph.
By transitivity, the probability $\Pr[j'\succeq S_0|\rho]$ must be identical across all vertices $\rho$ in a connected component, for all $j'\in S$.
Note that it was important for the graph to only include vertices for the \textit{realizable} prefixes $\rho$ with $\Pr[\rho]>0$; otherwise there can only be very few\footnote{
If all prefixes were included, then there would be exactly three connected components: one for vertex $(j)$, one for all the prefixes $\rho$ with $\rho_{\echar}=j$ and $\rho_{\pre}=N\setminus S\setminus\{j\}$, and one for all other prefixes satisfying $\rho_{\echar}=j$ and $\rho\cap S=\emptyset$.
}
connected components.
Let $C$ denote the number of components in this graph.

Consider $\rho,\rho'$ in different components of this graph.
Since $\rho$ and $\rho'$ are not adjacent, it must be that one of them contains the other (otherwise there would be an edge between $\rho$ and $\rho'$), say $\rho\supseteq\rho'$.
We claim that then any other prefix $\varrho$ adjacent to $\rho$ must also satisfy $\varrho\supseteq\rho'$.
This is easy to see---since $\varrho$ is non-adjacent to $\rho'$, either $\varrho\supseteq\rho'$ or $\varrho\subseteq\rho'$, but if $\varrho\subseteq\rho'$, then $\varrho\subseteq\rho$ by transitivity, contradicting the fact that $\varrho$ is adjacent to $\rho$.
By propagating this argument throughout the connected component containing $\rho$, we establish that all prefixes in this component contain $\rho'$.
Similarly, we can argue that all prefixes in the connected component of $\rho'$ are contained within $\rho$ (and also contained within any $\varrho$ in the component of $\rho$).

Therefore, we can take a single representative $\rho_c$ from each of the components $c=1,\ldots,C$, and know that for any $c,c'$, either $\rho_c\subseteq\rho_{c'}$ or $\rho_{c'}\subseteq\rho_c$ (or both).
This defines a \textit{total preorder} on the connected components---a binary relation $\subseteq$ which is reflexive, transitive, and any pair of elements are comparable, but the relation is not necessarily antisymmetric ($\rho\subseteq\rho'$ and $\rho'\subseteq\rho$ does not imply that $\rho=\rho'$ as ordered prefixes).
Nonetheless, we can take the equivalence classes with respect to this preorder to get a total order.

Suppose that there are $T$ equivalence classes, where $T\le C$, and let $\cT(t)$ denote set of all prefixes lying in connected components belonging to equivalence class $t$, ordered so that whenever $t>t'$, and $\rho\in\cT(t),\rho'\in\cT(t')$, we have that $\rho\supseteq\rho'$ and $\rho\nsubseteq\rho'$.  This implies that $\rho\supsetneq\rho'$ as sets.
In conjunction with~\eqref{eqn:graphConnComp}, this completes the proof of condition~\ref{cond:one} in \Cref{lem:tiers}.

To establish condition~\ref{cond:two}, we claim that: an equivalence class either consists of a single component, or all of its components are isolated vertices.  To show this, suppose for contradiction that an equivalence class includes two components, one which includes two vertices.  Let $\rho,\varrho$ denote adjacent prefixes in the same component and $\rho'$ denote a prefix in another component.
Since the components are in the same equivalence class, we have that $\rho\subseteq\rho'\subseteq\varrho$, but this implies that $\rho\subseteq\varrho$, contradicting the adjacency of $\rho$ and $\varrho$.
Having shown our claim, we can now see that equivalence classes consisting of a single component satisfy condition~\eqref{tier:allSame} (where we again used~\eqref{eqn:graphConnComp}), while equivalence classes consisting of isolated vertices satisfy condition~\eqref{tier:incomparable}.
This completes the proof of \Cref{lem:tiers}.
\Halmos\endproof

\subsection{Example Illustrating History-monotone Futures and Tier Decomposition} \label{sec:egTiers}

\begin{figure}
\caption{Example illustrating the details in the definition of history-monotone futures.}
\label{fig:monFutures}
\begin{center}
\begin{tikzpicture}
\node[draw,rectangle] (Start) at (-\hordist,3.75*\vertdist) {root};

\node[draw,circle] (B) at (0,6.5*\vertdist) {$B$};
\node[draw,circle,dashed] (B0) at (\hordist,6.5*\vertdist) {$0$};
\node[draw,circle] (C) at (0,4.75*\vertdist) {$C$};
\node[draw,circle] (D) at (0,1.75*\vertdist) {$D$};

\node[draw,circle] (CB) at (\hordist,5.5*\vertdist) {$B$};
\node[draw,circle] (CD) at (\hordist,4*\vertdist) {$D$};
\node[draw,circle] (DB) at (\hordist,2.5*\vertdist) {$B$};
\node[draw,circle] (DC) at (\hordist,1*\vertdist) {$C$};

\node[draw,circle] (CBA) at (2*\hordist,6*\vertdist) {$A$};
\node[draw,circle,dashed] (CBA0) at (3*\hordist,5.5*\vertdist) {$0$};
\node[draw,circle,dashed] (CB0) at (2*\hordist,5*\vertdist) {$0$};
\node[draw,circle] (CDB) at (2*\hordist,4*\vertdist) {$B$};
\node[draw,circle] (DBA) at (2*\hordist,3*\vertdist) {$A$};
\node[draw,circle,dashed] (DBA0) at (3*\hordist,2.5*\vertdist) {$0$};
\node[draw,circle,dashed] (DB0) at (2*\hordist,2*\vertdist) {$0$};
\node[draw,circle] (DCB) at (2*\hordist,1*\vertdist) {$B$};

\node[draw,circle] (CDBA) at (3*\hordist,4.5*\vertdist) {$A$};
\node[draw,circle,dashed] (CDBA0) at (4*\hordist,4*\vertdist) {$0$};
\node[draw,circle,dashed] (CDB0) at (3*\hordist,3.5*\vertdist) {$0$};
\node[draw,circle] (DCBA) at (3*\hordist,1*\vertdist) {$A$};
\node[draw,circle,dashed] (DCBA0) at (4*\hordist,1*\vertdist) {$0$};

\draw[->](Start)--node[fill=white]{\footnotesize $\nicefrac{1}{3}\quad$}(B);
\draw[->](Start)--node[fill=white]{\footnotesize $\nicefrac{1}{3}\quad$}(C);
\draw[->](Start)--node[fill=white]{\footnotesize $\nicefrac{1}{3}\quad$}(D);

\draw[->](C)--node[fill=white]{\footnotesize $\nicefrac{1}{2}$}(CB);
\draw[->](C)--node[fill=white]{\footnotesize $\nicefrac{1}{2}$}(CD);
\draw[->](D)--node[fill=white]{\footnotesize $\nicefrac{1}{2}$}(DB);
\draw[->](D)--node[fill=white]{\footnotesize $\nicefrac{1}{2}$}(DC);

\draw[->](CB)--node[fill=white]{\footnotesize $\nicefrac{1}{2}$}(CBA);
\draw[->](CB)--node[fill=white]{\footnotesize $\nicefrac{1}{2}$}(CB0);
\draw[->](CD)--node[fill=white]{\footnotesize $1$}(CDB);
\draw[->](DB)--node[fill=white]{\footnotesize $\nicefrac{1}{2}$}(DBA);
\draw[->](DB)--node[fill=white]{\footnotesize $\nicefrac{1}{2}$}(DB0);
\draw[->](DC)--node[fill=white]{\footnotesize $1$}(DCB);

\draw[->](CDB)--node[fill=white]{\footnotesize $\nicefrac{3}{4}$}(CDBA);
\draw[->](CDB)--node[fill=white]{\footnotesize $\nicefrac{1}{4}$}(CDB0);
\draw[->](DCB)--node[fill=white]{\footnotesize $1$}(DCBA);

\draw[->](B)--node[fill=white]{\footnotesize $1$}(B0);
\draw[->](CBA)--node[fill=white]{\footnotesize $1$}(CBA0);
\draw[->](CDBA)--node[fill=white]{\footnotesize $1$}(CDBA0);
\draw[->](DBA)--node[fill=white]{\footnotesize $1$}(DBA0);
\draw[->](DCBA)--node[fill=white]{\footnotesize $1$}(DCBA0);
\end{tikzpicture}
\end{center}
\end{figure}

\begin{example}[History-Monotone Futures]\label{eg:monTierFut}
Consider the list distribution over four items $A,B,C,D$ whose tree diagram is shown in \Cref{fig:monFutures}.
We verify that it has history-monotone futures.

There are four realizable prefixes with endpoint $A$---$(CBA),(CDBA),(DBA),(DCBA)$.  \Cref{def:monotoneTiered} is satisfied since all of their futures are empty, dominating each other.

There are five realizable prefixes with endpoint $B$---$(B),(CB),(DB),(CDB),(DCB)$.
Since $A$ is the only item that can appear in the future, domination is fully determined by the probabilities of visiting $A$ from these prefixes, which are $0,1/2,1/2,3/4,1$ respectively.
We verify \Cref{def:monotoneTiered} using the ``equivalent'' condition stated at the end.
$(CB)$ and $(DB)$ have futures which are \textit{not dominated by} $(B)$, which causes no violation, since their respective histories are $\{C\}$ and $\{D\}$, which contain $\emptyset$.
Meanwhile, $(CDB)$ and $(DCB)$ have futures which are not dominated by the previous ones, which also causes no violation, since for both $(CDB)$ and $(DCB)$ the history is $\{C,D\}$, which contains the previous ones.
Finally, $(DCB)$ has a future which is not dominated by $(CDB)$, which again causes no violation, since its history $\{C,D\}$ weakly contains (more specifically, is identical to) the history for $(CDB)$.

There are two realizable prefixes with endpoint $C$---$(C),(DC)$.  The future from $(DC)$ must dominate in order to satisfy \Cref{def:monotoneTiered}, which it does, because it is $(BA)$ w.p.~1 while the future from $(C)$ (ignoring item $D$) is $(BA)$ w.p.~5/8 and $(B)$ w.p.~3/8.
Note it was important here to ignore $D$ in the definition of domination.
%; otherwise domination would be much more difficult to satisfy.

There are two realizable prefixes with endpoint $D$---$(D),(CD)$.
The future from $(CD)$ must dominate in order to satisfy \Cref{def:monotoneTiered}, which it does,
because ignoring item $C$, both futures are $(BA)$ w.p.~3/4 and $(B)$ w.p.~1/4.
Note that \Cref{def:monotoneTiered} would be violated if the transition probability from prefix $(CDB)$ to $A$ was reduced below 3/4.\Halmos
\end{example}

\Cref{eg:monTierFut} can also be used to illustrate the details in the tier decomposition \Cref{lem:tiers}.
Specifically, consider the case where $S=\{A\},j=B$, and recall that there were five realizable prefixes with endpoint $j$---$(B),(CB),(DB),(CDB),(DCB)$---none of which intersect $S$.
The graph described in the proof of \Cref{lem:tiers} will have four connected components, with an edge between $(CB),(DB)$ along with three isolated vertices.
$(CDB),(DCB)$, will later be put into the same tier, resulting in
$$
\cT(1)=\{(B)\},\qquad
\cT(2)=\{(CB),(DB)\},\qquad
\cT(3)=\{(CDB),(DCB)\}.
$$
Property~\eqref{cond:one} in \Cref{lem:tiers} is immediately verified from the calculations in \Cref{eg:monTierFut}.
Meanwhile, tier $t=2$ satisfies property~\eqref{tier:incomparable} while tier $t=3$ satisfies property~\eqref{tier:allSame}.
Note that it is possible for two prefixes in a tier of type~\eqref{tier:allSame} to have different futures; in this example, $\Pr[A\succeq\{A\}_0|(CDB)]=3/4$ which is different from $\Pr[A\succeq\{A\}_0|(DCB)]=1$.

\subsection{Proof of \Cref{thm:mr}} \label{sec:mrPf}

By \Cref{prop:x_to_f,prop:f_to_phi}, $\OPT^x\le\OPT^f\le\OPT^{\phi}$, and hence the revenue of an optimal IC mechanism $x$ is upper-bounded by the revenue of an optimal monotone stopping policy $\phi$ taking values in $\{0,1\}$.
We show that at least one of these optimal policies $\phi$ corresponds to offering an assortment.
Suppose for contradiction this is not the case, and take an optimal policy $\phi$ which
\textit{maximizes} the number of 1-entries, i.e. maximizes $|\{j\in N,H\subseteq N\setminus\{j\}:\phi_j(H)=1\}|$.

Let $S$ denote the set of $j$ for which $\phi_j$ is the all-1 function.
Since $\phi$ does not correspond to offering an assortment, it must be that $S\neq N$.
%; otherwise $\phi$ corresponds to offering the grand assortment $ N$.
For items $j$ not in $S$, the value of $\phi_j(H)$ is inconsequential on histories $H$ that intersect $S$, because $\phi$ would have already stopped on any item in $S$.
Since $\phi$ maximizes the number of 1-entries, we can WOLOG assume that $\phi_j(H)=1$ for all $j\notin S$ and $H\subseteq N\setminus\{j\}$ such that $H\cap S\neq\emptyset$; note that this does not violate the monotonicity of $\phi$.

We now look for an item $j\notin S$ and history $H\cap S=\emptyset$ for which we can modify $\phi_j(H)$ from 0 to 1 while preserving optimality, which would contradict $\phi$ having a maximum number of 1-entries.
We will frequently use the identity from \Cref{lem:revAdj} to adjust for the externality to assortment $S$, which we duplicate below for convenience:
\begin{align}
\Rev[\phi]-\Rev[S] &=\sum_{\rho:\rho\cap S=\emptyset}r^{S}(\rho)\Pr[\rho]\phi_{\rho_{\echar}}(\rho_{\pre})\prod_{k=1}^{|\rho|-1}(1-\phi_{\rho_k}(\rho_{1:k-1})). \label{eqn:finalEqn}
\end{align}
To simplify notation, we let $\Omega^S$ denote the set $\{\rho\in\Omega:\rho\cap S=\emptyset\}$.
For item $j\notin S$ and history $H\subseteq N\setminus S\setminus\{j\}$, let $\Omega^S(j,H)$ denote the subset of prefixes in $\Omega^S$ which first contain the elements in $H$ in any order, followed by $j$.  That is, $\Omega^S(j,H)=\{\rho\in\Omega^S:\rho_{1:|H|}=H,\rho_{|H|+1}=j\}$.

Recall that for items $j\notin S$, function $\phi_j$ is not 1 everywhere, but $\phi_j(H)=1$ whenever $H\cap S\neq\emptyset$.
Consequently, for each such $j$, the collection of histories $\{H\subseteq N\setminus S\setminus\{j\}:\phi_j(H)=0\}$ is non-empty; arbitrarily choose a history $H_j$ from this collection which is \textit{set-wise maximal}.
By this we mean that there does not exist another history $H'$ from the same collection which contains $H_j$,
implying that changing $\phi_j(H_j)$ from 0 to 1 would not violate the monotonicity of $\phi_j$.
We now note that $\phi_j(H_j)$ only appears in the RHS of~\eqref{eqn:finalEqn} if $\rho\in\Omega^S(j,H_j)$:
in the form of $\phi_{\rho_{\echar}}(\rho_{\pre})$ if $\rho$ has length exactly $|H_j|+1$, 
and in the form of $(1-\phi_{\rho_k}(\rho_{1:k-1}))$ if $\rho$ has length greater than $|H_j|+1$.
In either case, if $\Pr[\rho]=0$ for all $\rho\in\Omega^S(j,H_j)$, then changing $\phi_j(H_j)$ from 0 to 1 would have no impact on $\Rev[\phi]$, which means that we have found a feasible and optimal perturbation of $\phi$ which increases the number of 1-entries, leading to a contradiction.

Therefore, we can proceed assuming that for every $j\notin S$, the history $H_j$ chosen has at least one prefix $\rho\in\Omega^S(j,H_j)$ with $\Pr[\rho]>0$.
This allows us to define
\begin{align} \label{eqn:maximinAdjPrice}
j^*\in\argmax_{j\notin S}\min_{\rho\in\Omega^S(j,H_j),\Pr[\rho]>0}r^{S}(\rho);\qquad r^*=\min_{\rho\in\Omega^S(j^*,H_{j^*}),\Pr[\rho]>0}r^{S}(\rho)
\end{align}
where we note that the set $j\notin S$ is also non-empty because $S\neq N$.
We now argue that
\begin{align} \label{eqn:greaterThanCannib}
r^*&\ge r^S(\rho) &\forall \rho:\rho\in\Omega^S,\Pr[\rho]>0,\phi_{\rho_{\echar}}(\rho_{\pre})=1.
\end{align}

To establish~\eqref{eqn:greaterThanCannib}, let $j=\rho_{\echar}$.
Since $r^*$ was defined using a \textit{maximin} in~\eqref{eqn:maximinAdjPrice}, it must be at least the minimum $S$-adjusted price of a realizable prefix $\rho'\in\Omega^S(j,H_j)$, i.e.\ $r^*\ge\min_{\rho'\in\Omega^S(j,H_j),\Pr[\rho']>0}r^{S}(\rho')$.
Thus,~\eqref{eqn:greaterThanCannib} would be established by showing that for any realizable $\rho'\in\Omega^S(j,H_j)$, we have $r^{S}(\rho')\ge r^S(\rho)$.
Since both $\rho$ and $\rho'$ are realizable prefixes in $\Omega^S$ which share the same endpoint $j$, we can apply the tier decomposition for such prefixes, where we let $\rho\in\cT^S_j(t)$ and $\rho'\in\cT^S_j(t')$ for tiers $t,t'$.  We show that $r^{S}(\rho')\ge r^S(\rho)$ in all of the following cases.
\begin{itemize}
\item If $t>t'$, then $r^{S}(\rho)\le r^S(\rho')$ follows immediately from \Cref{cor:monAdjustedPrice}.
\item If $t<t'$, then by \Cref{lem:tiers}, we have $\rho'\supsetneq\rho$.  Since $\rho_{\echar}=\rho'_{\echar}=j$, this implies that $\rho'_{\pre}\supsetneq\rho_{\pre}$.  However, recall that $\rho,\rho'$ were chosen so that $\phi_j(\rho_{\pre})=1$ (by definition in~\eqref{eqn:greaterThanCannib}) and $\phi_j(\rho'_{\pre})=0$ (because $\rho'_{\pre}$ equals $H_j$ as a set, and $H_j$ was chosen so that $\phi_j(H_j)=0$).
Since $\rho'_{\pre}\supsetneq\rho_{\pre}$, the monotonicity of $\phi_j$ would be violated, and hence this case cannot occur.
\item If $t=t'$, then as long as $\rho\neq\rho'$ when viewed as sets, \Cref{cor:monAdjustedPrice} states that $r^{S}(\rho)\le r^S(\rho')$.
Otherwise, we have $\phi_j(\rho_{\pre})=\phi_j(\rho'_{\pre})$, which again contradicts the facts $\phi_j(\rho_{\pre})=1,\phi'_j(\rho_{\pre})=0$.
\end{itemize}

Having completed the proof of~\eqref{eqn:greaterThanCannib}, we now argue that $r^*$ is also non-negative.
While adjusted revenues $r^S(\rho)$ could generally be negative, if $r^*<0$, then~\eqref{eqn:greaterThanCannib} implies that $r^S(\rho)<0$ for all realizable prefixes $\rho$ within $\Omega^S$ which policy $\phi$ stops on.
However, in this case it would be impossible for the RHS of~\eqref{eqn:finalEqn} to be positive, which means that $\Rev[\phi]\le\Rev[S]$, contradicting the presumption that an assortment $S$ could not be optimal.

Finally, equipped with~\eqref{eqn:greaterThanCannib} and the fact that $r^*\ge0$, we argue that 
changing $\phi_{j^*}(H_{j^*})$ from 0 to 1, which is a feasible perturbation preserving the monotonicity of $\phi_{j^*}$ (since $H_{j^*}$ was set-wise maximal), does not decrease $\Rev[\phi]$.
To argue this, we manipulate the RHS of~\eqref{eqn:finalEqn} back into a summation over lists $\ell$.
Let $\kappa^S(\ell)$ denote the minimum $k$ for which $\ell_k\in S$, with $\kappa^S(\ell)=|\ell|+1$ if $\ell\cap S=\emptyset$.  Then
\begin{align}
\Rev[\phi]-\Rev[S]
&=\sum_{\rho:\rho\cap S=\emptyset}r^{S}(\rho)\left(\sum_{\ell:\ell_{1:|\rho|}=\rho}p(\ell)\right)\phi_{\rho_{\echar}}(\rho_{\pre})\prod_{k=1}^{|\rho|-1}(1-\phi_{\rho_k}(\rho_{1:k-1})) \nonumber
\\ &=\sum_{\ell\in\cL}p(\ell)\sum_{k<\kappa^S(\ell)}r^{S}(\ell_{1:k})\phi_{\ell_k}(\ell_{1:k-1})\prod_{k'=1}^{k-1}(1-\phi_{\ell_{k'}}(\ell_{1:k'-1})). \label{eqn:finalEqnPath}
\end{align}
$\phi_{j^*}(H_{j^*})$ only has an impact on the RHS of~\eqref{eqn:finalEqnPath} for sample paths $\ell$ with $p(\ell)>0$, $\ell_{1:|H_{j^*}|}=H_{j^*}$, $\ell_{|H_{j^*}|+1}=j^*$, and $\phi_{\ell_{k'}}(\ell_{1:k'-1})=0$ for all $k'=1,\ldots,|H_{j^*}|$.
On these sample paths, changing $\phi_{j^*}(H_{j^*})$ from 0 to 1 gains a revenue of $r^{S}_{j^*}(H_{j^*})$, but could lose a revenue of $r^S(\ell_{1:k})$ for some $k>|H_{j^*}|+1$ such that $\phi_{\ell_k}(\ell_{1:k-1})=1$.
However, the lost revenue $r^S(\ell_{1:k})$ must be no more than the gained revenue $r^{S}_{j^*}(H_{j^*})$, because $r^{S}_{j^*}(H_{j^*})\ge r^*$ by definition of $r^*$ in~\eqref{eqn:maximinAdjPrice} while $r^*\ge r^S(\ell_{1:k})$ since $\ell_{1:k}$ is a valid $\rho$ in~\eqref{eqn:greaterThanCannib} (in these assertions, we have made use of the fact that $p(\ell)>0$ and hence all of its prefixes are realizable).
%the condition $\Pr[(\ell_k,\ell_{1:k-1})]>0$ holds because $p(\ell)>0$.
Alternatively, on sample paths without a $k>|H_{j^*}|+1$ for which $\phi_{\ell_k}(\ell_{1:k-1})=1$, we have only gained a revenue of $r^{S}_{j^*}(H_{j^*})$, which is non-negative.
Therefore, for all $\ell$ we have shown that the change on the RHS of~\eqref{eqn:finalEqnPath} is non-negative, thereby establishing that the perturbed $\phi$ must also be an optimal policy, contradicting the presumption that the original $\phi$ had a maximum number of 1-entries among optimal policies.
We have reached a contradiction in all cases, completing the proof of our main result \Cref{thm:mr}.

\section{Proofs from \Cref{sec:assortment}} \label{sec:defPfsAssortment}

\proof{Proof of \Cref{prop:mc}.}
Take realizable prefixes $\rho,\rho'$ with the same endpoint.
We show that regardless of any containment relations between $\rho_{\pre}$ and $\rho'_{\pre}$, as long as assortment $S$ is disjoint from both $\rho$ and $\rho'$, it will be the case that $\Pr[j\succeq S_0|\rho]=\Pr[j\succeq S_0|\rho']$ for any $j\in S$.
To see this, consider the future from $\rho$.
This is generated by a random path on the Markov Chain starting from state $\rho_{\echar}$, where any newly-visited items not on $\rho$ are added to the list.
The future from $\rho'$ is generated by an identically-distributed path, except the newly-visited items on $\rho'$ are excluded instead.
However, these differences do not affect the event $j\succeq S_0$, since $j\in S$ and $S$ is disjoint from both $\rho$ and $\rho'$.
Therefore, $\Pr[j\succeq S_0|\rho]=\Pr[j\succeq S_0|\rho']$ for all $j\in S$, satisfying the condition~\eqref{eqn:domFuture} for domination as equality and completing the proof of \Cref{prop:mc}.
\Halmos\endproof

\proof{Proof of \Cref{prop:elimByAspects}.}
Take realizable prefixes $\rho,\rho'$ with the same endpoint $j$, and suppose that $j$ is in nest $i$.
By the rules of drawing the balls, $\rho$ and $\rho'$ (when viewed as sets) can only differ on items in $N_i$.
Take any assortment $S$ disjoint from $\rho$ and $\rho'$, and we would like to show that
$\Pr[j'\succeq S_0|\rho]=\Pr[j'\succeq S_0|\rho']$ for all $j'\in S$.
Clearly if $j'\in N_i$ then this is true, since $S$ is disjoint from any items in $N_i$ which have already appeared in $\rho$ or $\rho'$.
On the other hand, if $j'\notin N_i$, then this is also true, because either both probabilities are 0 (if $S$ contains any items in $N_i$), or both probabilities are calculated from an identical starting point where the items in nest $N_i$ have all been drawn (recall that $\rho$ and $\rho'$ do not setwise differ outside of $N_i$).
\Halmos\endproof

\proof{Proof of \Cref{prop:mixture}.}
Take realizable prefixes $\rho,\rho'$ with $\rho_{\echar}=\rho'_{\echar}$ and $\rho_{\pre}\nsubseteq\rho'_{\pre}$.
Note that this implies $\rho_{\pre}\neq\emptyset$, so $\rho$ has length at least 2.
Therefore, the future conditional on $\rho$ did not change from before.
On the other hand, the future conditional on $\rho'$ either did not change (if $|\rho'|\ge2$), or $\Pr[j\succeq S_0|\rho']$ decreased for every $j\in S$ (if $|\rho'|=1$).
In either case, the future from $\rho$ continues to dominate the future from $\rho'$, verifying the definition of history-monotone futures and completing the proof.
\Halmos\endproof

\proof{Proof of \Cref{prop:NLnotMC}.}
WOLOG normalize $w_0$ to 1.
Suppose there are 3 items, which have weight $w>0$, in the same nest, which has dissimilarity parameter $\gamma\in(0,1)$.
Let $N=\{1,2,3\}$ denote the set of 3 items.
For a Markov Chain choice model to capture the Nested Logit choice probabilities in~\eqref{eqn:nestedLogitChoiceProbs}, its parameters $\lambda_1,\lambda_2,\lambda_3,\rho_{12},\rho_{13},\rho_{21},\rho_{23},\rho_{31},\rho_{32}$ must satisfy the following constraints.

First, $\lambda_j$ must equal the probability of item $j$ being chosen from the full assortment $N$, for all $j\in N$.  Therefore,
\begin{align*}
\lambda_j &=\frac{(3w)^\gamma}{1+(3w)^\gamma}\cdot\frac{1}{3} &\forall j=1,2,3.
\end{align*}

Second, for any assortment consisting of two distinct items $j,j'\in N$, note that the probability of $j$ being chosen from assortment $\{j,j'\}$ is $\lambda_j+\lambda_{j''}\sigma_{j'',j}$, where $j''$ is the item that is not $j$ or $j'$.  Therefore,
\begin{align*}
\frac{(2w)^{\gamma}}{1+(2w)^{\gamma}}\cdot\frac{1}{2} &=\lambda_j+\lambda_{j''}\sigma_{j'',j}
\\ \frac{(2w)^{\gamma}}{1+(2w)^{\gamma}}\cdot\frac{1}{2} &=\frac{(3w)^\gamma}{1+(3w)^\gamma}\cdot\frac{1}{3}(1+\sigma_{j'',j})
\end{align*}
which implies that
\begin{align*}
\sigma_{j'',j} &=\frac{(3w)^{-\gamma}+1}{(2w)^{-\gamma}+1}\cdot\frac{3}{2}-1 &\forall j\neq j''\in\{1,2,3\}.
\end{align*}

Finally, for a singleton assortment $\{j\}$, the Nested Logit probability of $j$ being chosen is
\begin{align} \label{eqn:3itemNL}
\frac{w^{\gamma}}{1+w^{\gamma}}.
\end{align}
This must equal the probability of visiting $j$ before terminal state 0 in the Markov Chain, which can be computed via taking a sum over the probability of visiting $j$ for the first time after transitioning exactly $k$ times (to a non-zero state) in the Markov Chain:
\begin{align}
&\frac{(3w)^\gamma}{1+(3w)^\gamma}\cdot\frac{1}{3}+\sum_{k=1}^{\infty}2\frac{(3w)^\gamma}{1+(3w)^\gamma}\cdot\frac{1}{3}\left(\frac{(3w)^{-\gamma}+1}{(2w)^{-\gamma}+1}\cdot\frac{3}{2}-1\right)^k
\nonumber \\ &=\frac{(3w)^\gamma}{1+(3w)^\gamma}\cdot\frac{1}{3}\left(1+2\frac{1}{(\frac{(3w)^{-\gamma}+1}{(2w)^{-\gamma}+1}\cdot\frac{3}{2}-1)^{-1}-1}\right). \label{eqn:ugly}
\end{align}

However, it can be checked that this expression does not equal~\eqref{eqn:3itemNL} for general values of $w$ and $\gamma$ (it can also be verified that~\eqref{eqn:ugly} \textit{does} equal~\eqref{eqn:3itemNL} when $\gamma=1$, which is the special case of MNL).  As a concrete example, taking $w=1$,~\eqref{eqn:ugly} can be re-written as
\begin{align*}
\frac{3^{\gamma-1}}{1+3^\gamma}\left(1+2\frac{1}{(\frac{3^{1-\gamma}+3}{2^{1-\gamma}+2}-1)^{-1}-1}\right)
=\frac{3^{\gamma-1}}{1+3^\gamma}\left(1+\frac{2}{\frac{2^{1-\gamma}+2}{3^{1-\gamma}-2^{1-\gamma}+1}-1}\right)
\end{align*}
which only equals $\frac{1}{2}$ (the value of~\eqref{eqn:3itemNL}) when $\gamma=1$.
Therefore, a Nested Logit choice model with 3 items that have equal weights and are in the same nest is not captured by a Markov Chain, completing the proof.
\Halmos\endproof

\proof{Proof of \Cref{prop:NL3itemRep}.}
When there are only 3 items, the following identities are sufficient for a list distribution defined through its probabilistic tree (refer to \Cref{def:qAndTree}) to be consistent with a given set of choice probabilities.
For brevity, we will omit the inner parentheses when prefixes are expressed inside the $q(\cdot)$ function.
\begin{align*}
q(j) &=\Pr[j\succeq N_0] &\forall j\in N
\\ q(j)+q(j')q(j'j) &=\Pr[j\succeq (N\setminus\{j'\})_0] &\forall j'\neq j\in N
\\ q(j)+q(j')q(j'j)+q(j'')q(j''j)+q(j')q(j'j'')q(j'j''j)+q(j'')q(j''j')q(j''j'j) &=\Pr[j\succeq \{j\}_0] &\forall j\in N
\end{align*}
In the third set of identities, $j'$ and $j''$ refer to the two items other than $j$.
Note that this system has $3+6+6=15$ variables but only $3+6+3=12$ equations.
We will allow for a full specification of the $q$-probabilities by imposing that $q(j'j''j)=q(j''j'j)$ for all $j\in N$, effectively adding 3 equations.
We show this always results in a list distribution with history-monotone futures.

To verify history-monotone futures for list distributions on 3 items, one only needs to check that
\begin{align} \label{eqn:3itemSuffCond}
\Pr[j\succeq\{j\}_0|(j''j')]\ge\Pr[j\succeq\{j\}_0|(j')]
\end{align}
for all permutations of the items $j,j',j''$.
(Note that in~\eqref{eqn:3itemSuffCond} we have assumed $S=\{j\}$; if $|S|\ge 2$, then there cannot be two distinct prefixes in $N\setminus S$ with the same endpoint, and hence the condition holds vacuously.)
We now proceed to check this for an arbitrary arrangement of $j,j',j''$, which for brevity we relabel as $A,B,C$ respectively.

Note that the LHS of~\eqref{eqn:3itemSuffCond} equals $q(CBA)$, while the RHS of~\eqref{eqn:3itemSuffCond} equals $q(BA)+q(BC)q(BCA)$.
The goal is to show that $q(CBA)\ge q(BA)+q(BC)q(BCA)$ when the $q$-values are defined through the identities above and the choice probabilities comes from a Nested Logit choice model.
To solve for $q(CBA)$, which is the same as $q(BCA)$, we can write:
\begin{align*}
q(A)+q(B)q(BA)+q(C)q(CA)+(q(B)q(BC)+q(C)q(CB))q(CBA) &=\Pr[A\succeq\{A\}_0]
\\ (q(A)+q(B)q(BA))+(q(A)+q(C)q(CA))+(q(B)q(BC)+q(C)q(CB))q(CBA) &=\Pr[A\succeq\{A\}_0]+q(A)
\\ \Pr[A\succeq\{A,C\}_0]+\Pr[A\succeq\{A,B\}_0]+(q(B)q(BC)+q(C)q(CB))q(CBA) &=\Pr[A\succeq\{A\}_0]+\Pr[A\succeq N_0]
\end{align*}
This allows us to deduce:
\begin{align} \label{eqn:qCBA}
q(CBA)
&=\frac{\Pr[A\succeq\{A\}_0]-\Pr[A\succeq\{A,B\}_0]-\Pr[A\succeq\{A,C\}_0]+\Pr[A\succeq\{A,B,C\}_0]}{\Pr[C\succeq\{A,C\}_0]-\Pr[C\succeq\{A,B,C\}_0]+\Pr[B\succeq\{A,B\}_0]-\Pr[B\succeq\{A,B,C\}_0]}
\end{align}
Meanwhile, we can write:
\begin{align} \label{eqn:qBAfrac}
\frac{q(BA)}{1-q(BC)}=\frac{q(B)q(BA)}{q(B)-q(B)q(BC)}
=\frac{\Pr[A\succeq\{A,C\}_0]-\Pr[A\succeq\{A,B,C\}_0]}{\Pr[B\succeq\{A,B,C\}_0]-(\Pr[C\succeq\{A,C\}_0]-\Pr[C\succeq\{A,B,C\}_0])}
\end{align}
To prove~\eqref{eqn:3itemSuffCond}, or equivalently $q(CBA)\ge q(BA)+q(BC)q(CBA)$, it suffices to prove that the expression on the RHS of~\eqref{eqn:qCBA} is at least the expression on the RHS of~\eqref{eqn:qBAfrac}.

Since there is a single nest, the Nested Logit choice probability for an item $j$ being chosen from an assortment $S$ can be expressed as (setting $w_0=1$ and ignoring the nest subscript $i$):
\begin{align*}
\frac{w_j}{(\sum_{j'\in S}w_{j'})^{1-\gamma}+(\sum_{j'\in S}w_{j'})}.
\end{align*}
Moreover, all of the assortments in expressions~\eqref{eqn:qCBA} and~\eqref{eqn:qBAfrac} include item $A$.  Therefore, we let $d(S)$ denote the denominator $(w_A+\sum_{j'\in S}w_{j'})^{1-\gamma}+(w_A+\sum_{j'\in S}w_{j'})$ where $S$ is a subset of $\{B,C\}$.
We also hereafter let $a,b,c$ be shorthand for the weights $w_A,w_B,w_C$ respectively.
Then, the desired relationship $\eqref{eqn:qCBA}\ge\eqref{eqn:qBAfrac}$ is equivalent to
\begin{align} \label{eqn:6290}
\frac{\frac{a}{d(\emptyset)}-\frac{a}{d(B)}-\frac{a}{d(C)}+\frac{a}{d(BC)}}{\frac{b}{d(B)}-\frac{b}{d(BC)}+\frac{c}{d(C)}-\frac{c}{d(BC)}}\ge\frac{\frac{a}{d(C)}-\frac{a}{d(BC)}}{\frac{b}{d(BC)}+\frac{c}{d(BC)}-\frac{c}{d(C)}}
\end{align}
where we have omitted braces and commas in the input to the function $d$.
Since the denominators are non-negative, we can cross-multiply and~\eqref{eqn:6290} is equivalent to
\begin{align}
\left(\frac{1}{d(\emptyset)}-\frac{1}{d(B)}-\frac{1}{d(C)}+\frac{1}{d(BC)}\right)\left(\frac{b+c}{d(BC)}-\frac{c}{d(C)}\right)\ge\left(\frac{1}{d(C)}-\frac{1}{d(BC)}\right)\left(\frac{b}{d(B)}+\frac{c}{d(C)}-\frac{b+c}{d(BC)}\right). \label{eqn:2908}
\end{align}

After expanding~\eqref{eqn:2908} and canceling terms that appear on both sides, we are left with
\begin{align*}
\frac{b+c}{d(\emptyset)d(BC)}+\frac{c}{d(B)d(C)}\ge\frac{b}{d(B)d(C)}+\frac{c}{d(\emptyset)d(C)}+\frac{c}{d(B)d(BC)}.
\end{align*}
Multiplying both sides by $d(\emptyset)d(B)d(C)d(BC)$ and substituting in the definition of $d$, we get
\small
\begin{align*}
&(b+c)((a+b)^{1-\gamma}+a+b)((a+c)^{1-\gamma}+a+c)+c(a^{1-\gamma}+a)((a+b+c)^{1-\gamma}+a+b+c)
\\ &\ge
b(a^{1-\gamma}+a)((a+b+c)^{1-\gamma}+a+b+c)+c((a+b)^{1-\gamma}+a+b)((a+b+c)^{1-\gamma}+a+b+c)+c(a^{1-\gamma}+a)((a+c)^{1-\gamma}+a+c).
\end{align*}
\normalsize
Note that after expanding, any products without a $(\cdot)^{1-\gamma}$ term cancel out on both sides.
Hence, it suffices to prove
\small
\begin{align*}
&b(a+b)^{1-\gamma}(a+c)^{1-\gamma}
+b(a+b)^{1-\gamma}(a+\cancel{c})
+b(a+b)(a+c)^{1-\gamma}
\\ &+c(a+b)^{1-\gamma}(a+c)^{1-\gamma}
+\cancel{c(a+b)^{1-\gamma}(a+c)}
+c(\cancel{a}+b)(a+c)^{1-\gamma}
+ca^{1-\gamma}(a+b+c)^{1-\gamma}
+ca^{1-\gamma}(\cancel{a}+\cancel{b}+\cancel{c})
+\cancel{ca(a+b+c)^{1-\gamma}}
\\ &\ge
ba^{1-\gamma}(a+b+c)^{1-\gamma}
+ba^{1-\gamma}(a+b+\cancel{c})
+ba(a+b+c)^{1-\gamma}
\\&+c(a+b)^{1-\gamma}(a+b+c)^{1-\gamma}
+c(a+b)^{1-\gamma}(\cancel{a}+\cancel{b}+\cancel{c})
+c(\cancel{a}+b)(a+b+c)^{1-\gamma}
+ca^{1-\gamma}(a+c)^{1-\gamma}
+\cancel{ca^{1-\gamma}(a+c)}
+\cancel{ca(a+c)^{1-\gamma}}
\end{align*}
\normalsize
which after cancellations is reduced to
\small
\begin{align*}
&b(a+b)^{1-\gamma}(a+c)^{1-\gamma}
+b(a+b)^{1-\gamma}(a)
+b(a+b)(a+c)^{1-\gamma}
+c(a+b)^{1-\gamma}(a+c)^{1-\gamma}
+cb(a+c)^{1-\gamma}
+ca^{1-\gamma}(a+b+c)^{1-\gamma}
\\ &\ge
ba^{1-\gamma}(a+b+c)^{1-\gamma}
+ba^{1-\gamma}(a+b)
+ba(a+b+c)^{1-\gamma}
+c(a+b)^{1-\gamma}(a+b+c)^{1-\gamma}
+cb(a+b+c)^{1-\gamma}
+ca^{1-\gamma}(a+c)^{1-\gamma}.
\end{align*}
\normalsize

Now, we bifurcate the products on both sides depending on whether they contain one or two $(\cdot)^{1-\gamma}$ terms.
We separately show that the LHS is at least the RHS for both types of terms.
That is, we show
\begin{align}
&b(a+b)^{1-\gamma}(a+c)^{1-\gamma}+c(a+b)^{1-\gamma}(a+c)^{1-\gamma}+ca^{1-\gamma}(a+b+c)^{1-\gamma} \nonumber
\\ &\ge
ba^{1-\gamma}(a+b+c)^{1-\gamma}+c(a+b)^{1-\gamma}(a+b+c)^{1-\gamma}+ca^{1-\gamma}(a+c)^{1-\gamma}; \label{eqn:twoGamma}
\\ &ba(a+b)^{1-\gamma}+b(a+b)(a+c)^{1-\gamma}+cb(a+c)^{1-\gamma} \nonumber
\\ &\ge
b(a+b)a^{1-\gamma}+ba(a+b+c)^{1-\gamma}+cb(a+b+c)^{1-\gamma}. \label{eqn:oneGamma}
\end{align}
Doing so would suffice to establish~\eqref{eqn:3itemSuffCond} and hence complete the proof of the \namecref{prop:NL3itemRep}.

Having separated the original inequality into~\eqref{eqn:twoGamma} and~\eqref{eqn:oneGamma}, note that these new inequalities are \textit{homogeneous}, in that the total degree of every term is the same---in~\eqref{eqn:twoGamma}, the total degree of every term is $3-2\gamma$; in~\eqref{eqn:oneGamma}, the total degree of every term is $3-\gamma$.  Therefore, during the proof of both~\eqref{eqn:twoGamma} and~\eqref{eqn:oneGamma}, we can WOLOG assume that $a+b+c=1$.
Also for brevity, we will let $\delta:=1-\gamma\in[0,1)$.

We now rewrite~\eqref{eqn:twoGamma} as
\begin{align}
&& (b+c)(a+b)^\delta(a+c)^\delta+ca^\delta &\ge ba^\delta+c(a+b)^\delta+ca^\delta(a+c)^\delta \nonumber
\\ \Longleftrightarrow && (b+c)(1+b/a)^\delta(1-b)^\delta+c &\ge b+c(1+b/a)^\delta+c(1-b)^\delta \nonumber
\\ \Longleftrightarrow && b\left((1+b/a)^\delta(1-b)^\delta-1\right) &\ge c\left((1+b/a)^\delta+(1-b)^\delta-(1+b/a)^\delta(1-b)^\delta-1\right) \nonumber
\\ \Longleftrightarrow && \frac{b}{c} &\ge\frac{\left((1+b/a)^\delta-1\right)\left(1-(1-b)^\delta\right)}{\left((1+b/a)^\delta(1-b)^\delta-1\right)} \label{eqn:1084}
\end{align}
Our goal is to show that the maximum of the RHS of~\eqref{eqn:1084} over $\delta\in[0,1]$ occurs at $\delta=1$, given any values of $a,b$ satisfying $b/a\ge b$ (which is true since $a<1$).  We will equivalently show that
\begin{align*}
\log\left((1+b/a)^\delta-1\right)+\log\left(1-(1-b)^\delta\right)-\log\left((1+b/a)^\delta(1-b)^\delta-1\right)
\end{align*}
is maximized at $\delta=1$.  Taking the derivative with respect to $\delta$, we get
\begin{align*}
\frac{(1+b/a)^\delta\log(1+b/a)}{(1+b/a)^\delta-1}-\frac{(1-b)^\delta\log(1-b)}{1-(1-b)^\delta}-\frac{(1+b/a)^\delta(1-b)^\delta\log((1+b/a)(1-b))}{(1+b/a)^\delta(1-b)^\delta-1}.
\end{align*}
We want to show that this derivative is non-negative, i.e.
\begin{align}
&& \frac{\log(1+b/a)}{1-(1+b/a)^{-\delta}}
&\ge\frac{\log(1-b)}{(1-b)^{-\delta}-1}+\frac{\log(1+b/a)+\log(1-b)}{1-(1+b/a)^{-\delta}(1-b)^{-\delta}} \nonumber
\\ \Longleftrightarrow && \frac{\log(1+b/a)}{1-(1+b/a)^{-\delta}}-\frac{\log(1+b/a)}{1-((1+b/a)(1-b))^{-\delta}}
&\ge\frac{\log(1-b)}{(1-b)^{-\delta}-1}+\frac{\log(1-b)}{1-((1+b/a)(1-b))^{-\delta}} \nonumber
\\ \Longleftrightarrow && \log(1+b/a)\frac{(1+b/a)^{-\delta}-((1+b/a)(1-b))^{-\delta}}{(1-(1+b/a)^{-\delta})(1-((1+b/a)(1-b))^{-\delta})}
&\ge\log(1-b)\frac{(1-b)^{-\delta}-((1+b/a)(1-b))^{-\delta}}{((1-b)^{-\delta}-1)(1-((1+b/a)(1-b))^{-\delta})} \nonumber
\\ \Longleftrightarrow && \log(1+b/a)\frac{1-(1-b)^{-\delta}}{(1+b/a)^\delta-1}
&\ge\log(1-b)\frac{1-(1+b/a)^{-\delta}}{1-(1-b)^\delta} \nonumber
\\ \Longleftrightarrow && \log(\frac{1}{1-b})\frac{1-(1+b/a)^{-\delta}}{1-(1-b)^\delta}
&\ge\log(1+b/a)\frac{(1-b)^{-\delta}-1}{(1+b/a)^\delta-1} \nonumber
\\ \Longleftrightarrow && \frac{(1-(1+b/a)^{-\delta})((1+b/a)^\delta-1)}{\log(1+b/a)}
&\ge\frac{(1-(\frac{1}{1-b})^{-\delta})((\frac{1}{1-b})^{\delta}-1)}{\log(\frac{1}{1-b})} \label{eqn:9088}
\end{align}
where we emphasize that in the third line, $1-((1+b/a)(1-b))^{-\delta}\ge1-((1+\frac{b}{1-b})(1-b))^{-\delta}=0$, and in the final equivalence, all terms moved to the other side are also non-negative, and hence we never had to flip the inequality.
Finally, to establish~\eqref{eqn:9088}, note that it suffices to show the function $\frac{(1-y^{-\delta})(y^\delta-1)}{\log y}=\frac{y^\delta+y^{-\delta}-2}{\log y}$ is increasing over $y\ge1$, since $1+b/a\ge1+\frac{b}{1-b}=\frac{1}{1-b}\ge1$.
To show that $\frac{y^\delta+y^{-\delta}-2}{\log y}$ is increasing in $y$, we take the derivative with respect to $y$, which can be checked to equal
\begin{align*}
\frac{y^\delta\log y^\delta-y^\delta-\frac{\log y^\delta}{y^\delta}-\frac{1}{y^\delta}+2}{y(\log y)^2},
\end{align*}
whose numerator can be seen to be non-negative over $[1,\infty)$ using a univariate plot of the variable $y^\delta$.
This completes the proof of~\eqref{eqn:twoGamma}.

To show the other inequality~\eqref{eqn:oneGamma}, we cancel out $b$ from both sides, and then similarly rewrite it using $a+b+c=1$, $\delta=1-\gamma$:
\begin{align}
&& a(a+b)^\delta+(a+b)(a+c)^\delta+c(a+c)^\delta &\ge (a+b)a^\delta+a+c \nonumber
\\ \Longleftrightarrow && a(a+b)^\delta+(1-b)^\delta &\ge (a+b)a^\delta+1-b \nonumber
\\ \Longleftrightarrow && (1-b)^\delta-(1-b) &\ge (a+b)a^\delta-a(a+b)^\delta \label{eqn:1255}
\end{align}
Our goal is to show that holding $b$ and $\delta$ fixed, the RHS of~\eqref{eqn:1255} is increasing in $a$.  This would complete the proof because the maximum possible value of $a$ is $1-b$, in which case the RHS equals $(1-b)^\delta-(1-b)$, identical to the LHS.
We take the derivative of the RHS of~\eqref{eqn:1255} with respect to $a$ and aim to show it is non-negative:
\begin{align}
&& a^\delta+(a+b)\delta a^{\delta-1}-(a+b)^\delta-a\delta(a+b)^{\delta-1} &\ge0 \nonumber
\\ \Longleftrightarrow && 1+(1+b/a)\delta-(1+b/a)^\delta-a\frac{\delta(1+b/a)^{\delta}}{a+b} &\ge0 \nonumber
\\ \Longleftrightarrow && 1+y\delta-y^\delta-\delta y^{\delta-1} &\ge 0 \label{eqn:8882}
\end{align}
where in the final equivalence we have let $y:=1+b/a$, a quantity ranging over $[1,\infty)$.
To show~\eqref{eqn:8882}, we claim that its minimum value over $[1,\infty)$ is achieved at $y=1$, in which case it equals 0.
To see why, we take the derivative of the LHS with respect to $y$:
\begin{align*}
1+\delta-\delta y^{\delta-1}-\delta(\delta-1) y^{\delta-2}
=1+\delta(1-\frac{1}{y^{1-\delta}})+\delta(1-\delta) y^{\delta-2}
\end{align*}
which is clearly non-negative over $[1,\infty)$.  Therefore, inequality~\eqref{eqn:8882} always holds, which implies~\eqref{eqn:1255}, which in turn implies~\eqref{eqn:oneGamma}.
Having established~\eqref{eqn:twoGamma} and~\eqref{eqn:oneGamma}, this completes the entire proof.
\Halmos\endproof

\proof{Proof of \Cref{prop:NL4itemRep}.}
WOLOG normalize $w_0$ to 1.  Let $w$ denote the common weight of the items and $\gamma$ denote the dissimilarity parameter.  According to~\eqref{eqn:nestedLogitChoiceProbs}, the probability of any item being chosen when it is offered in an assortment of size $k$ is
\begin{align} \label{eqn:8582}
P_k &:=\frac{(kw)^{\gamma}}{1+(kw)^{\gamma}}\cdot\frac{1}{k}=\frac{1}{k(1+(kw)^{-\gamma})} &\forall k=1,\ldots,n.
\end{align}

We now define a list distribution which is consistent with the Nested Logit choice probabilities $P_k$ and has history-monotone futures.
We define this list distribution by the transition probabilities $q$ for its tree diagram (see \Cref{def:qAndTree}).
Our tree will be symmetric, and we let $q_k$ denote the probability of transitioning to the node for any unvisited item after visiting exactly $k-1$ previous item nodes, for all $k=1,\ldots,n$.
Note that the probability of transitioning to the terminal node after visiting exactly $k-1$ previous item nodes will be $1-(n-k+1)q_k$.

By symmetry, our tree is completely specified by the probabilities $q_1,\ldots,q_n$.
In order to be consistent with the choice probabilities $P_k$, they must satisfy the following system of equations:
\begin{align} \label{eqn:NLsymmetricGeneralIdentity}
P_k &=\sum_{k'=0}^{n-k}q_{k'+1}\prod_{k''=1}^{k'}(n-k-k''+1)q_{k''} &\forall k=1,\ldots,n.
\end{align}
To explain the RHS, note that there are $n-k$ items not in the assortment.
Any of the $k$ items in the assortment gets chosen if and only if it appears on the list following exactly $k'$ distinct ordered items not in the assortment, with $k'$ ranging from 0 to $n-k$ inclusive.

\textbf{Solving for the transition probabilities in the tree.}
We now inductively establish the identities
\begin{align} \label{eqn:NLnice}
q_{m+1}\prod_{m'=1}^mq_{m'} &=\frac{1}{m!}\sum_{m'=0}^m(-1)^{m-m'}\binom{m}{m'}P_{n-m'} &\forall m=0,\ldots,n-1.
\end{align}
To do so, note that the base case $m=0$ says $q_1=P_n$ which follows from~\eqref{eqn:NLsymmetricGeneralIdentity} with $k=n$.
Now, suppose we have verified identify~\eqref{eqn:NLnice} for all values smaller than a particular $m\in\{1,\ldots,n-1\}$.
Using~\eqref{eqn:NLsymmetricGeneralIdentity} with $k=n-m$, we get
\begin{align*}
P_{n-m} &=m!q_{m+1}\prod_{m'=1}^mq_{m'}+\sum_{k'=0}^{m-1}q_{k'+1}\prod_{k''=1}^{k'}(n-k-k''+1)q_{k''}
\\ &=m!q_{m+1}\prod_{m'=1}^mq_{m'}+\sum_{k'=0}^{m-1}\frac{1}{k'!}\sum_{m'=0}^{k'}(-1)^{k'-m'}\binom{k'}{m'}P_{n-m'}\left(\prod_{k''=1}^{k'}(m-k''+1)\right)
\\ &=m!q_{m+1}\prod_{m'=1}^mq_{m'}+\sum_{m'=0}^{m-1}P_{n-m'}\sum_{k'=m'}^{m-1}(-1)^{k'-m'}\frac{1}{k'!}\binom{k'}{m'}\left(\prod_{k''=1}^{k'}(m-k''+1)\right)
\\ &=m!q_{m+1}\prod_{m'=1}^mq_{m'}+\sum_{m'=0}^{m-1}(-1)^{m-1-m'}P_{n-m'}\sum_{k'=m'}^{m-1}(-1)^{m-1-k'}\binom{m}{k'}\binom{k'}{m'}
\\ &=m!q_{m+1}\prod_{m'=1}^mq_{m'}-\sum_{m'=0}^{m-1}(-1)^{m-m'}P_{n-m'}\binom{m}{m'}
\end{align*}
where the second equality applies the induction hypothesis,
the third equality switches sums,
and the final equality uses the Principle of Inclusion-Exclusion on the number of ways to choose a size-$m'$ subset from $m$ elements by first selecting $k'$ elements to choose from.  This completes the induction and establishes~\eqref{eqn:NLnice}.
We can now solve for the values of $q_1,\ldots,q_n$ via taking the quotient of adjacent identities in this family.

\textbf{Showing that the $q$-values exhibit a monotonicity property.}
Our goal is to show that the resulting values of $q_k$ satisfy the following monotonicity property, which will later be useful for establishing history-monotone futures:
\begin{align}\label{eqn:qMonProperty}
\frac{1}{q_k} &\ge\frac{1}{q_{k+1}}+1 &\forall k=1,\ldots,n-1.
\end{align}

We verify this using brute force when $n=4$.  Using~\eqref{eqn:NLnice}, we can algebraically express $q_1,q_2,q_3,q_4$:
\begin{align*}
q_1 &=P_4
\\ q_2q_1 &=P_3-P_4
\\ q_3q_2q_1 &=\frac{P_2-2P_3+P_4}{2}
\\ q_4q_3q_2q_1 &=\frac{P_1-3P_2+3P_3-P_4}{6}
\end{align*}
which implies:
\begin{align*}
\frac{1}{q_1} &=\frac{1}{P_4}
\\ \frac{1}{q_2} &=\frac{P_4}{P_3-P_4}
\\ \frac{1}{q_3} &=\frac{2(P_3-P_4)}{P_2-2P_3+P_4}
\\ \frac{1}{q_4} &=\frac{3(P_2-2P_3+P_4)}{P_1-3P_2+3P_3-P_4}
\end{align*}
The three inequalities to prove in~\eqref{eqn:qMonProperty} for $n=4$ are:
\begin{align*}
\frac{1}{P_4} &\ge\frac{P_4}{P_3-P_4}+1
\\ \frac{P_4}{P_3-P_4} &\ge\frac{2(P_3-P_4)}{P_2-2P_3+P_4}+1
\\ \frac{2(P_3-P_4)}{P_2-2P_3+P_4} &\ge\frac{3(P_2-2P_3+P_4)}{P_1-3P_2+3P_3-P_4}+1
\end{align*}
which are respectively equivalently to:
\begin{align}
\frac{1}{P_4} &\ge\frac{1}{P_3}+1 \label{eqn:grind1}
\\ \frac{2}{P_3} &\ge\frac{1}{P_2}+\frac{1}{P_4} \label{eqn:grind2}
\\ \frac{4}{P_1P_3}+\frac{4}{P_2P_4} &\ge\frac{3}{P_1P_4}+\frac{3}{P_2P_3}+\frac{1}{P_1P_2}+\frac{1}{P_3P_4} \label{eqn:grind3}
\end{align}

Substituting in the Nested Logit choice probabilities from~\eqref{eqn:8582}, the first inequality~\eqref{eqn:grind1} is equivalent to
$
4(1+(4w)^{-\gamma}) \ge 3(1+(3w)^{-\gamma})+1
$
or
$4(4w)^{-\gamma}\ge3(3w)^{-\gamma}$ which is clearly true for all $w>0$ and $\gamma\in(0,1]$.
Meanwhile, the second inequality~\eqref{eqn:grind2} is equivalent to
$
6(1+(3w)^{-\gamma})\ge4(1+(4w)^{-\gamma})+2(1+(2w)^{-\gamma})
$
or
$6(3w)^{-\gamma}\ge4(4w)^{-\gamma}+2(2w)^{-\gamma}$.
Since $w>0$, we can cancel out $w^{-\gamma}$ from both sides and the inequality is equivalent to $6(3)^{-\gamma}\ge4(4)^{-\gamma}+2(2)^{-\gamma}$, which can be verified to be true over $\gamma\in(0,1]$.
Finally, the third inequality~\eqref{eqn:grind3} is equivalent to
\begin{align*}
&12(1+\frac{1}{w^{\gamma}})(1+\frac{1}{(3w)^{\gamma}})+32(1+\frac{1}{(2w)^{\gamma}})(1+\frac{1}{(4w)^{\gamma}})
\\ &\ge 12(1+\frac{1}{w^{\gamma}})(1+\frac{1}{(4w)^{\gamma}})+18(1+\frac{1}{(2w)^{\gamma}})(1+\frac{1}{(3w)^{\gamma}})
+2(1+\frac{1}{w^{\gamma}})(1+\frac{1}{(2w)^{\gamma}})+12(1+\frac{1}{(3w)^{\gamma}})(1+\frac{1}{(4w)^{\gamma}}).
\end{align*}
The constant terms on both sides are equal to 44.
The coefficient in front of $\frac{1}{w^\gamma}$ on the LHS is $12+\frac{32}{2^\gamma}+\frac{12}{3^\gamma}+\frac{32}{4^\gamma}$, while the same coefficient for the RHS is $14+\frac{20}{2^\gamma}+\frac{30}{3^\gamma}+\frac{24}{4^\gamma}$.
The difference $12+\frac{32}{2^\gamma}+\frac{12}{3^\gamma}+\frac{32}{4^\gamma}-(14+\frac{20}{2^\gamma}+\frac{30}{3^\gamma}+\frac{24}{4^\gamma})$ can be verified to be non-negative over $\gamma\in(0,1]$.
Similarly, the coefficient in front of $\frac{1}{w^{2\gamma}}$ on the LHS is $\frac{12}{3^\gamma}+\frac{32}{2^\gamma4^\gamma}$, while the same coefficient for the RHS is $\frac{12}{4^\gamma}+\frac{18}{2^\gamma3^\gamma}+\frac{2}{2^\gamma}+\frac{12}{3^\gamma4^\gamma}$.
The difference $\frac{12}{3^\gamma}+\frac{32}{2^\gamma4^\gamma}-(\frac{12}{4^\gamma}+\frac{18}{2^\gamma3^\gamma}+\frac{2}{2^\gamma}+\frac{12}{3^\gamma4^\gamma})$ can be verified to be non-negative over $\gamma\in(0,1]$.
All in all, since the LHS has a higher coefficient for all terms $1,\frac{1}{w^\gamma},\frac{1}{w^{2\gamma}}$, which are non-negative since $w>0$, this completes the proof that the desired inequalities~\eqref{eqn:qMonProperty} indeed hold in the special case of $n=4$.

\textbf{Completing the verification of history-monotone futures.}
Denote the set of items using $N=\{A,B,C,D\}$.
To verify history-monotone futures for list distributions on 4 items, we consider two possibilities: $|S|=1$ and $|S|=2$.
(If $|S|\ge 3$, then there cannot be two distinct prefixes in $N\setminus S$ with the same endpoint, and hence the condition holds vacuously.)
If $|S|=1$, then WOLOG let $S=\{A\},\rho_{\echar}=B$.  The possible prefixes in $N\setminus\{A\}$ with $\rho_{\echar}=B$ are $(B),(CB),(DB),(CDB),(DCB)$.
We will show for $j=A$ that
\begin{align} \label{eqn:1235}
\Pr[j\succeq S_0|(B)]\le\Pr[j\succeq S_0|(CB)]=\Pr[j\succeq S_0|(DB)]\le\Pr[j\succeq S_0|(CDB)]=\Pr[j\succeq S_0|(DCB)].
\end{align}
On the other hand, if $|S|=2$, then WOLOG let $S=\{A,B\},\rho_{\echar}=C$.
The possible prefixes in $N\setminus S$ with $\rho_{\echar}=C$ are $(C)$ and $(DC)$.
We will show for $j=A$ (the $j=B$ case holds symmetrically) that
\begin{align} \label{eqn:2089}
\Pr[j\succeq S_0|(C)]\le\Pr[j\succeq S_0|(DC)].
\end{align}

To show~\eqref{eqn:1235}, substituting $j=A$ and $S=\{A\}$, note that
$\Pr[A\succeq \{A\}_0|(CDB)]=\Pr[A\succeq \{A\}_0|(DCB)]=q_4$.
Meanwhile, $\Pr[A\succeq \{A\}_0|(CB)]=q_3+q_3q_4$, because starting from prefix $(CB)$, either $A$ can be visited right away, or it can be visited after $D$.
By symmetry, we also have $\Pr[A\succeq \{A\}_0|(DB)]=q_3+q_3q_4$.
Note that
\begin{align*}
q_3+q_3q_4=(1-q_3)\frac{1}{1/q_3-1}+q_3\frac{1}{1/q_4}
\end{align*}
which is indeed at most $q_4$ since it is a convex combination of $q_4$ with a probability that is smaller, because $1/q_3-1\ge 1/q_4$ according to~\eqref{eqn:qMonProperty}.
Similarly, $\Pr[A\succeq \{A\}_0|(B)]=q_2+2q_2(q_3+q_3q_4)$, because starting from prefix $(B)$, either $A$ can be visited right away, or $C$ or $D$ could be visited next both of which imply the probability of event $A\succeq \{A\}_0$ is $q_3+q_3q_4$.
Note that
\begin{align*}
q_2+2q_2(q_3+q_3q_4)=(1-2q_2)\frac{1}{1/q_2-2}+2q_2(q_3+q_3q_4)
\end{align*}
which is indeed at most $q_3+q_3q_4$ since it is a convex combination of $q_3+q_3q_4$ with a probability that is smaller, because $1/q_2-2\ge1/q_3-1$ according to~\eqref{eqn:qMonProperty}.
This establishes all of the desired relationships in~\eqref{eqn:1235}.

To show~\eqref{eqn:2089}, substituting $j=A$ and $S=\{A,B\}$, note that $\Pr[A\succeq\{A,B\}_0|(DC)]=q_3$ while $\Pr[A\succeq\{A,B\}_0|(C)]=q_2+q_2q_3=(1-q_2)\frac{1}{1/q_2-1}+q_2q_3$.  Again we use the fact that $1/q_2-1\ge1/q_3$ according to~\eqref{eqn:qMonProperty}, which shows $\Pr[A\succeq\{A,B\}_0|(C)]\le\Pr[A\succeq\{A,B\}_0|(DC)]$.
This completes the proof that the list distribution defined has history-monotone futures.
\Halmos\endproof

\section{Proofs from \Cref{sec:generalLists}} \label{sec:defPfsGeneral}

\proof{Proof of \Cref{prop:subm}.}
Fix any monotone submodular function $f:2^N\to[0,1]$, and define $x$ according to the statement of the \namecref{prop:subm}.
We must show that constraints~\eqref{constr:IC}--\eqref{constr:nonNeg} for IC mechanisms are satisfied.
The non-negativity constraints~\eqref{constr:nonNeg} follow from $f$ being increasing, and the constraints~\eqref{constr:sumToOne} follow from $f$ having a range of at most 1.
To establish the IC constraints~\eqref{constr:IC}, take lists $\ell,\ell'$ and index $k\le|\ell|$.
By the definition of $x$, the RHS of~\eqref{constr:IC} can be expressed as
\begin{align}\label{eqn:5289}
\big(f(\ell_{1:i_1})-f(\ell_{1:i_1-1})\big)+\big(f(\ell_{1:i_2})-f(\ell_{1:i_2-1})\big)+\cdots+\big(f(\ell_{1:i_{k'}})-f(\ell_{1:i_{k'}-1})\big)
\end{align}
where $i_1<\ldots<i_{k'}$ are the indices $i$ such that $\ell'_i\in\ell_{1:k}$, with $k'\le k$.
Applying the condition of submodularity,~\eqref{eqn:5289} can be upper-bounded by
\begin{align*}
\big(f(\ell_{i_1})-f(\emptyset)\big)+\big(f(\{\ell_{i_1},\ell_{i_2}\})-f(\ell_{i_1})\big)+\cdots
+\big(f(\{i_1,\ldots,i_{k'}\})-f(\{i_1,\ldots,i_{k'-1}\})\big)
&=f(\{i_1,\ldots,i_{k'}\})-f(\emptyset)
\\ &\le f(\ell_{1:k})-f(\emptyset)
\end{align*}
where $f(\{i_1,\ldots,i_{k'}\})\le f(\ell_{1:k})$ because the indices $i_1,\ldots,i_{k'}$ are contained within $\ell_{1:k}$ and $f$ is increasing.
The final expression equals the LHS of~\eqref{constr:IC}, completing the proof of the forward direction.

For the converse direction, only the submodularity of $f$ has not already been established in \Cref{prop:x_to_f}.
To establish submodularity, it suffices to show that $f(S)+f(S\cup\{j,j'\})\le f(S\cup\{j\})+f(S\cup\{j'\})$ for all subsets $S\subseteq N$ and distinct items $j,j'\in N\setminus S$.
So take any $S\subseteq N$ and distinct items $j,j'\in N\setminus S$.
Let $\ell$ denote the list consisting of the items in $S$ in any fixed order, followed by $j$.
Let $\ell'$ denote the list consisting of the items in $S$ in the same fixed order, followed by $j'$, and then $j$.
If $\ell$ reports truthfully, then their probability of receiving an item in $S\cup\{j\}$ is $f(S\cup\{j\})$, by identity~\eqref{eqn:diffF} which was established in \Cref{prop:x_to_f}.
On the other hand, if $\ell$ misreports $\ell'$, then their probability of receiving an item in $S\cup\{j\}$ is $f(S)+x_{j}(\ell')=f(S)+f(S\cup\{j',j\})-f(S\cup\{j'\})$, again making use of identity~\eqref{eqn:diffF}.
Substituting $k=|S|+1$ into the IC constraints~\eqref{constr:IC}, we see that the former probability must be lower-bounded by the latter, i.e.\ $f(S\cup\{j\})\ge f(S)+f(S\cup\{j',j\})-f(S\cup\{j'\})$, which suffices for establishing submodularity.
\Halmos\endproof

\proof{Proof of \Cref{thm:topKGap}.}
It is immediate to see that the revenue of a top-2 lottery is
\begin{align*}
\sum_{j=1}^n M^{-j}\cdot\frac{O(M^{j-1})+M^j}{2}
&=\sum_{j=1}^n(1/2+O(M^{-1}))
\end{align*}
which is at least $n/2$.
On the other hand, consider any assortment $S$ with $|S|=k$, for some $k\in\{1,\ldots,n\}$.
Conditional on the most expensive item on the buyer's list being $j$ for some $j$, if $j\in S$, then the assortment will sell item $j$ to the buyer as long as their first choice does not also lie in $S$, which occurs w.p.~$1-\frac{|S\cap\{1,\ldots,j-1\}|}{j-1}$.
Therefore, the revenue of assortment $S$ equals
\begin{align*}
\sum_{j=1}^n M^{-j}\left(O(M^{j-1})+M^j\bI(j\in S)(1-\frac{|S\cap\{1,\ldots,j-1\}|}{j-1})\right)
&=O(1/M)+\sum_{j\in S}(1-\frac{|S\cap\{1,\ldots,j-1\}|}{j-1})
\\ &\le O(1/M)+\sum_{k'=1}^k(1-\frac{k-k'}{n-k'})
\\ &= O(1/M)+(n-k)\sum_{k'=1}^k\frac{1}{n-k'}.
\end{align*}
We explain the inequality.  For all $k'=1,\ldots,k$, if $j$ is the $k'$'th-most-expensive item in $S$, then there must be exactly $k-k'$ elements in $|S\cap\{1,\ldots,j-1\}|$.  Meanwhile, the maximum possible index $j$ could be is $n+1-k'$, and hence $j-1\le n-k'$.

It remains to upper-bound the quantity $(n-k)\sum_{k'=1}^k1/(n-k')$ over all $k=1,\ldots,n$.
If $k<n-1$, then we upper-bound the sum by $\log\frac{n-1}{n-k-1}$, in which case the assortment revenue is at most
\begin{align*}
O(1/M)+(n-k)\log\frac{n-1}{n-k-1}
&=O(1/M)+\left(n(1-\frac{k}{n-1})+\frac{k}{n-1}\right)\log\frac{n-1}{n-k-1}
\\ &=O(1/M)+n(1-\frac{k}{n-1})\log\frac{1}{1-\frac{k}{n-1}}+O(\log n)
\\ &\le n/e+O(\log n)
\end{align*}
where the inequality holds because the maximum of function $(1-a)\log\frac{1}{1-a}$ over $a\in[0,1)$ is $1/e$, occurring at $a=1-1/e$.
On the other hand, if $k\ge n-1$, then it is easy to see that the assortment revenue cannot be better than $O(\log n)$.
Therefore, assortments can obtain no more than $\frac{n/e+O(\log n)}{n/2}\approx2/e$ times the revenue of the top-2 lottery as $n\to\infty$, completing the proof.
\Halmos\endproof

\proof{Proof of \Cref{thm:bdedListLength}.}
Given the mechanism $x$, define function $f$ as in \Cref{prop:x_to_f}.
Following the same reasoning as in the proof of \Cref{thm:budgetAdditive}, $f(\ell_{1:k})-f(\ell_{1:k-1})$ is the probability of the mechanism selling item $\ell_k$ to type $\ell$.
Therefore, it suffices to show that the random assortment sells each type $\ell$ each item $\ell_k$ with probability at least $\frac{2}{eL}(f(\ell_{1:k})-f(\ell_{1:k-1}))$, for $\ell\in\cL$ and $k\le|\ell|$.

The probability that the random assortment sells item $\ell_k$ to type $\ell$ is $\varphi(f_{\ell_k})\prod_{k'=1}^{k-1}(1-\varphi(f_{\ell_{k'}}))$.
If $p(\ell)>0$ and $f(\ell_{1:k})-f(\ell_{1:k-1})>0$, then we can establish the inequalities
\begin{align*}
\frac{\varphi(f_{\ell_k})\prod_{k'=1}^{k-1}(1-\varphi(f_{\ell_{k'}}))}{f(\ell_{1:k})-f(\ell_{1:k-1})}
&\ge\frac{\varphi(f(\ell_{1:k})-f(\ell_{1:k-1}))\prod_{k'=1}^{k-1}(1-\varphi(f(\ell_{1:k-1})))}{f(\ell_{1:k})-f(\ell_{1:k-1})}
\\ &\ge\frac{\varphi(1-f(\ell_{1:k-1}))}{1-f(\ell_{1:k-1})}(1-\varphi(f(\ell_{1:k-1})))^{ L-1}.
\end{align*}
Indeed, the first inequality holds because: function $\varphi$ is non-decreasing; $f_{\ell_k}=f(\ell_k)\ge f(\ell_{1:k})-f(\ell_{1:k-1})$ since $f$ is submodular (by \Cref{prop:subm}); and $f(\ell_{k'})\le f(\ell_{1:k-1})$ for all $k'=1,\ldots,k-1$ since $f$ is increasing.
Subsequently, the second inequality holds because: the function $\varphi(a)/a$ is non-increasing in $a$; $f(\ell_{1:k})\le1$; and $k\le|\ell|\le L$ (since $p(\ell)>0$).

Now, let $a=f(\ell_{1:k-1})$ and we claim that the expression $\frac{\varphi(1-a)}{1-a}(1-\varphi(a))^{ L-1}$ is at least $\frac{2}{ L}(1-\frac{1}{ L})^{ L-1}$ for all $a\in[0,1)$.
To verify this, one must check two cases: $a\le1/2$ and $a>1/2$.
In the first,
\begin{align*}
\frac{\varphi(1-a)}{1-a}(1-\varphi(a))^{ L-1}=\frac{1}{ L\cdot(1-a)}(1-\frac{2a}{ L})^{ L-1}
\end{align*}
whose RHS can be seen to be a decreasing function over $a\in[0,1/2]$ for all integers $ L\ge2$.  Therefore, the expression is lower-bounded by its value at $a=1/2$, which is $\frac{2}{ L}(1-\frac{1}{ L})^{ L-1}$.
In the second case, $\frac{\varphi(1-a)}{1-a}(1-\varphi(a))^{ L-1}=\frac{2(1-a)/ L}{1-a}(1-\frac{1}{ L})^{ L-1}=\frac{2}{ L}(1-\frac{1}{ L})^{ L-1}$ for all $a\in(1/2,1)$.

We have shown that the randomized assortment sells each realizable type $\ell$ their $k$'th most-preferred item $\ell_k$ with probability at least $\frac{2}{ L}(1-\frac{1}{ L})^{ L-1}(f(\ell_{1:k})-f(\ell_{1:k-1}))$.
Since $(1-\frac{1}{ L})^{ L-1}$ is lower-bounded by $1/e$ for all values of $L$, this completes the proof of the theorem.
\Halmos\endproof

\end{APPENDICES}

% References here (outcomment the appropriate case)

% CASE 1: BiBTeX used to constantly update the references
%   (while the paper is being written).
%\bibliographystyle{informs2014} % outcomment this and next line in Case 1
%\bibliography{bibliography} % if more than one, comma separated

% CASE 2: BiBTeX used to generate mypaper.bbl (to be further fine tuned)
%\input{mypaper.bbl} % outcomment this line in Case 2

%If you don't use BiBTex, you can manually itemize references as shown below.

\end{document}